\documentclass{aa}  

\usepackage{graphicx}
\usepackage{txfonts}
\usepackage{natbib}
\usepackage{soul}
\usepackage{xcolor}
\usepackage{multicol}
\usepackage{graphicx,graphics,rotating,amssymb,amsmath}
\usepackage{xcolor}
\usepackage{multirow}
\usepackage{longtable}
\usepackage{booktabs}
\usepackage{subfig}
\usepackage{float,capt-of}
\usepackage{natbib}
\usepackage[english]{babel}
\usepackage{morefloats}
\usepackage{color}
\usepackage{booktabs}
\usepackage{sidecap}
\usepackage{epsfig,color}
\usepackage{wrapfig}
\usepackage{lscape}
\usepackage{ulem}
\usepackage{url}
\usepackage{tablefootnote}
\usepackage{txfonts}
\usepackage{hyperref}

\newcommand\kms{km~s$^{-1}$}
\newcommand\alphaco{$\alpha_{\rm CO}$}
\newcommand\co{CO(1-0)}

\newcommand{\xmm}{{\it {XMM-Newton}}}

\newcommand{\nhsym}{N_{\mbox{\scriptsize H}}}
\newcommand{\nustar}{{\it {NuSTAR}}}
\newcommand{\swift}{{\it {Swift}}}
\newcommand{\XMM}{{\it {XMM}}}
\newcommand{\errUD}[2]{\ensuremath{^{+#1}_{-#2}}}

\newcommand{\logxi}{erg\,cm\,s$^{-1}$}
\newcommand{\nh}{cm$^{-2}$}

\begin{document} 

\title{Another X-ray UFO without a momentum-boosted molecular outflow}
\subtitle{ALMA CO(1-0) observations of the galaxy pair IRAS~05054+1718}

\author{Francesca Bonanomi 
        \inst{1,2,3}\thanks{francesca.bonanomi@univie.ac.at}
        \and
        Claudia Cicone\inst{4}
        \and 
        Paola Severgnini\inst{2}
        \and 
        Valentina Braito\inst{5,6}
        \and 
        Cristian Vignali\inst{7,8}
        \and
        James N. Reeves \inst{6,5}
        \and
        Mattia Sirressi\inst{9}
        \and
        Isabel Montoya Arroyave\inst{4}
        \and
        Roberto Della Ceca\inst{2}
        \and
        Lucia Ballo\inst{10}
        \and
        Massimo Dotti\inst{3}}

   \institute{Department of Astrophysics, University of Vienna, T\"urkenschanzstrasse 17, 1180 Vienna, Austria 
        \and
        INAF - Osservatorio Astronomico di Brera, Via Brera 28, 20121 Milano, Italy 
   \and
        Department of Physics G. Occhialini, University of Milano-Bicocca, Piazza della Scienza 3, I-20126 Milan, Italy 
        \and
        Institute of Theoretical Astrophysics, University of Oslo, P.O. Box 1029, Blindern, 0315 Oslo, Norway 
        \and
        INAF, Osservatorio Astronomico di Brera, Via Bianchi 46 I-23807 Merate (LC), Italy
        \and
        Department of Physics, Institute for Astrophysics and Computational Sciences, The Catholic University of America, Washington, DC 20064, USA
        \and
        Dipartimento di Fisica e Astronomia ``Augusto Righi", Alma Mater Studiorum, Universit\`a degli Studi di Bologna, Via Gobetti 93/2, 40129 Bologna,
Italy
        \and 
        INAF-Osservatorio di Astrofisica e Scienza delle Spazio di Bologna, OAS, Via Gobetti 93/3, 40129 Bologna, Italy
        \and
        Department of Astronomy, Oskar Klein Centre, Stockholm University, AlbaNova University Centre, SE-106 91 Stockholm, Sweden 
    \and  European Space Astronomy Centre (ESA/ESAC), E-28691 Villanueva de la Canada, Madrid, Spain
             }

 \date{Received XX-XXX-20XX / Accepted: XX-XXX-20XX }
 
 \abstract{We present Atacama Large Millimetre/submillimetre Array (ALMA) \co~observations of the nearby infrared luminous (LIRG) galaxy pair IRAS~05054+1718 (also known as CGCG~468-002), as well as a new analysis of X-ray data of this source collected between 2012 and 2021 using the Nuclear Spectroscopic Telescope Array (NuSTAR), Swift, and the XMM-Newton satellites.
 The western component of the pair, NED01, hosts a Seyfert 1.9 nucleus that is responsible for launching a powerful X-ray ultra-fast outflow (UFO). Our X-ray spectral analysis suggests that the UFO could be variable or multi-component in velocity, ranging from $v/c\sim -0.12$ (as seen in \swift) to $v/c\sim -0.23$ (as seen in \nustar), and constrains its momentum flux to be $\dot p^{X-ray}_{out} \sim (4\pm2)\times 10^{34}$ g cm s$^{-2}$. 
The ALMA CO(1-0) observations, obtained with an angular resolution of $2.2\arcsec$,
although targeting mainly NED01, also include   the eastern component of the pair, NED02, a less-studied LIRG with no clear evidence of an active galactic nucleus (AGN). 
We study the CO(1-0) kinematics in the two galaxies using the 3D-BAROLO code. In both sources we can model the bulk of the CO(1-0) emission with rotating disks and, after subtracting the  best-fit models, we detect compact residual emission at S/N=15 within $\sim3$ kpc of the centre. A molecular outflow in NED01, if present, cannot be brighter than such residuals, implying an upper limit on its outflow rate of $\dot{M}^{mol}_{out} \lesssim 19\pm14~M_{\odot}~yr^{-1}$ and on its momentum rate of $\dot p^{mol}_{out} \lesssim (2.7\pm2.4) \times 10^{34}$~g~cm~s$^{-1}$. Combined with the revised energetics of the X-ray wind, we derive an upper limit on the momentum rate ratio of $\dot{p}^{mol}_{out}/\dot{p}^{X-ray}_{out}<0.67$. We discuss these results in the context of the expectations of AGN feedback models, and we propose that the X-ray disk wind in NED01 has not significantly impacted the molecular gas reservoir (yet), and we can constrain its effect to be much smaller than expectations of AGN `energy-driven' feedback models. We also consider and discuss the hypothesis of asymmetries of the molecular disk not properly captured by the 3D-BAROLO code.
Our results highlight the challenges in testing the predictions of popular AGN disk-wind feedback theories, even in the presence of good-quality multi-wavelength observations.}


\titlerunning{Another X-ray UFO without a prominent molecular outflow}
\authorrunning{F. Bonanomi, C. Cicone, P. Severgnini, V. Braito, C. Vignali et al.}
\maketitle


\section{Introduction}

Galaxy formation and evolution is a complex process involving several different physical phenomena acting simultaneously on different physical and temporal scales. Gas is a key player in this picture, feeding star formation and the accretion onto the central supermassive black hole (SMBH), and is in turn affected by feedback mechanisms. The feedback can manifest through powerful winds that are able to blow away the gas from the centre of the galaxy, quenching star formation and starving the BH of fuel \citep{2020Veilleux}. Active galactic nucleus (AGN) feedback processes play a fundamental role in galaxy growth and evolution;  they are thought to be at the origin of the M$_{\rm BH}-\sigma_\star$ relation \citep{2010King,1998Silk-Rees} and to prevent the overgrowth of massive galaxies \citep{2012Bower,2014Hopkins}.

In the hard X-ray spectrum, hot (T$\sim10^6-10^7$ K) ultra-fast outflows (UFOs) have been observed in $\sim$40\% of the bright nearby local AGN population \citep{2010Tombesi,Tombesi2012,Gofford2013,Gofford2015}. These winds, developed from the AGN accretion disk ($\le$ 1 pc), are observed through the detection of blueshifted (velocities up to $v\sim 0.3$c) absorption lines associated with  highly ionised iron  transitions in the hard X-ray spectrum \citep{2003Reeves,2010Tombesi}.

Massive galaxy-scale cold ($T_{kin}\sim10-100$~K) molecular outflows with velocities between hundreds of  and a few thousand  \kms{} have been observed in the last decade \citep{2010Feruglio, 2010Fisher}. These winds can be detected by P Cygni profiles of the OH molecule in the far-IR regime \citep{2010Fisher,2010Feruglio,2011Sturm,2013Veilleux} as well as blue- and redshifted high-velocity wings in the CO, HCN, or HCO$^+$ profiles using interferometric observations in the millimetre band \citep{2012Aalto,2012Cicone,2014Cicone}.

The theoretical model that is usually invoked to explain large-scale outflows launched by AGNs is the blast-wave scenario \citep{1998Silk-Rees,2010King,2012Faucher-Giguere}. According to this model, a nuclear wind arises from the accretion disk of an AGN and impacts on the interstellar medium (ISM) producing a forward shock and a reverse shock. The forward shock propagates through the unperturbed ISM producing a large-scale outflow. This outflow could be either energy- or momentum-driven, depending on whether cooling of the reverse shock is efficient. If it is, the energy is conserved and outflow propagates adiabatically (energy-driven), showing a momentum boost with respect to the X-ray wind 
($\dot{p}_{out}^{\rm ISM}/\dot{p}_{out}^{\rm X-ray}>>1$; \citealt{2012Faucher-Giguere}).
Otherwise, the energy is radiated away and only momentum is transferred to the ISM 
($\dot{p}_{out}^{\rm ISM}/\dot{p}_{out}^{\rm X-ray}\sim1$; \citealt{2010King}).

Which model is most favoured by observations is a highly debated question. Simultaneous observations of X-ray winds and large-scale outflows are needed to test their predictions. The momentum rate versus  wind velocity diagram is a widely used tool to visualise and compare the properties of different outflows \citep{2015Feruglio,2015Tombesi}. \citet{2019Smith} recently summarised the momentum rate versus the wind velocity for a sample of  ten objects with observed X-ray UFOs and large-scale galactic outflows (see also Fig.~\ref{fig:smithplot}, this work). Some of these sources, such as the luminous quasar PDS~456 \citep{2019Bischetti} and the ultra-luminous infrared galaxy (ULIRG\footnote{ULIRGs are galaxies that are extremely bright in the infrared, i.e. L$_{\rm IR}\ge10^{12}$ L$_\odot$, \citep{2003Sanders}.}) IRAS F11119+3257 \citep{2015Tombesi,2017Veilleux}, seem to favour the momentum-driven scenario, while other objects, such  as the ULIRG Mrk 231 \citep{2015Feruglio} and the Seyfert 1 galaxy IRAS 17020+4544 \citep{2015Longinotti, 2018Longinotti}, show large-scale outflows whose momentum rate is boosted compared to the X-ray wind. Finally, other sources, such as the multiple-lensed quasar SDSS~J1353+1138 \citep{2021Tozzi}, do not appear to favour either  of the   AGN feedback models. Overall, as clearly drawn by \citet{2019Smith}, the picture is much more complex than expected from AGN blast-wave feedback models.

Furthermore, most sources in the sample explored by \citet{2019Smith} are ULIRGs. These objects have an intense star formation, whose contribution to feedback processes is hard to distinguish from the AGN contribution. Testing the prediction of the blast-wave scenario in galaxies with a more moderate star formation activity is necessary to overcome this issue. The work by \citet{2019Sirressi} on the local Seyfert 2 galaxy  MCG-03-58-007, with a star formation rate (SFR) $\sim$20 M$_\odot$ yr$^{-1}$, was a first step towards this direction. These authors detected a compact H$_2$ component that, if interpreted as an outflow, would present a momentum rate equal to $\sim$40\% of that of the X-ray UFO. Our study on the LIRG and galaxy pair IRAS~05054+1718 (also known as CGCG 468-002) also fits into this context. The main target of this work is the western component of the pair (hereafter NED01), a local LIRG hosting a Seyfert 1.9 nucleus. The source shows a moderate SFR of 5-10 M $_\odot$ yr$^{-1}$ \citep{2014Delooze,2015Pereira} and hosts a powerful X-ray wind \citep{2015Ballo}, being a suitable candidate to test the AGN feedback scenario reducing the possible contamination from star formation-driven outflows.

Our aim is to investigate the presence of a large-scale molecular outflow in NED01 and in the companion NED02, by studying the distribution and the kinematics of the molecular gas. The latter is the phase of the ISM that is most tightly connected to star formation \citep{2002Wong,2008Bigiel,2008Leroy}, as stars form primarily in molecular clouds   \citep[e.g.][]{2010Lada,2014Andre}. 
We use  carbon monoxide (CO) as a tracer of cold molecular hydrogen ($H_2$), because it is the second most abundant molecule after $H_2$, and its rotational transitions are easily observable at submillimetre--millimetre wavelengths. The lowest J levels of CO can be easily excited at molecular cloud temperatures \citep[$T\sim10$ K,][]{2007Omont} and so the CO(1-0) and CO(2-1) lines can be used to estimate the total molecular gas mass in galaxies. 
Since these lines are optically thick at typical molecular cloud conditions, their luminosity is not proportional to the H$_2$ gas column density, but   a \co$-$to$-H_2$ conversion factor (\alphaco) needs to be assumed. The estimate of this value is not straightforward as it depends on the physical state of the gas and needs to be calibrated using multiple molecular line tracers, which are often difficult to detect in extragalactic sources. For the Milky Way ISM and for normal  star-forming galaxies, a value of \alphaco = 4.3~M$_\odot$ (K km s$^{-1}$ pc$^2$) is widely accepted \citep{2013Bolatto}. For different ISM environments, such as the massive molecular outflows discovered in local starbursts and AGN host galaxies \citep[see review by][]{2020Veilleux}, the \alphaco~parameter is very poorly constrained. In this work we   assume that molecular outflows have an  \alphaco=2.1$\pm$1.2 M$_\odot$ (K km s$^{-1}$ pc$^2$), which is the value measured by \citet{2018Cicone2} on the molecular outflow of the well-studied local ULIRG NGC~6240. We  use \alphaco=4.3$\pm$1.3 M$_\odot$ (K km s$^{-1}$ pc$^2$), a value typically assumed when treating the molecular ISM of isolated galaxies like the Milky Way \citep{2013Bolatto}, to evaluate the molecular mass of the galaxy disk.
 
This paper is organised as follows. The selected targets are presented in Sect.~\ref{sec:targets}. In Sect.~\ref{sec:observations} we describe the new high-sensitivity Atacama Large Millimetre/submillimetre Array (ALMA) \co{} observations used in this work. The analysis performed on the data is reported in Sect.~\ref{sec:analysis}.
In Sect.~\ref{sec:kinematics} we model the kinematics of the \co{} emission using the  3D-Based Analysis of Rotating Object via Line Observations (3D-BAROLO) software for the two targets. In Sects.~\ref{sec:X-ray} and \ref{sec:X-ray_analysis} we present the observations and the analysis of the new X-ray  datasets, and in Sect.~\ref{sec:X-ray_energetics} we derive the energetics of the X-ray wind. The interpretation of the results is discussed in Sect.~\ref{sec:discussion}, where we test different hypothesis in the AGN-driven feedback scenario. In Sect.~\ref{sec:summary} we summarise our results and conclusions.

Throughout the paper we adopt a standard $\Lambda$CDM cosmological model with $H_0=67.8$~\kms~Mpc$^{-1}$, $\Omega_{\Lambda}=0.692$, and  $\Omega_M=0.308$ \citep{2016PlanckCollaboration}. At the distance of IRAS~05054+1718 NED01 ($z=0.0178$, revised in this work), the physical scale is 0.373~kpc~arcsec$^{-1}$. 

\section{Target description}\label{sec:targets}

\begin{figure}
        \centering
        \includegraphics[clip=true,trim=0.8cm 0.6cm 0.8cm 0.6cm,width=\columnwidth]{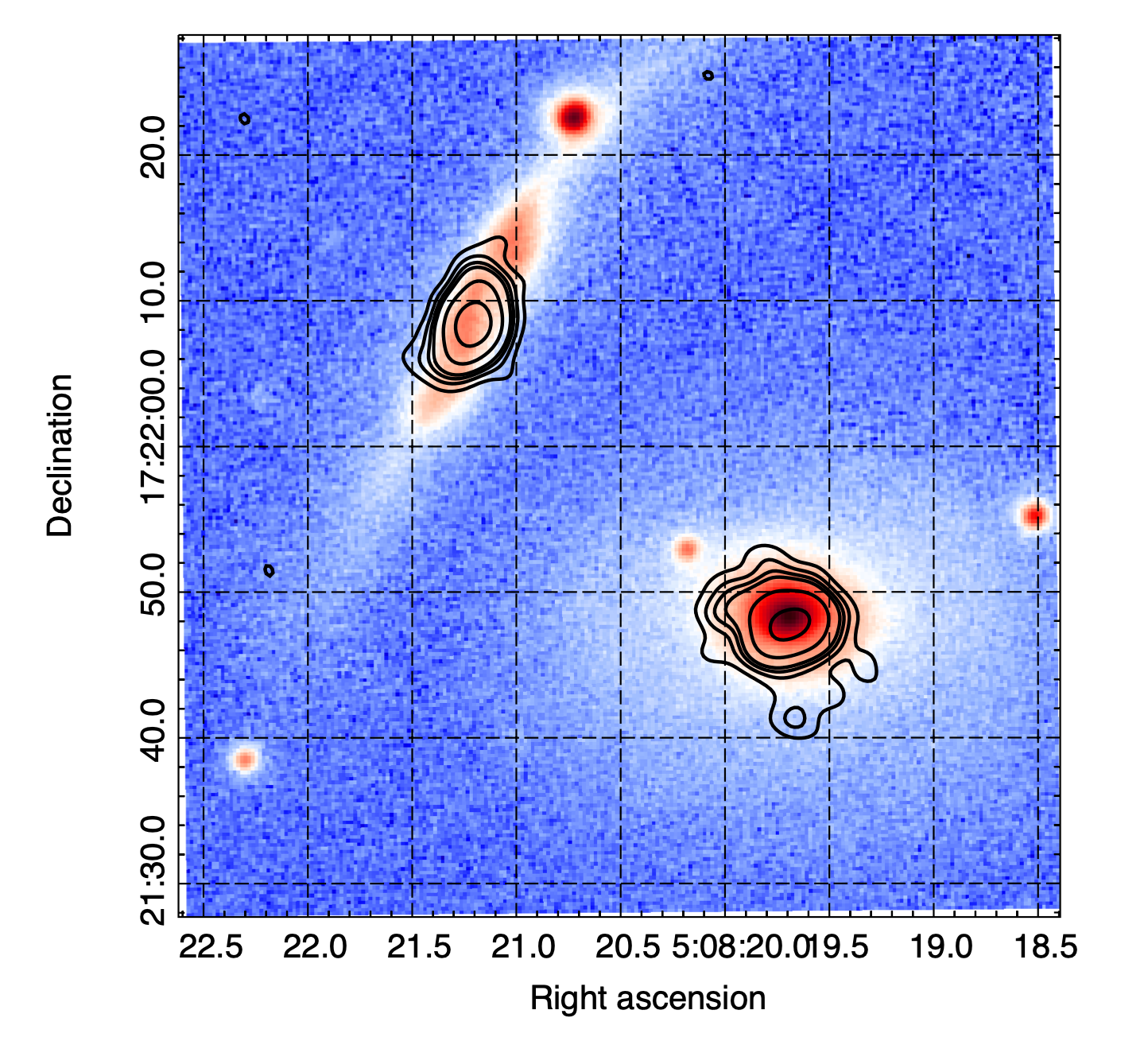}\\
        \caption{ALMA \co~map (black contours) overlayed onto the g-band optical image from the Pan-STARR Survey~1. The ALMA \co~emission was averaged over a spectral range corresponding to $v=[-490, +230]$~\kms\ with respect to the systemic redshift of NED01, which includes the \co~ emission from both members of the galaxy pair. The contours correspond to the (3, 6 , 9, 12, 24, 50)$\times\sigma_{RMS}$ levels, with $\sigma_{RMS}= 0.2$ mJy~beam$^{-1}$ being the average rms of the ALMA \co~map (not corrected for the primary beam).}
        \label{fig:source}
\end{figure}

The western and eastern pair members, in this work indicated respectively as NED01 and NED02 (see also \citealt{2015Pereira}), have a projected distance of $\sim29.6''\sim11$ kpc. Figure~\ref{fig:source} shows the ALMA \co~contours overlayed onto a g-band image from the Pan-STARR Survey 1 \citep{Panstar}. According to \citet{2013Stierwalt}, the system is in an early merger stage after a first encounter between the two galaxies since their disks are still symmetric, but show signs of tidal tales. 

The western galaxy NED01, at $z=0.0178\pm0.0004$\footnote{CO-based redshift measured in this work, see Sect.~\ref{sec:z-calc}}, hosts a Seyfert 1.9 nucleus, and it is classified as a LIRG 
($\rm log(L_{IR(8-1000\mu m)}/L_\odot$)=10.6, \citealt{2015Pereira}).
Because of the presence of the AGN, its SFR is not well constrained in the literature, with values ranging between 5 M$_{\rm \odot}$ yr$^{-1}$ \citep{2014Delooze,2015Pereira} and $\sim$10 M$_{\rm \odot}$ yr$^{-1}$ \citep{2010Howell}. Based on the ratio of the SFR to the BH accretion rate ($\rm \log(SFR/\dot{m}_{BH})\sim2$), obtained from the [Ne{\sc ii}]15.56~$\mu m$ and [O{\sc iii}]$\lambda5007$ gas velocity dispersion, the stellar velocity dispersion, and the 8-1000~$\mu m$ IR-luminosity, \cite{2013Alonso-Herrero} suggested that NED01 is transitioning from a H{\sc ii}-dominated to a Seyfert-dominated LIRG.

NED01 represents an interesting case study for the effects of AGN feedback on galaxies.  \citet{2015Ballo} detected a deep absorption trough at $E\sim7.8$ keV (2.1$\sigma$ significance) in its {\it Swift}-XRT (X-ray telescope) spectrum, which has been interpreted as a  highly ionised (log$\xi\sim3$ erg cm$^{-2}$ s$^{-1}$),
high column density ($N_{\rm H}\sim10^{23}$ cm$^{-2}$), 
and ultra-fast ($v_{\rm out}=(0.11\pm0.03) c$) disk wind.

The companion NED02 is also a LIRG ($L_{IR}=10^{11}L_\odot$, \citealt{2015Pereira}) and has a measured redshift of $z=0.016812\pm0.000003.$\footnote{CO-based redshift, measured in this work, see Sect.~\ref{sec:z-calc}.} NED02 was classified as a composite galaxy according to the BPT classification by \citet{2015Pereira}, but no evidence for AGN emission has been detected 
to date \citep[see e.g.][]{2012Alonso-Herrero}.
The SFR estimates for this galaxy range between SFR$_{\rm (1-10) Myr}\sim15$ M$_{\rm \odot}$ yr$^{-1}$ \citep{2015Pereira} and SFR$\sim20$  M$_{\rm \odot}$ yr$^{-1}$ \citep{2010Howell,2014Delooze}.

\section{ALMA CO(1-0) observations}\label{sec:observations}

The ALMA Band~3 (84.0-116.0 GHz) observations of IRAS 05054+1718 were carried out in Cycle 5 (Project code: 2016.1.00694.S, PI: P. Severgnini). The primary target was NED01 (corresponding to the phase centre of the interferometric dataset), but the field of view and spectral bandwidth of the data also cover  the CO(1-0) emission from the companion NED02, and so we include the latter in our analysis. We only use  data from the two scheduling blocks that have passed the ALMA data quality assurance (QA0), which are also the only datasets that were delivered to the PI, with observing dates 5 and 6 March 2017. According to the QA0 report, the total observing time including overheads for the two combined valid execution blocks was 100 min, and the total time on target was 62 min. The 40 ALMA   12m antennas were arranged in the most compact configuration (C40-1), with baselines ranging from 14 m to 310 m. The precipitable water vapour (PWV) varied from 3mm to 8mm, wind speed was 3.3~m~s$^{-1}$, and  humidity $\sim50$\%. 
The quasar J0423-0120 was used for flux calibration, J0510+1800 was instead used for band-pass response, phase calibration, and pointing, and both sources in addition to the main target were used for atmospheric calibration and radiometric phase correction.

We employed four spectral windows, two for each side band of the ALMA correlator. Two adjacent high-resolution, 1.875 GHz wide spectral windows (960 channels each, channel width of 1953.13 kHz, corresponding to 5.2~\kms) were centred at sky frequencies of $113.179$~GHz and $111.438$~GHz in order to sample both the \co~line and the $N=1$ spin-doublet transition  of CN, which have rest-frame frequencies of $\nu^{rest}_{CO(1-0)}=115.2712$~GHz and $\nu^{rest}_{CN(1-0)}\simeq113.4910$~GHz.\footnote{Frequency of the expected brightest component of the CN(1-0) line group, see also \citet{2020Cicone}.} Two additional low-resolution 2~GHz wide spectral windows (128 channels, 15.625 MHz channel width, corresponding to $\sim50$~\kms) were centred at sky frequencies of 101.190~GHz and 99.387~GHz to probe the 3~mm continuum. 
In this work, we focus on the \co~line data, and we postpone the analysis of the CN(1-0) line to a future publication. Through  the spectral line modelling described in Sect.~\ref{sec:z-calc}, we found the \co~lines of NED01 and NED02 to be respectively centred at (sky) frequencies of $\nu=113.2578$~GHz and $\nu=113.3651$~GHz, which we used to compute new estimates of the systemic redshift of the two galaxies. Except for Sect.~\ref{sec:z-calc}, where we worked with the initial datacubes not corrected for the right redshift, the rest of the analysis presented in this paper was performed on two separate datacubes (one for NED01 and one for NED02) corrected for their new CO-based redshift estimates. 

The data were calibrated by running the version 5.4.0 of the Common Astronomy Software Applications (CASA) package calibration pipeline \citep{CASA}. For the cleaning and other analysis steps we used CASA software version 5.6.1-8.
We combined the measurement sets of the two execution blocks by using the task \texttt{concat}, after having pre-selected with \texttt{split} the \co~line spectral windows relevant to our target. An analysis of the continuum at $100$~GHz, conducted using the two line-free spectral windows (see further details in Appendix~\ref{sec:appendix_continuum}), shows a clear detection of both NED01 and NED02, with respective continuum peak flux densities equal to $S_{cont}^{NED01} = 0.71\pm0.03$~mJy~beam$^{-1}$ and $S_{cont}^{NED02} = 2.26\pm0.05$~mJy~beam$^{-1}$. For this reason, before proceeding with the analysis of the CO kinematics, we subtracted the continuum from the CO(1-0) spectral windows in the {\it uv} visibility plane using the task \texttt{uvcontsub}. We selected a zeroth-order polynomial fit to the continuum channels adjacent to the CO(1-0) line in the 112.26-113.0~GHz and 113.57-114.12~GHz sky frequency ranges (corresponding to $v\in[-2650,-690]$~\kms\ and $v\in[820, 2280]$~\kms\ with respect to the \co~line centre). We then worked exclusively on the continuum-subtracted CO(1-0) line data.

The cleaning procedure for modelling the true sky brightness distribution of the source out of the {\it uv} visibility data was performed using the task \texttt{tclean}, by selecting the automasking algorithm (\texttt{auto-multithresh} parameter), which creates a different mask for every channel, minimising negative sidelobes. We used Briggs weighting with the robust parameter set equal to zero, and a cell size of $0.2''$. 
The synthesised beam size of the resulting cleaned datacube changes slightly with spectral channels, with a median value of $2.59''\times2.25''$, corresponding to an average spatial resolution of $\rm 0.97~kpc\times0.84~kpc$. 
We adopted the native spectral resolution of 5.17~\kms\ for the channel size. We selected a cleaning threshold equal to our first estimate of the average line rms of $2.5$~mJy~beam$^{-1}$ per channel. In order to account for the bias of the primary beam (PB) pattern on the image, we divided the cleaned datacube by the PB response using the task \texttt{impbcor}. 

The mean rms \co~line sensitivity for source NED01 as measured in the cleaned and PB-corrected cube is 2.2~mJy~beam$^{-1}$ per 5.2~\kms~channel. This value was calculated with the task \texttt{imstat}, by selecting the central $40''$ portion of the field of view, and so it is adequate to characterise the noise fluctuations of the \co~data for NED01. We   also verified that the noise follows a Gaussian distribution, so we adopted the mean rms value across the whole \co~spectral range on which our analysis is focused.
For the companion NED02 instead, given its proximity to the edge of the PB's FWHM, the \co~line sensitivity is lower, with an average 1$\sigma$ rms value of 4.0~mJy~beam$^{-1}$ per 5.2~\kms~channel. For this source, the 3D-BAROLO kinematic analysis (presented in Sect.~\ref{sec:NED02_BBarolo}) will be conducted on a portion of the datacube centred on NED02 and not corrected for the PB.
Table~\ref{tab:obs-feature} summarises the main observational parameters.

\begin{table}[tbp]
        \centering 
        \caption{Description of observations}
        \small
        \label{tab:obs-feature}      
        \begin{tabular}{ll}        
        \toprule
         Instrument & ALMA Band~3 \\
        Target  IDs & IRAS05054+1718 (pair) \\
                                                & CGCG468-002 NED01 and NED02 \\
        RA, Dec (ICRS) & 05:08:19.700, +17:21:48.100\tablefootnote{Coordinates of ALMA observations, centred on the component NED01.} \\
        $\nu_{\rm rest}^{\rm CO(1-0)}$ &   115.2712 GHz \\
    Spectral resolution & 1953.125 kHz = 5.17~\kms \\
    Angular resolution  & $2.59''\times2.25''$ \\
    Max recoverable scale & $28''$ \\
    Field of view & $55''\times 55''$ \\
    $1\sigma$ rms (5.2~\kms) & 2.2 mJy~beam$^{-1}$ (NED01) \\
    $1\sigma$ rms (5.2~\kms) & 4.0 mJy~beam$^{-1}$ (NED02) \\
\midrule                       
\end{tabular}
\end{table}

\section{Analysis of the CO(1-0) line emission}\label{sec:analysis}

\subsection{CO-based redshift estimates}\label{sec:z-calc}

\begin{figure*}[tbp]
        \centering
        \includegraphics[width=.9\columnwidth]{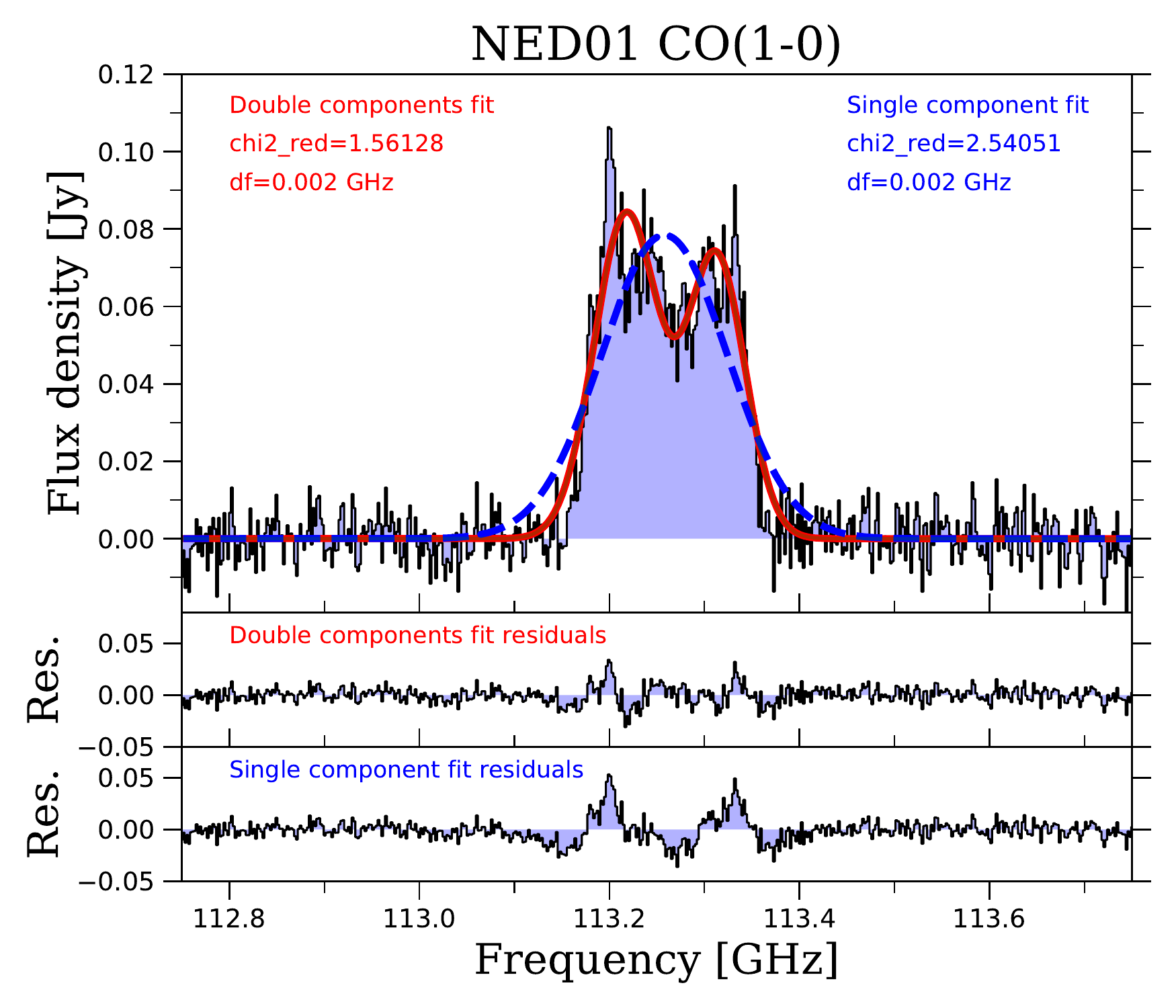}\quad
        \includegraphics[width=.9\columnwidth]{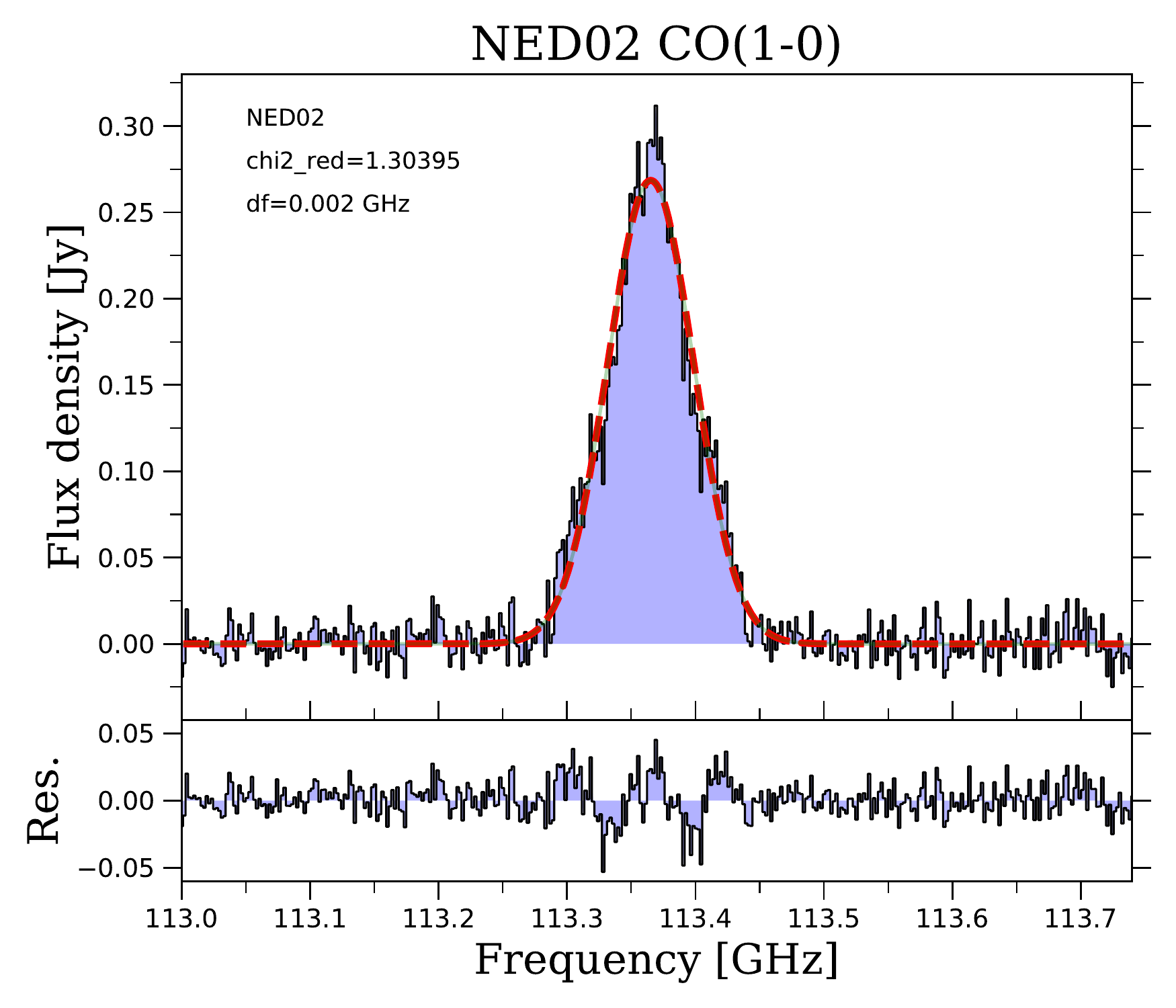}\\
        \includegraphics[width=.9\columnwidth]{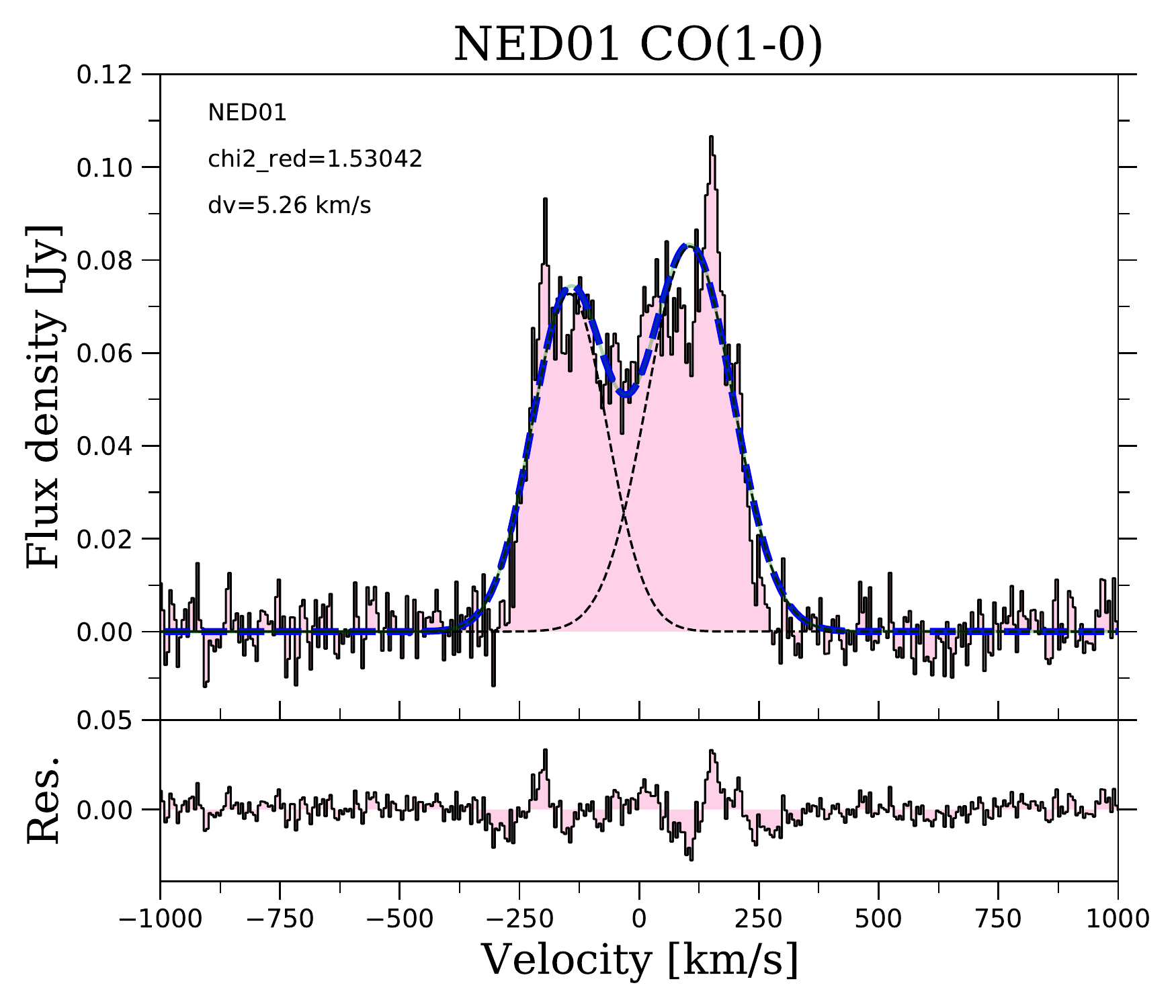}\quad
        \includegraphics[width=.9\columnwidth]{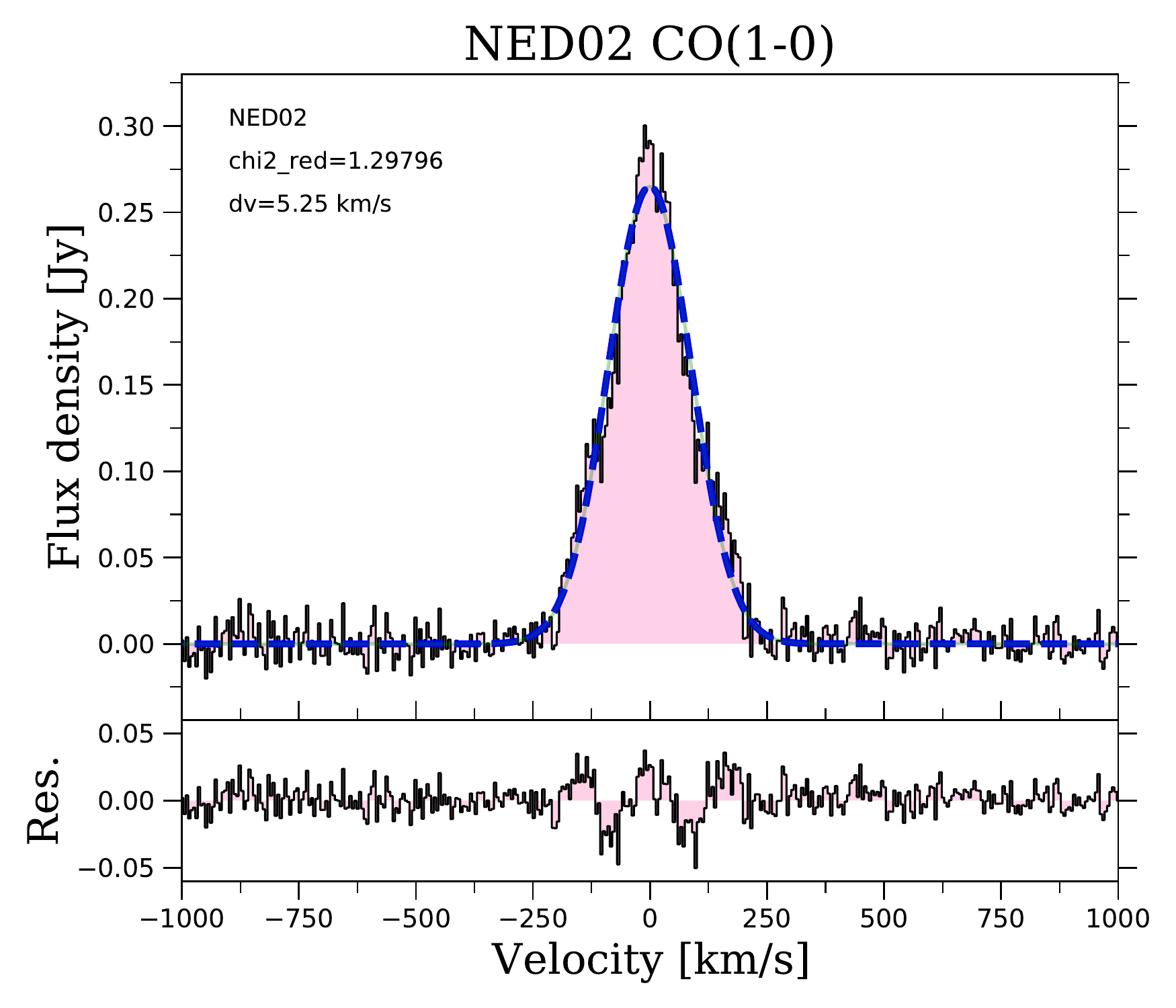}\\
        \caption{ALMA CO(1-0) continuum-subtracted spectra of the interacting galaxy pair IRAS 05054+1718: NED01 (left panels) and NED02 (right panels). The top and bottom panels display the same spectral data, but with different units on the x-axes. In particular, the top panels report the \co~flux density as a function of observed (sky) frequency, and were used to refine the systemic redshift estimates for the two galaxies (see Sect.~\ref{sec:z-calc}). The bottom panels show the spectra corrected for redshift, where the \co~flux density is reported as a function of optical velocity along the line of sight.
        The \co~line spectrum of NED01 was extracted from an elliptical aperture maximising the \co~flux, with size $23''\times14''$, centred at RA=05$^h$08$^m$19.858$^s$, Dec=+17$^\circ$21$'$45.898$''$. The \co~spectrum of NED02 was extracted from an $10''\times12''$ elliptical aperture centred at RA=05$^h$08$^m$21.212$^s$, Dec=+17$^\circ$22$'$08.660$''$.
        The best-fit spectral parameters are listed in Table~\ref{tab:total_co_spectral_fits}.
}\label{fig:total_co_spectra}
\end{figure*}

\begin{table*}[tbp]
        \centering 
        \caption{Best-fit parameters of the \co~spectral fits shown in Fig.~\ref{fig:total_co_spectra}}
                \small
        \label{tab:total_co_spectral_fits}     
        \begin{tabular}{lccclccc}
                \toprule
                  \multicolumn{4}{c}{\bf NED01} & \multicolumn{4}{c}{\bf NED02} \\
                        Fit     & $\nu_0$  [GHz] & $S_{peak}$ [mJy] & $\sigma_\nu$ [MHz]  & Fit & $\nu_0$ [GHz]  & $S_{peak}$ [mJy] & $\sigma_\nu$ [MHz] \\                         
                \midrule                        
                1-Gauss & 113.2578 (0.0009) & 78.5 (0.9) & 66.2 (0.9) & 1-Gauss & 113.3651 (0.0003) & 269 (2) & 33.5 (0.3)  \\
                \multirow{2}{*}{2-Gauss} & 113.2176 (0.0010) & 83.9 (1.3) &  33.7 (1.0) & & &  & \\
                                                                                                                & 113.3125 (0.0011) & 72.7 (1.4) &  30.3 (1.0) & & & & \\
                \midrule
                Fit     & $v_0$  [\kms] & $S_{peak}$ [mJy] & $\sigma_v$ [\kms] & Fit & $v_0$  [\kms] & $S_{peak}$ [mJy] & $\sigma_v$ [\kms] \\ 
                \midrule 
                \multirow{2}{*}{2-Gauss} & $-146$ (3) & 72.7 (1.3) & 79 (2) & 1-Gauss & $-0.4$ (0.8) & 265 (2)  & 88.4 (0.8) \\      
                                                                                                        & 106 (2) & 82.9 (1.2) & 90 (2) & & & & \\
                \midrule
        \end{tabular}

        \begin{flushleft}
        {\it Notes:} The top and bottom parts of the table list respectively the best-fit parameters of the top and bottom panels of Fig.~\ref{fig:total_co_spectra}. For each Gaussian component, we report its centre, peak \co~flux density, and standard deviation ($\sigma$), from which it is possible to compute the full width at half maximum (FWHM = 2.3548$\sigma$). Uncertainties, always reported within  brackets, correspond to 1$\sigma$ statistical errors and do not include the absolute flux calibration error. 
        \end{flushleft}
\end{table*}

\subsubsection{NED01}

\begin{figure}[tbp]
        \centering
        \includegraphics[clip=true,trim=2.5cm 3.5cm 3cm 2.5cm,width=.8\columnwidth]{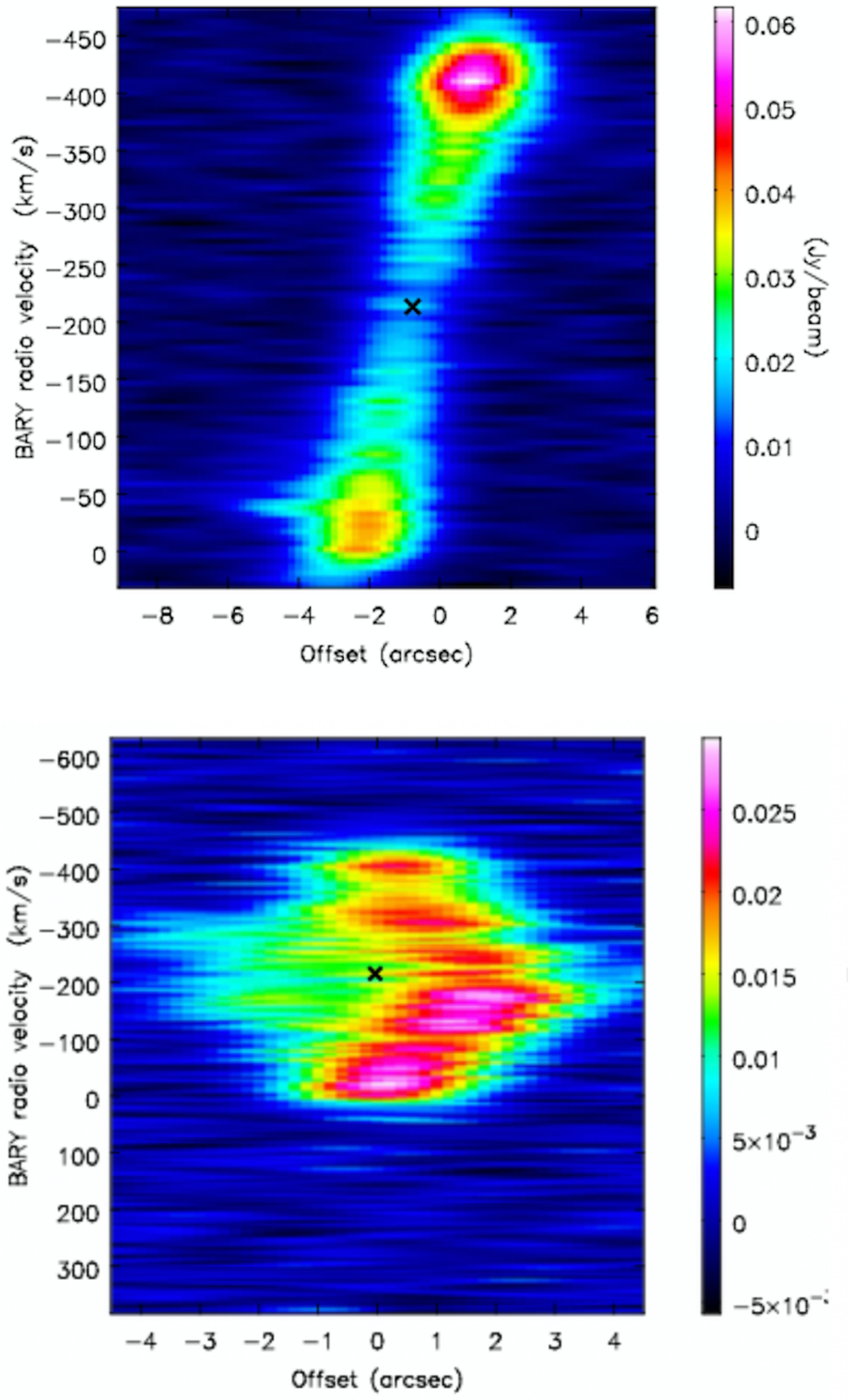}
        \caption{\co~ PV diagrams of NED01, extracted from a datacube where velocities are calculated with respect to the previously known redshift of the source ($z=0.0184$). The PV diagrams confirm that our new redshift estimate of $z=0.0178\pm0.0004$, indicated with a black cross, closely matches  the kinematic centre of the \co~source. The upper panel shows the PV diagram obtained from a slit-like aperture along the axis connecting the blue and red peaks of the \co~emission, with a size of $\sim30''$ and a position angle of $95^\circ$ (measured anti-clockwise from the north direction). The PV diagram shown in the bottom panel was computed from a slit-like aperture with a size of $\sim9''$ orthogonal to the previous one.}
        \label{fig:p-v}
\end{figure}

The continuum-subtracted \co~spectrum of NED01, reported in the left panels of Fig.~\ref{fig:total_co_spectra}, shows a double-peaked emission line centred at $\nu=113.26$~GHz. This is slightly different from the (redshifted) \co~central frequency of 113.189 GHz that was expected from a previous redshift estimate of this source ($z=0.0184$), derived from the heliocentric velocity (error-weighted average of the optical and radio velocities) reported in the HyperLeda catalogue \cite[see][]{2014Makarov}. To refine the redshift estimate of NED01 we performed two spectral fits, one with a single-Gaussian function and one with two Gaussians, both displayed in the top left panel of Fig.~\ref{fig:total_co_spectra}, with results listed in Table~\ref{tab:total_co_spectral_fits}. We computed a new systemic redshift using the central frequency value derived from the single-Gaussian fit ($\nu=113.2578\pm0.0009$ GHz). However, the source shows a double-peaked profile with clear asymmetries. To take this into account, we conservatively assigned to such CO-based redshift an uncertainty equal to the average frequency difference between the two CO line peaks (as measured from the double-Gaussian fit) and the single-Gaussian fit peak frequency value. We obtained $z=0.0178\pm0.0004$, which is offset by $v\sim-210 \ kms^{-1}$ with respect to the previously known redshift.
We further checked that the new redshift estimate matches with the kinematic centre of NED01's host galaxy disk. The \co~position-velocity (PV) diagrams displayed in Fig.~\ref{fig:p-v} were computed in \texttt{CASA} from slit-like apertures with sizes of $\sim30''$ and $\sim9''$, aligned with the major and minor axes of the \co~disk of NED01 respectively (see kinematic modelling reported in Sect~5). The new redshift estimate (obtained through the spectral analysis described above) is shown as a black cross at the centre of the two PV diagrams, hence confirming the correspondence with the rotational centre of the \co~disk.

\subsubsection{NED02}

A similar analysis aimed at refining the systemic redshift of the host galaxy was performed on the companion source, NED02.
The total continuum-subtracted \co~spectrum of NED02, displayed in the right panel of Fig.~\ref{fig:total_co_spectra}, presents a single peak. We modelled it using a single-Gaussian function, whose best-fit parameters are reported in Table~\ref{tab:total_co_spectral_fits}.
The previous redshift estimate for this source, $z=0.016842$, derived from the heliocentric velocity reported in the 2MASS \textit{Redshift Survey} \citep{2012Huchra}, would have produced a \co~emission line peaked at 113.1385~GHz. Instead, our spectral analysis of the new ALMA \co~observations of NED02 shows that the \co~ line is centred at a higher frequency of $113.3651\pm 0.0003$ GHz. We used this value and its associated uncertainty to compute a new systemic redshift of NED02 equal to $z=0.016812\pm0.000003$, which is offset by $\sim50$ \kms\ with respect to the previously known redshift.

\subsection{CO(1-0) morphology}\label{sec:morphology}

\begin{table}[tbp]
        \caption{Molecular gas mass estimates}
        \label{tab:mol_masses}     
        \begin{tabular}{lccc}
                \toprule
                Source  & $S_{\rm CO(1-0)}dv$ & $L^{\prime}_{\rm CO(1-0)}$ & $M_{mol}$  \\
                & [Jy~\kms] & [10$^8$~K~\kms~pc$^2$] & [10$^9$ M$_{\odot}$] \\
                (1)             &  (2)                  & (3)                                                           & (4) \\
                \midrule 
                NED01 & 33.1 (1.4) & 5.1 (0.2) & 2.1 (0.7) \\ 
                NED02 & 59 (3) & 8.0 (0.4) & 3.4 (1.1) \\ 
                \midrule
        \end{tabular}
        
        \begin{flushleft}
                {\it Notes:} Col. (1): Source; Col. (2): Total \co~line flux, computed from the spectral line fit results (Table~\ref{tab:total_co_spectral_fits}). For NED01, the total flux was  computed by adding the contribution of the two Gaussian components employed in the spectral fit. Errors on the \co~integrated fluxes include a systematic calibration error of 5\%, typical for ALMA Band 3 observations. Col. (3): \co~line luminosity computed following the definition by \cite{1997Solomon}; Col. (4): Molecular gas mass computed using a CO-to-H$_2$ conversion factor of $\alpha_{CO}$ = $4.3\pm1.3$~M$_{\odot}$~K~\kms~pc$^2$ \citep{2013Bolatto}. 
        \end{flushleft}
\end{table}

The redshift-corrected \co~emission line spectra of NED01 and NED02, plotted as a function of line-of-sight velocity, are shown in the bottom panels of Fig.~\ref{fig:total_co_spectra}. The corresponding best-fit spectral parameters are listed in Table~\ref{tab:total_co_spectral_fits}. By using the \co~line fluxes obtained from the spectral fits, we computed the total \co~line luminosities of the two galaxies, and from these estimated their molecular line masses, by adopting a standard CO-to-H$_2$ conversion factor of $\alpha_{CO}$ = $4.3\pm1.3$~M$_{\odot}$~K~\kms~pc$^2$. These values are reported in Table~\ref{tab:mol_masses}.

\begin{figure*}[tbp]
        \centering
        \includegraphics[clip=true,trim=0.9cm 11.cm 8.8cm 8.cm,width=.65\columnwidth]{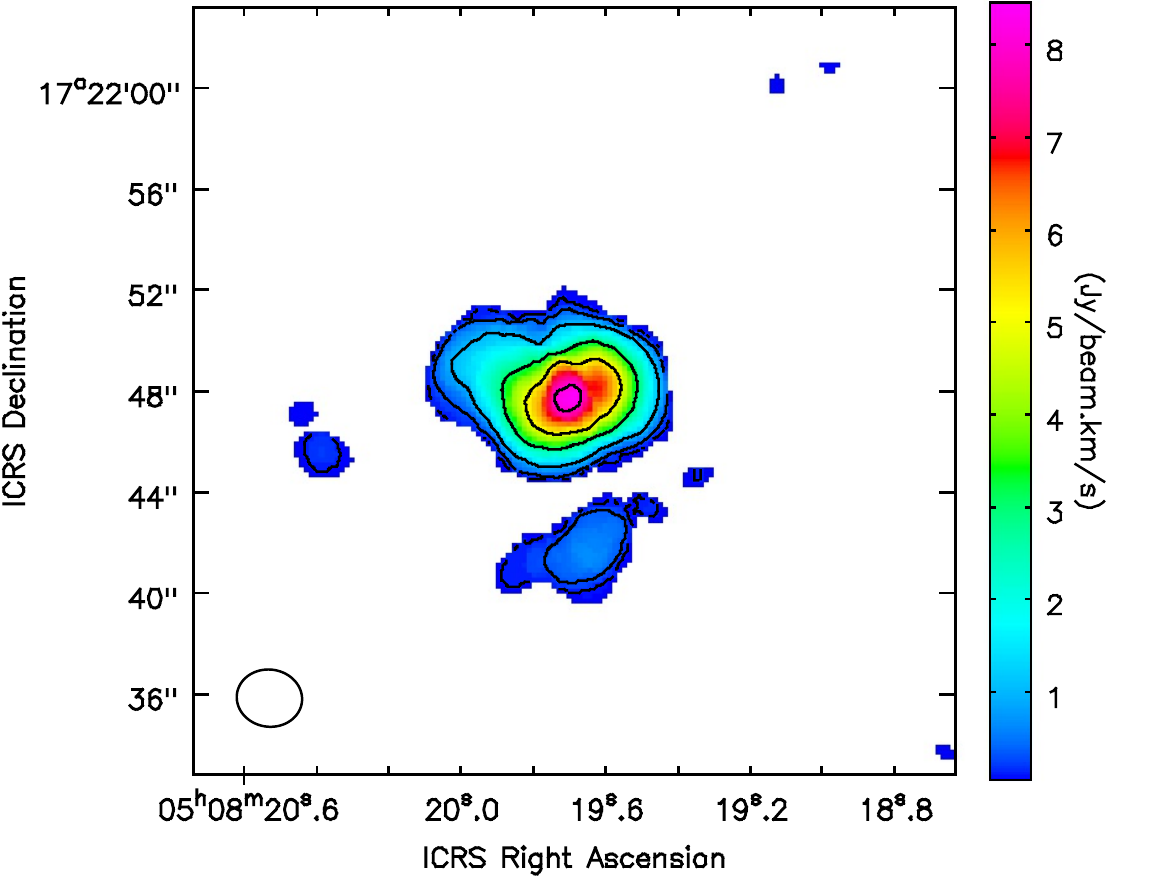}\quad
        \includegraphics[clip=true,trim=0.9cm 11.cm 8.8cm 8.cm,width=.65\columnwidth]{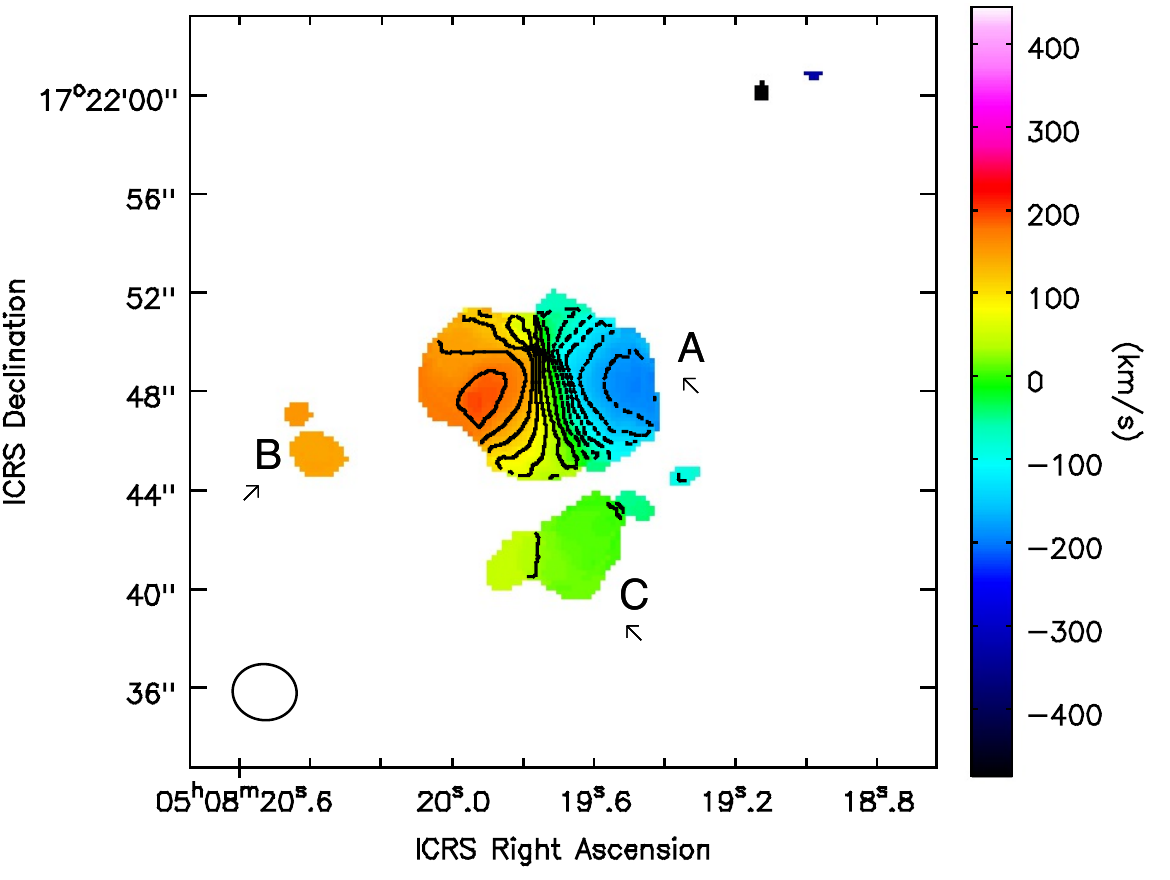}\quad
        \includegraphics[clip=true,trim=0.9cm 11.cm 8.8cm 8.cm,width=.65\columnwidth]{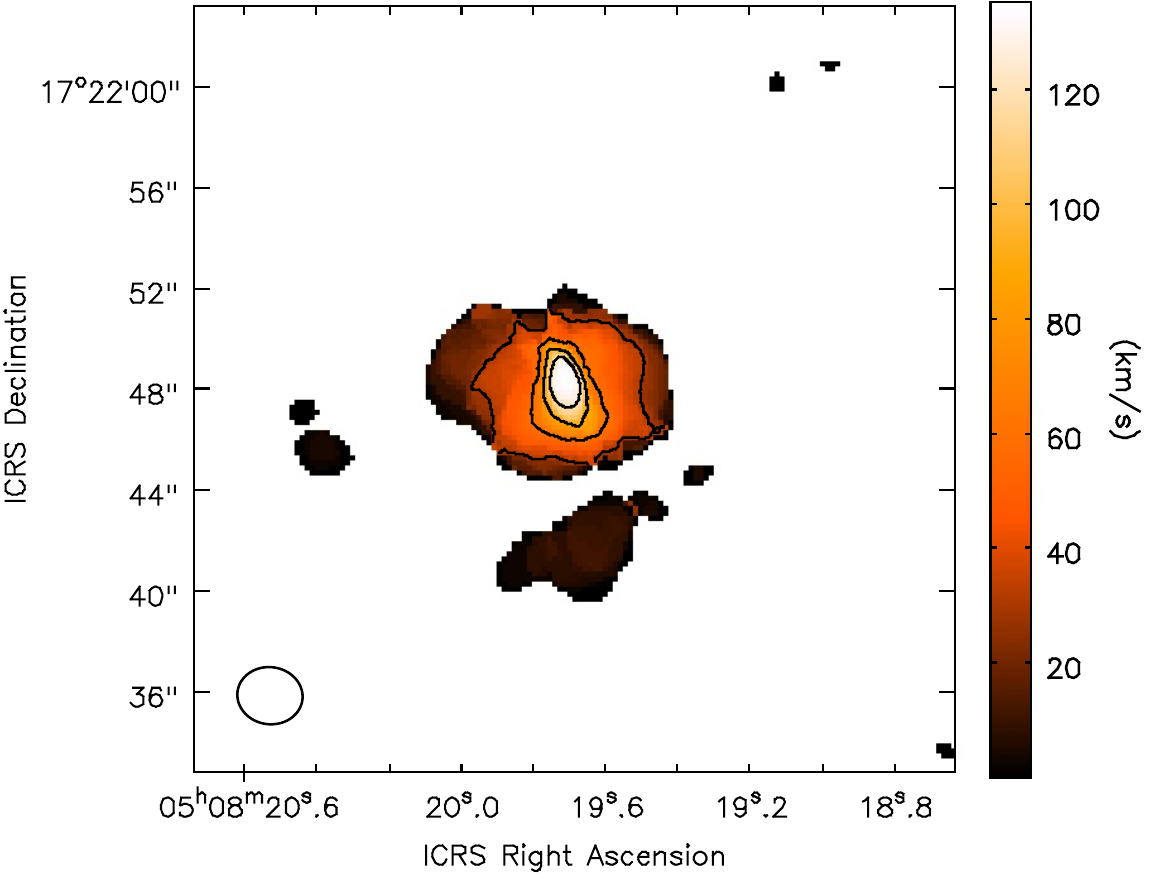}\\
        \caption{Moment maps of NED01 computed within a box of size 30$\arcsec$ and including pixels above a \co~intensity threshold of 0.01 Jy~beam$^{-1}$ and within $v\in[-500, 500]$~\kms. {\it Left:} Intensity (moment 0) map, with contours plotted at [0.1, 0.3, 1, 3, 5, 8]~Jy~beam$^{-1}$~\kms. {\it Centre:} velocity (moment 1) map, with contours plotted every 30~\kms; the letters indicate the three \co~components discussed in the main text, where A indicates the main rotating \co~disk. {\it Right:} Velocity dispersion (moment 2) map, with contours plotted at intervals of 30~\kms. }
        \label{fig:NED01_mom_maps}
\end{figure*}

\begin{figure*}[tbp]
        \centering
        \includegraphics[clip=true,trim=0.9cm 11.cm 8.8cm 8.cm,width=.65\columnwidth]{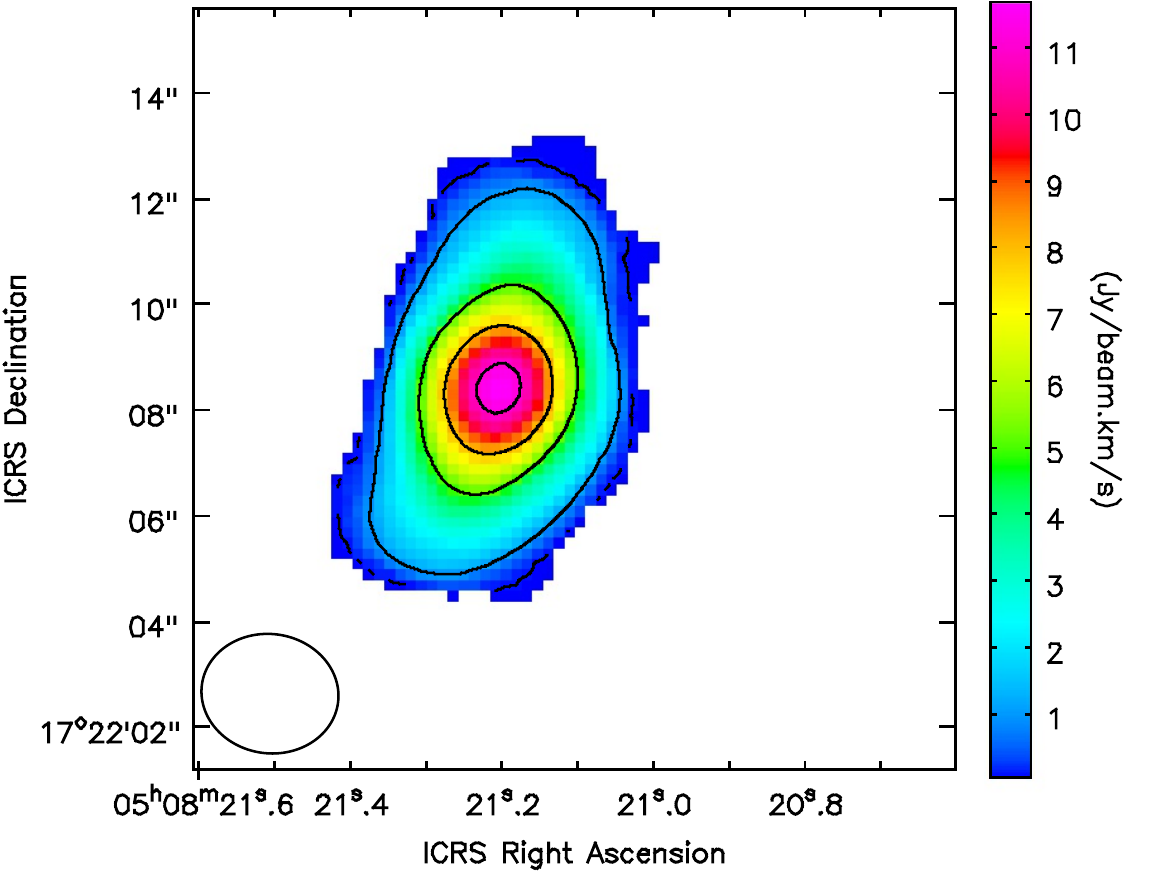}\quad
        \includegraphics[clip=true,trim=0.9cm 11.cm 8.8cm 8.cm,width=.65\columnwidth]{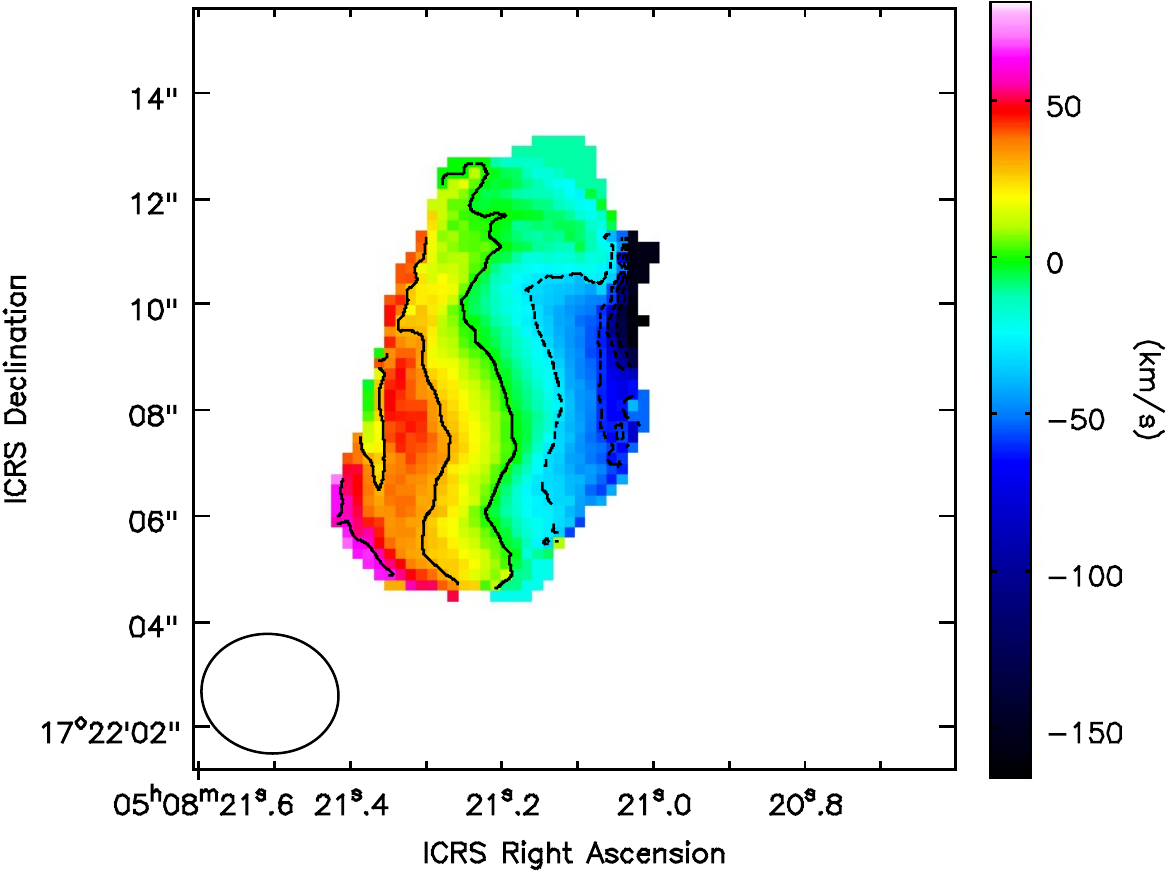}\quad
        \includegraphics[clip=true,trim=0.9cm 11.cm 8.8cm 8.cm,width=.65\columnwidth]{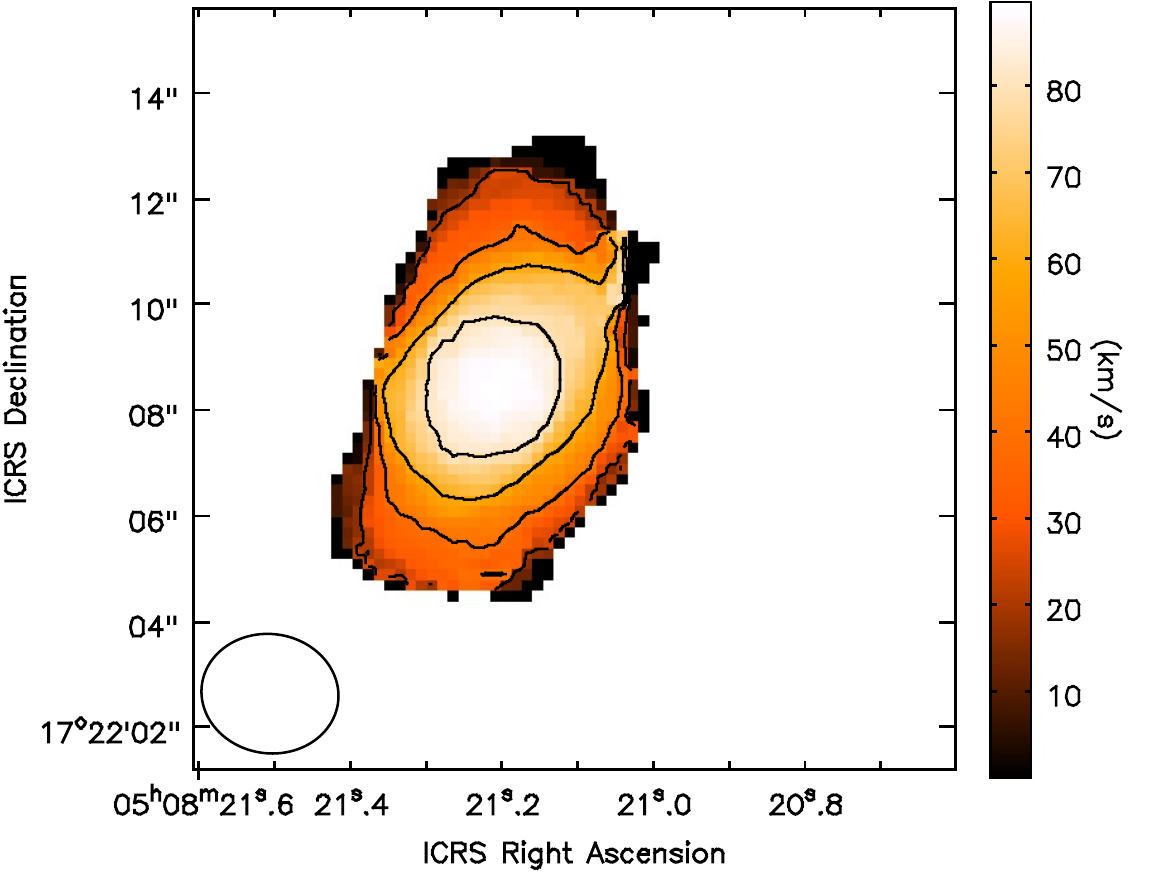}\\
        \caption{Moment maps of NED02 computed from the non-PB-corrected datacube within a box of size 14$\arcsec$ centred at  RA(ICRS) = 05:08:21.108; Dec(ICRS) = 17.22.08.489, including pixels above a \co~intensity threshold of 0.009 Jy~beam$^{-1}$ and within $v\in[-500, 500]$~\kms. {\it Left:} Intensity (moment 0) map, with contours plotted at [0.1, 1, 5, 8, 11]~Jy~beam$^{-1}$~\kms. {\it Centre:} Velocity (moment 1) map, with contours plotted every 30~\kms. {\it Right:} Velocity dispersion (moment 2) map, with contours plotted at intervals of 20~\kms. }
        \label{fig:NED02_mom_maps}
\end{figure*}

Figure~\ref{fig:NED01_mom_maps} displays the zeroth, first, and second moment maps of the source NED01, computed by applying an intensity threshold of 0.01 Jy~beam$^{-1}$ and within a box $30\arcsec$ in size   centred on the AGN. The velocity map shows clearly that the \co~gas kinematics in NED01 is dominated by ordered rotation. The  blueshifted emission arises from the western side of the galaxy, and the redshifted emission from the eastern side, following the typical pattern of a rotating disk. Therefore, from Fig.~\ref{fig:NED01_mom_maps} we can already infer that if a molecular outflow is present, it does not seem to impact the bulk of the \co~kinematics in this source. This was also evident from the \co~spectrum of NED01, which presents a clear double peak and no evidence for very high-velocity wings, which are typical of extreme molecular outflows detected in some local (U)LIRGs  \citep[see e.g.][and several references in  \citealt{2020Veilleux}]{2014Cicone}.

In addition to the central molecular disk (component A), the CO moment maps of NED01 reveal two apparently disconnected \co~emitting structures in this galaxy,  labelled B and C in Fig.~\ref{fig:NED01_mom_maps}. Component B is $\sim3.3$~kpc and component C is $\sim2.3$~kpc from the centre of NED01's main disk. Despite their apparent offset from the disk in the moment maps shown in Fig.~\ref{fig:NED01_mom_maps}, which results from the adoption of a sensitivity threshold, these two components are physically linked to the central disk, and also follow the same velocity pattern. This will be confirmed by the kinematic modelling with the BBarolo software presented in the next section. The \co~spectral line profiles of components B and C are single-peaked and narrow. Component B is redshifted, centred at a velocity of $v\simeq150$~\kms, with a velocity dispersion of $\sigma_v\simeq13$~\kms, and entrains a \co~flux of $0.65$~Jy~\kms, which, if adopting the same $\alpha_{\rm CO}$ as above,    corresponds to $\sim4\times10^7$~M$_\odot$ of molecular hydrogen gas. Component C is only slightly redshifted ($v\sim9$~\kms), and can be modelled with a single-Gaussian with $\sigma_v\simeq30$~\kms\ and an integrated \co~flux of $2$~Jy~\kms, corresponding to $1.3\times10^8$~M$_\odot$.
Figure~\ref{fig:source} shows that components B and C of the \co~emission from NED01 (detected respectively at 3 and 10$\sigma$ in the \co~channel map shown as contours in Fig.~\ref{fig:source}) do not correspond to any significant sub-structure in the optical continuum; instead,   they overlap with diffuse lower surface brightness stellar light. We can rule out the hypothesis that the CO~components B and C are not distinguishable in the g-band Pan-STARR image because of high dust extinction, since in this case we would expect to detect them in the ALMA 3~mm continuum map, which is not the case (see Appendix~\ref{sec:appendix_continuum}). We therefore suggest that the \co~components B and C are ISM substructures of the main galaxy disk, possibly tracing a spiral arm. 

The companion galaxy NED02 is located close to the edge of the PB, and so the moment maps, reported in Fig.~\ref{fig:NED02_mom_maps}, were computed from the datacube not corrected for the PB. The \co~emission in this edge-on galaxy (see Fig.~\ref{fig:source}) appears more compact than in NED01. The spider diagram shows a velocity gradient skewed towards blueshifted velocities, tracing a disturbed rotating disk. The modelling presented in the next section supports this interpretation.

\section{Modelling of the CO(1-0) kinematics}\label{sec:kinematics}

In this section we analyse the gas kinematics in IRAS 05054+1718. As shown by Fig.~\ref{fig:NED01_mom_maps}, the bulk of the \co~emission from NED01 traces a rotating molecular disk, while the disk of NED02 appears more disturbed (Fig.~\ref{fig:NED02_mom_maps}). In order to study the presence of \co~components that are not participating in the rotation and may trace the effect of AGN feedback on the large-scale ISM, we first model the disk kinematics, and then study any residual emission, similar to \cite{2019Sirressi}. For completeness, we perform the same analysis on the companion galaxy NED02, even though we do not have any evidence for the presence of an AGN in this source. A uniform and common analysis of the \co~gas kinematics of both members of the galaxy pair can give us insights into the role of galaxy interactions in shaping the cold ISM kinematics.

\subsection{Modelling of the CO(1-0)~disk in NED01}\label{sec:NED01_BBarolo}

We model the disk rotation in NED01 using the 3D-Based Analysis of Rotating Object via Line Observations \cite[3D-BAROLO,][also known as BBarolo]{BAROLO}. BBarolo identifies the set of geometrical and kinematic parameters that best fit the rotating gaseous disk observations, and uses these parameters to produce a mock datacube of the best-fit model. An important assumption of the BBarolo model, to consider when interpreting the results, is the hypothesis that the line emission is distributed in a geometrically thin disk whose kinematics is dominated by pure rotational motion. In other words, the model does not consider the presence of additional outflow components, tidal tails, or other features that do not belong to the main rotating disk. The disk model is built up by combining concentric rings with a user-defined width, and the comparison between models and data is performed ring by ring. 

As shown by Fig.~\ref{fig:NED01_mom_maps}, the \co~line emission from NED01 has a complex morphology. In order to study   the effect of adding the extended \co~line components in the BBarolo modelling, we run BBarolo on three regions with different sizes, and compare the results. The regions are as follows (see Fig.~\ref{fig:mommap1}): the \textit{small region}, which includes only component A, with size=$10''\times8''$ ($3.7\times3.0$~kpc); the \textit{medium region}, which includes components A and C, with size=$16''\times15''$ ($6.0\times5.6$~kpc); and the \textit{big region}, which includes all three components, with size=$35''\times35''$ ($13\times13$~kpc).

We produced the \co~datacubes that need to be given as an input to the software by cropping the continuum-subtracted ALMA CO(1-0) (clean) datacube with the CASA task \texttt{imsubimage} and then exporting it into a FITS file using the \texttt{exportfits} task. We selected a spectral range corresponding to $v$=[-600,+600] \ km s$^{-1}$. For each region, we also produced a residual map by subtracting the BBarolo disk model from the input datacube.

As parameters of the BBarolo model we kept the inclination ($\iota$, i.e. the angle of the disk with respect to the line of sight) and the position angle (PA) of the molecular gas disk fixed at all radii. 
Using the ALMA moment~1 map shown in Fig.~\ref{fig:mommap1}, we estimated $PA=95^\circ\pm10^\circ$ and $\iota=45^\circ\pm15^\circ$. The inclination was inferred  from the ratio of the minor to the major axis of the molecular disk. The fitting was performed considering only pixels with a $S/N>2.3$ to enable inclusion of the low-S/N extended features. 

\begin{figure}[tbp]
        \centering
        \includegraphics[clip=true,trim=0.cm 0.cm 0.cm 0.cm, width=\columnwidth]{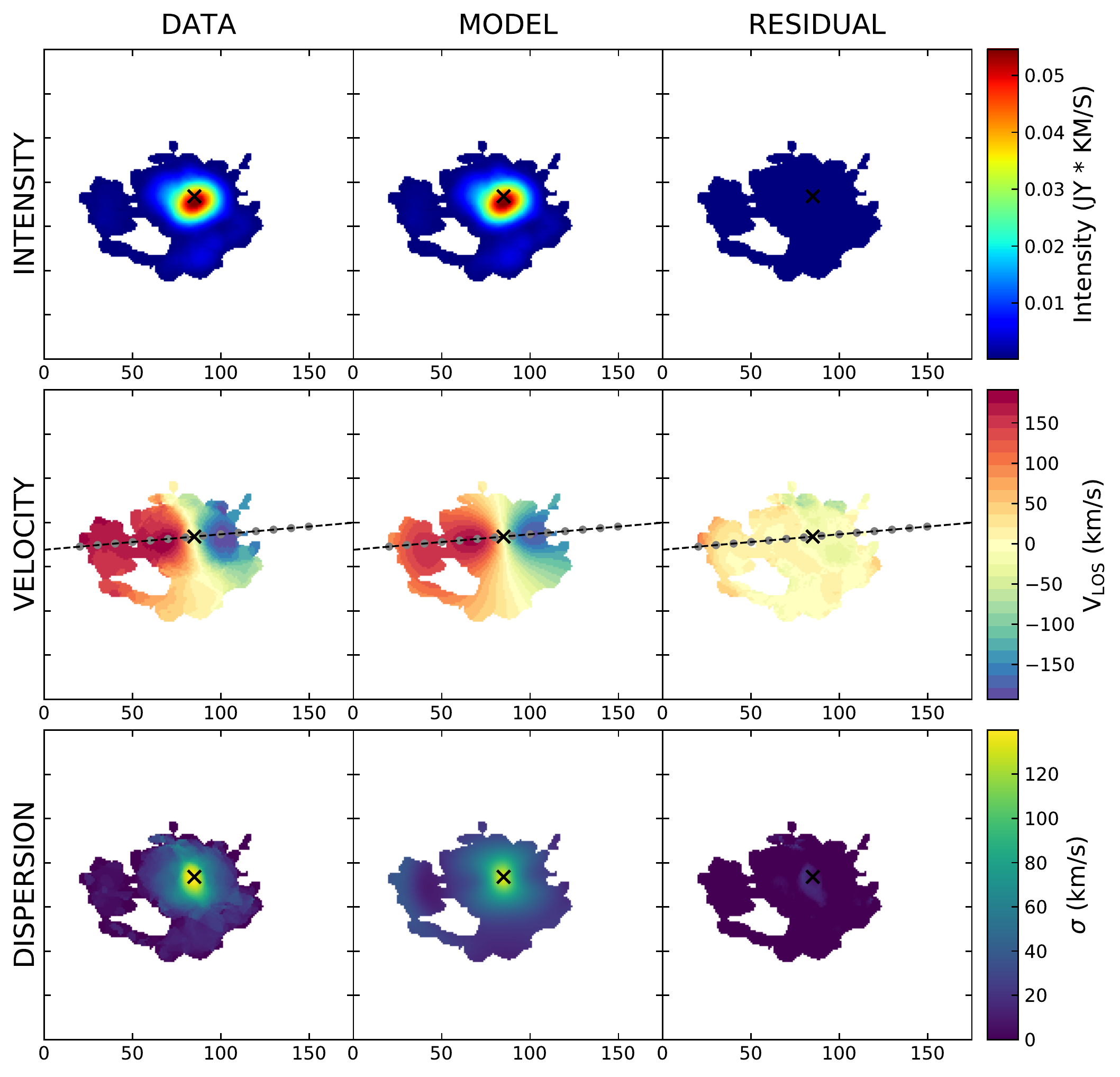}
        \caption{Moment maps showing the comparison between the data and the BBarolo best fit performed on the \textit{Big region} around NED01. Shown (from left to right) are  data, BBarolo model, and residual emission. The rows show (rom top to bottom) the moment~0 map (intensity), moment 1 map (line of sight velocity), and moment 2 map (velocity dispersion). The black cross indicates the rotating disk centre found by the fit, and the dashed black line shows the major axis of the disk model.}
        \label{fig:bigmom}
\end{figure}

\begin{figure}[tbp]
        \centering
        \includegraphics[clip=true,trim=0.cm 0.cm 0.cm 0.cm, width=.6\columnwidth]{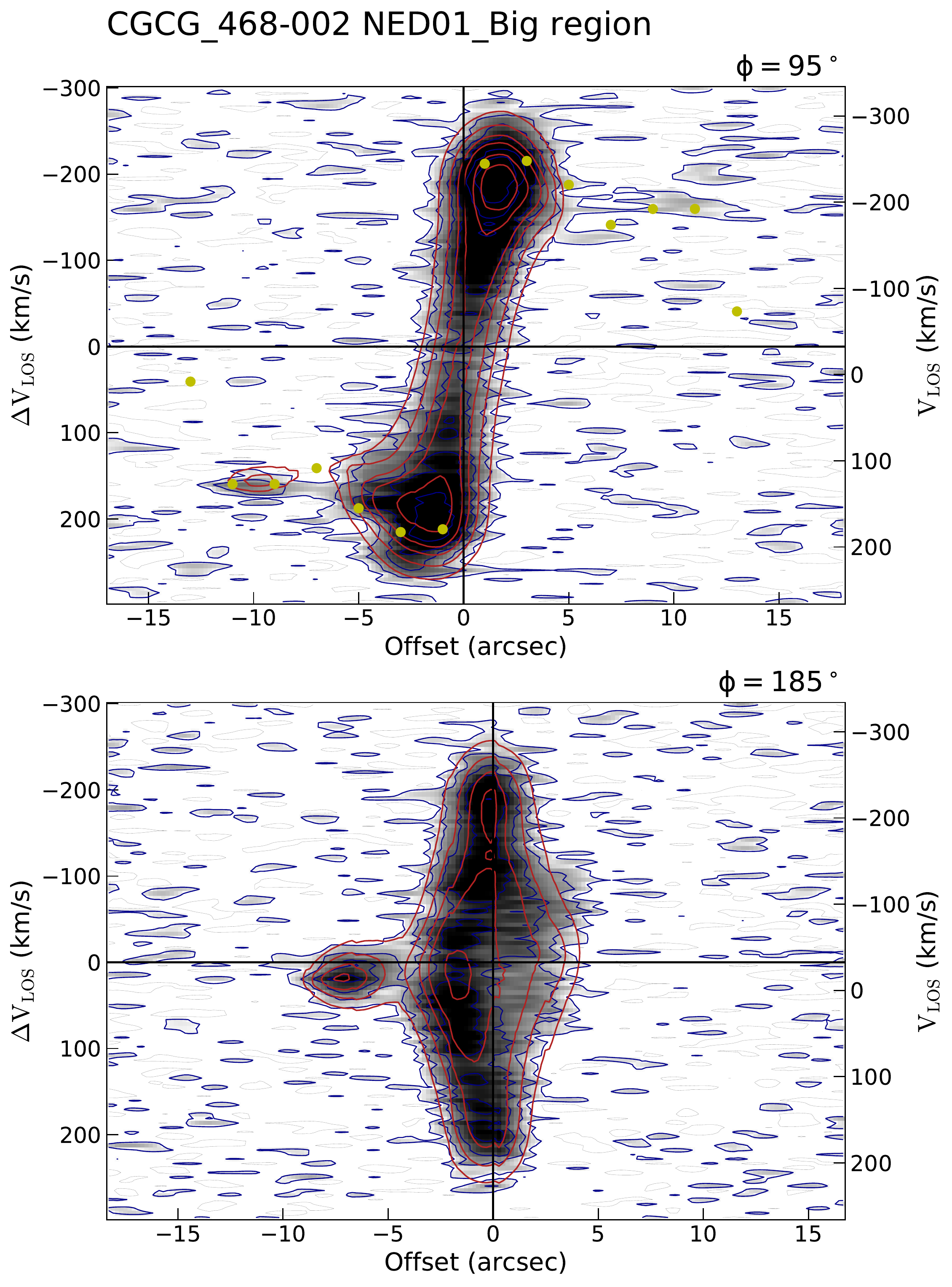}
        \caption{Position velocity diagrams showing the comparison between the data and the BBarolo best-fit model performed on the \textit{Big region} around NED01. The red contours show the disk model, and they are overplotted onto the observed ALMA \co~line data (displayed in grey, with blue contours at 3, 6, 9, 15, 20, 30, 40, and 50$\sigma$). The upper panel shows the PV diagram extracted along the major axis. The yellow data points indicate the best-fit rotational velocity of each ring. The lower panel shows the PV diagram computed along the minor axis.}
        \label{fig:bigpv}
\end{figure}

\begin{figure*}[tbp]
        \centering
        \includegraphics[clip=true,trim=0.5cm 0.5cm 0.5cm 0.cm,width=0.85\textwidth]{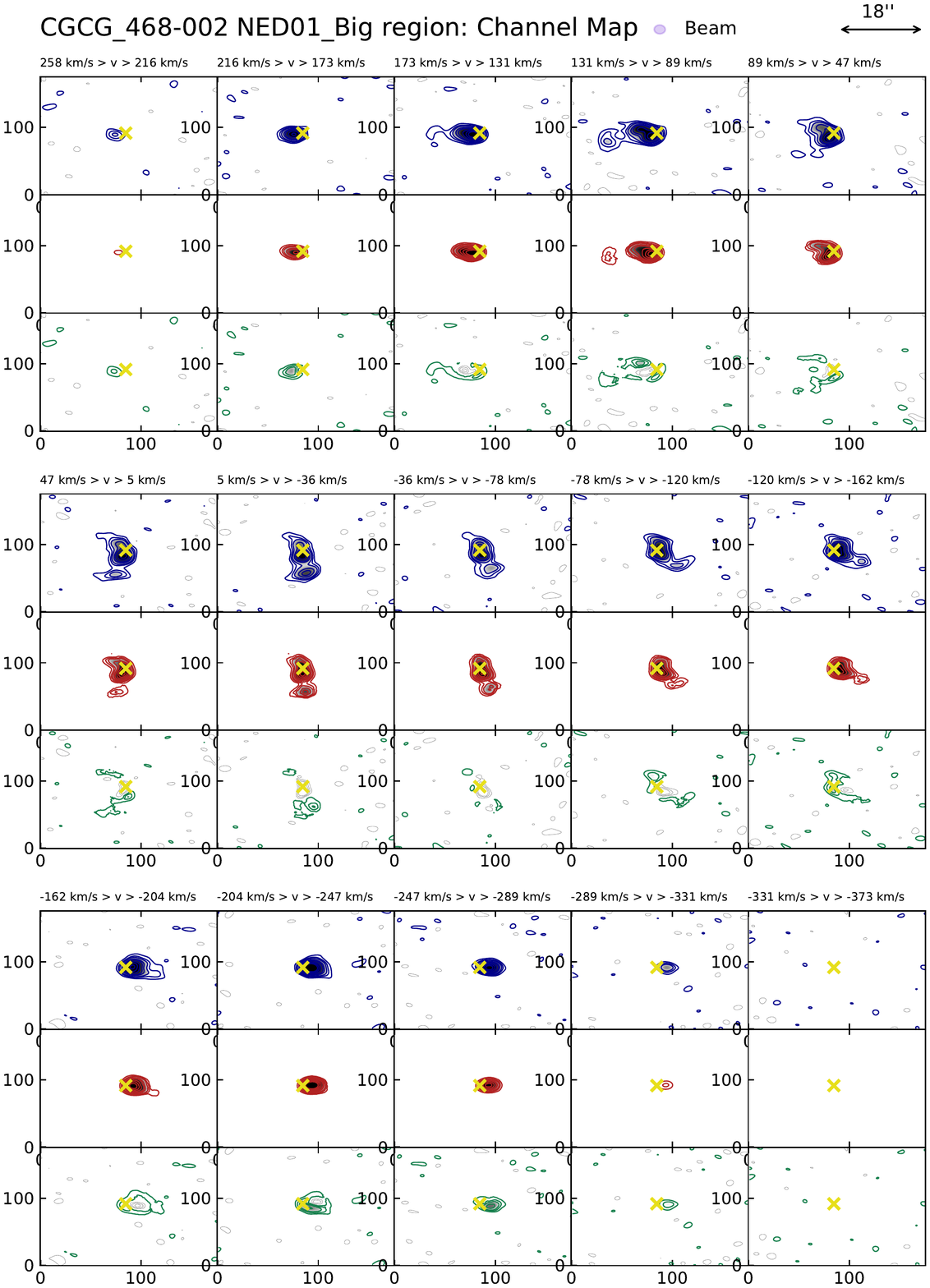}
        \caption{CO channel maps computed by integrating the \co~emission in channels of $\Delta v = 42$~\kms, showing the comparison between data and BBarolo best-fit disk model within the \textit{Big region} around  NED01. In each of the three panels, the top row (blue contours) displays the data, the central row the model (red contours), and the bottom row the residual emission (green contours). The contours are plotted at 3, 6, 9, 15, 20, 30, 40, 50$\sigma$. Symmetric negative contours are plotted in grey. The yellow cross indicates the centre of the disk model.}
        \label{fig:bigchann}
\end{figure*}

In the following we present the results obtained by using the \textit{Big region}, which includes all  three  \co~emission line components (A, B, and  C) observed in the moment~1 map of NED01, while the results from the other two regions are reported in Appendix~\ref{sec:appendix}. We focus on the {\it Big region} because, once it has been subtracted from the
ALMA datacube, this is the fit that  produces the lowest residual \co~emission. Additionally, the analysis of the {\it Big region} fit allows us to study the global \co~emission from NED01 as this region includes also components B and C. We therefore believe that the fit on the {\it Big region} is the most conservative one for our goal, which is to study residual emission whose kinematics is inconsistent with a rotating disk. 

We constructed the BBarolo model disk by using seven rings of width $2\arcsec$ (0.7 kpc), covering in total a region of $28\arcsec$ in diameter (10.4 kpc) around NED01. The results of this fit are shown in Figures~\ref{fig:bigmom}, \ref{fig:bigpv}, and \ref{fig:bigchann}.
Although the rotating disk clearly provides a very good fit for the bulk of \co~emission in and around NED01, there are statistically significant residuals. We note that the moment maps and the PV diagrams produced by BBarolo (Figures~\ref{fig:bigmom} and ~\ref{fig:bigpv}) are not optimised to visualise residual emission, because they use a colour scale that is fine-tuned to enhance the match between data and model. The bottom rows of the channel maps in Fig.~\ref{fig:bigchann}, displaying the residual \co~emission, show that positive (green contours) and negative  (grey contours) residuals with similar significance ($\sim3\sigma$) are present in each channel map with velocity $|v|>200$~\kms, consistent with low-level residual rotation that is not accounted for by the fit. However, at blueshifted velocities, within 
a range $-290<v<-200$~\kms, we detected only positive residuals at much higher significance ($S/N\ge15$), showing a compact morphology.

\begin{figure*}[tbp]
        \centering 
        \includegraphics[clip=true,trim=0cm 0cm 0cm 0cm,width=.64\columnwidth]{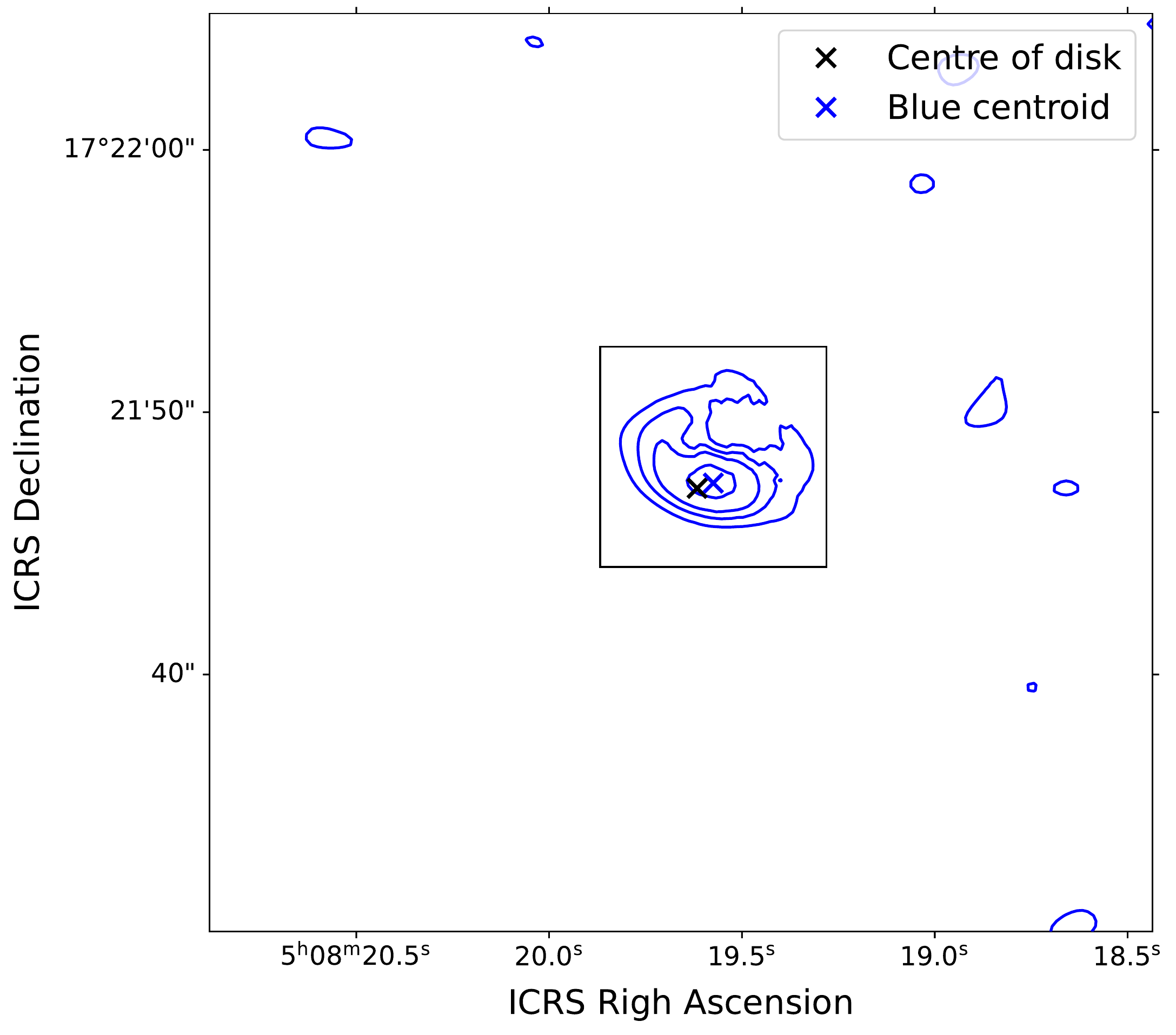}\quad
        \includegraphics[clip=true,trim=0cm 0.2cm 0.5cm 0.cm,width=.68\columnwidth]{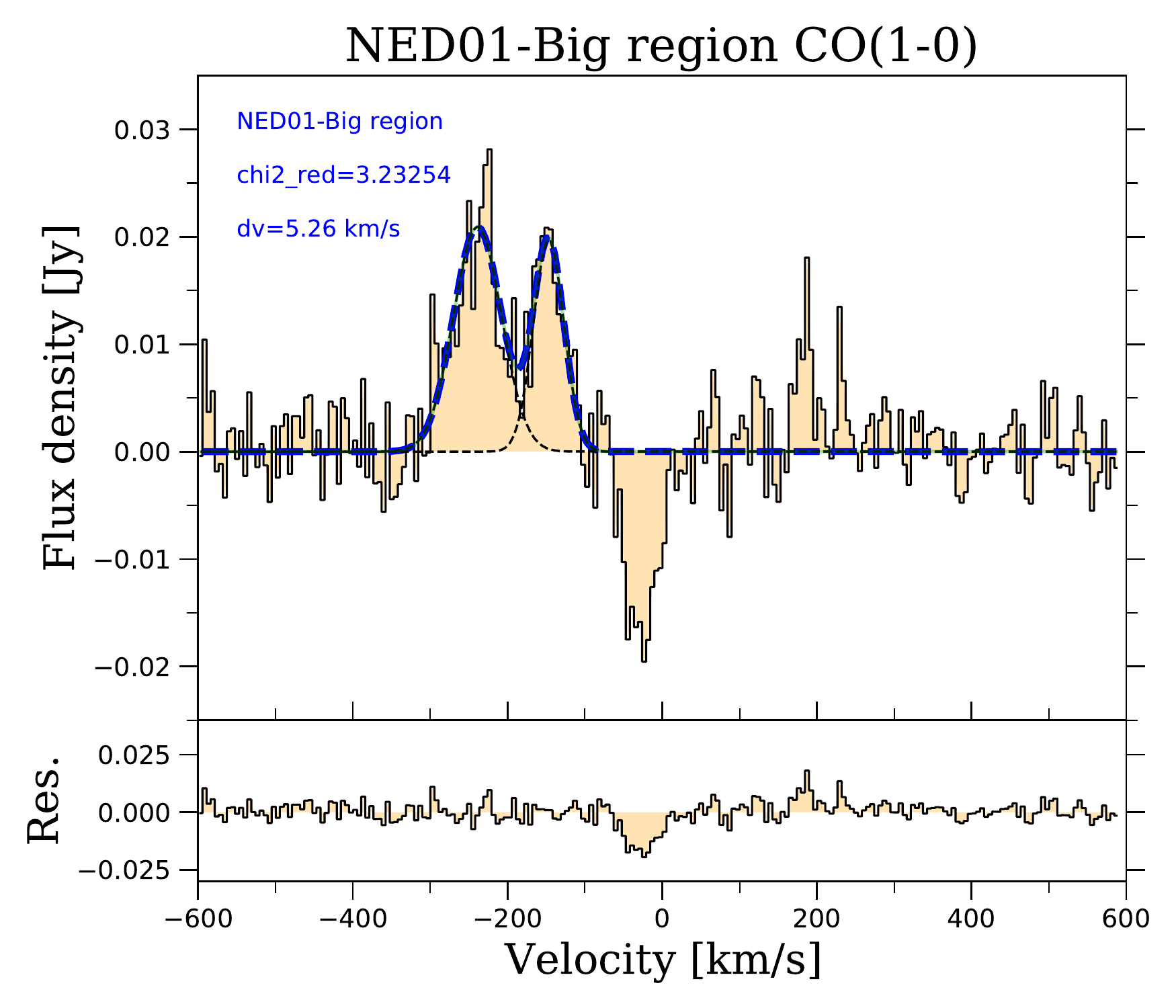}\quad
        \includegraphics[clip=true,trim=0cm 0cm 0cm 0cm,width=.64\columnwidth]{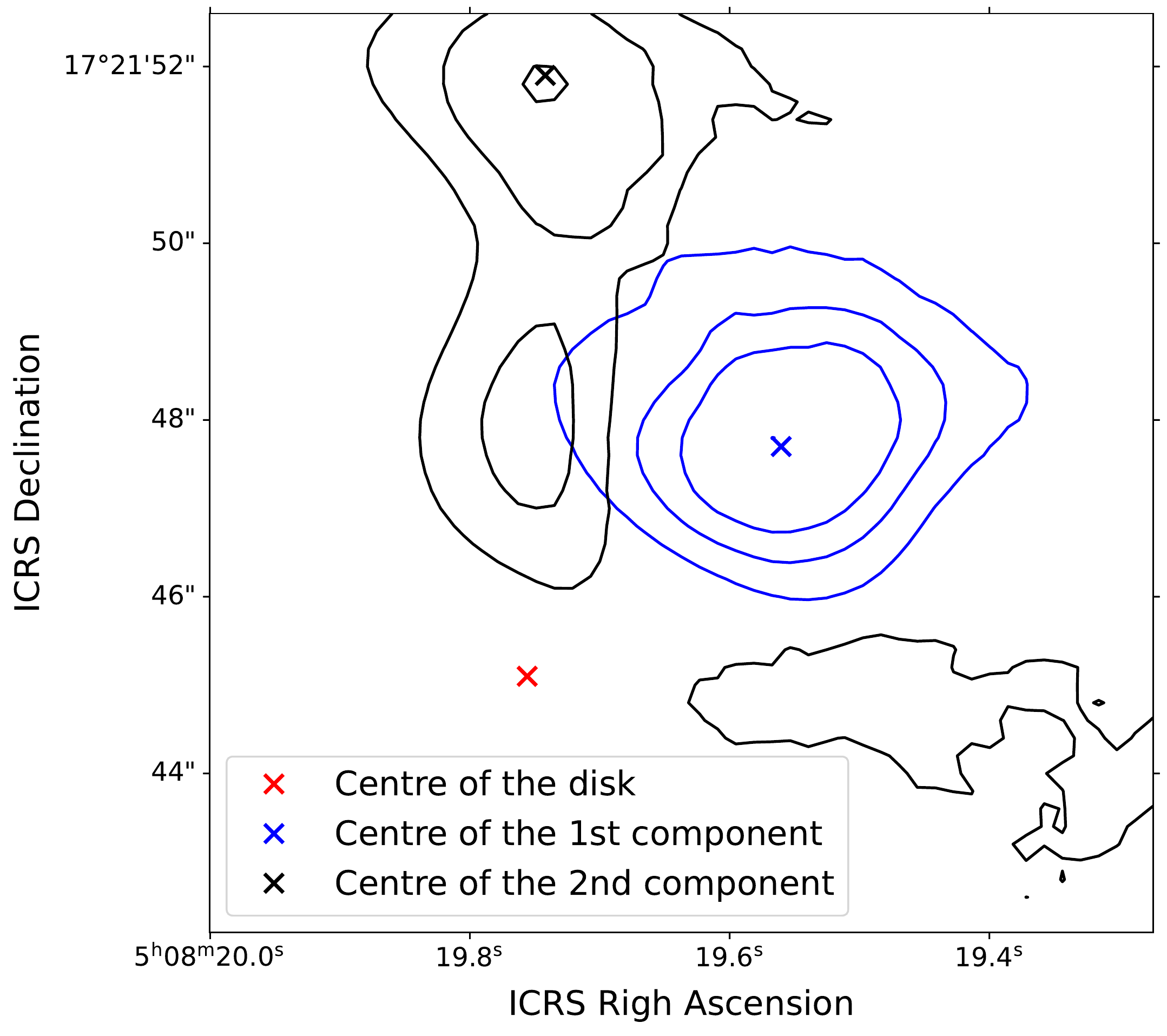}\\
        \caption{Residual \co~emission in NED01 after subtracting the best-fit disk model computed by BBarolo in the {\it Big region}. {\it Left:} Map of the blueshifted \co~residual emission integrated in the range $-290\lesssim v [\rm km~s^{-1}] \lesssim -200$. The blue contours are plotted at 3, 6, 9, 15, and 20$\sigma$. The black cross indicates the centre of the best-fit molecular disk. The blue cross indicates the peak of the \co~residual emission. The residual structure can be enclosed within a box of $8.4''\times8.4''$ ($\sim3$~kpc, see black box overlayed on the map).
        {\it Centre:} Spectrum of the \co~residual emission extracted from an aperture corresponding to the black box shown in the left panel. The two \co~peaks are fit with double-Gaussian functions whose parameters are reported in Table~\ref{tab:bigres1_fit}. {\it Right:} Map of the two blueshifted spectral components visible in the central panel, plotted with different colours. The first component (blue contours)  is the most blueshifted spectral peaks, integrated within $v\in[-400,-200]$~\kms; the second component (in black) corresponds to the secondary peak, integrated between $v\in[-200, 0]$~\kms.
        Contours are plotted at 3, 6, 9, 15, and 20$\sigma$.}
        \label{fig:residuals_bigregion}
\end{figure*}

Figure~\ref{fig:residuals_bigregion} displays the \co~residual emission in NED01 after subtracting the best-fit disk model computed by BBarolo. The left panel of Fig.~\ref{fig:residuals_bigregion}  
shows a contour map of such compact residual structure, obtained integrating the \co~residual flux within the velocity ranges where it is most prominent, that is  $-290\lesssim v [\rm km~s^{-1}] \lesssim -200$. This structure can be enclosed within a box of $8.4''\times8.4''$ ($\sim3$~kpc, see black box overlayed on the map). We find that a similar residual, compact, and  blueshifted structure is common to all three fits performed with BBarolo, and it is actually minimised in this one performed on the {\it Big region} compared to those using the smaller regions.
The spectrum reported in the central panel of Fig.~\ref{fig:residuals_bigregion}, extracted from the black squared aperture in the left panel, shows that the blueshifted \co~residual emission is spectrally resolved into two emission peaks, which we modelled with double-Gaussian functions (see best-fit parameters in Table~\ref{tab:bigres1_fit}). There is also a negative feature at the systemic velocity, common to all three BBarolo fits, which is probably an indication of overfitting of the rotational structure. By mapping separately the two blueshifted spectral components, we find that they trace two apparently independent, spatially offset structures, as shown in the right panel of Fig.~\ref{fig:residuals_bigregion}. Here  we plotted with different colours the first component of this residual \co~emission, which is the most blueshifted one integrated between $v\in[-400, -200]$~\kms\ (blue contours), and the second component, which was integrated between $v\in[-200, 0]$~\kms\ and displayed using the black contours. The first component peaks closer to the dynamic centre of the molecular disk, at a distance of $R=0.7\pm0.2$~kpc, while the second component peaks further away, at a distance of $R=1.5\pm0.2$~kpc. Table~\ref{tab:bigres1_fit} lists for each of the two components their spectral best-fit Gaussian parameters and corresponding \co~flux, \co~luminosity, molecular gas mass, and their physical distance from the best-fit BBarolo disk model. 

\begin{table}[tbp]
        \centering
        \caption{Best-fit Gaussian parameters of the \co~residual spectrum of NED01 shown in  Fig.~\ref{fig:residuals_bigregion}.}             
        \small 
        \label{tab:bigres1_fit}      
        \begin{tabular}{lcc}    
                \toprule
                Parameter & Comp. 1 & Comp. 2 \\
                \midrule
                $v_{cen}$ [\kms] & -239 (2) & -146.9 (1.7) \\
                $\sigma_v$ [\kms] & 31 (2) & 19.8 (1.8) \\
                $S_{\rm CO(1-0)}dv$ [Jy~\kms] & 1.63 (0.14) & 0.98 (0.11) \\
                $L^{\prime}_{\rm CO(1-0)}$ [10$^7$ K km s$^{-1}$ pc$^2$] & 2.5 (0.2) & 1.50 (0.17) \\
                $M_{mol}$$^{\dag}$ [10$^8$~M$_{\odot}$] & 0.5 (0.3) & 0.31 (0.18) \\   
                R [kpc] & 0.7 (0.2) & 1.5 (0.2) \\
                \bottomrule
        \end{tabular}
        
        \begin{flushleft}
                $^{\dag}$ Computed assuming $\alpha_{\rm CO} = 2.1 \pm 1.2$~M$_{\odot}$~(K~\kms~pc$^2$)$^{-1}$, which is the value with associated uncertainty estimated by \cite{2018Cicone2} for the molecular outflow in NGC~6240.
        \end{flushleft}
\end{table}

\subsection{Modelling of the CO(1-0)~disk in NED02}\label{sec:NED02_BBarolo}

Compared to NED01, the \co~moment maps of NED02 in Fig.~\ref{fig:NED02_mom_maps} show a more disturbed, higher inclination, molecular disk. The \co~emission detected by ALMA in this source is also more compact in size, and hence its morphology and kinematics are more affected by beam smearing effects. We used BBarolo to model this \co~disk, following a procedure similar to NED01. However, given the absence of significant extended features, we did not deem it necessary for NED02 to run BBarolo on regions of different sizes. 
The input 3D dataset for the BBarolo modelling was obtained by cropping the cleaned (non-PB-corrected) continuum-subtracted ALMA datacube around NED02 using a box size of $27.4''\times25.4''$ (i.e $9.4\times8.7$ kpc), and selecting a velocity range of $v\in[-1000,+1000]$~\kms, computed with respect to the systemic redshift of NED02. From the ALMA observations, we estimated a position angle of $PA=120^\circ\pm10^\circ$ and an inclination of $\iota=70^\circ\pm15^\circ$. We applied a S/N threshold of $S/N>3$ for each pixel included in the modelling. We used five rings, each of $2\arcsec$ width (0.7~kpc).

\begin{figure}
        \centering
        \includegraphics[clip=true,trim=0.cm 0.cm 0.cm 0.cm, width=\columnwidth]{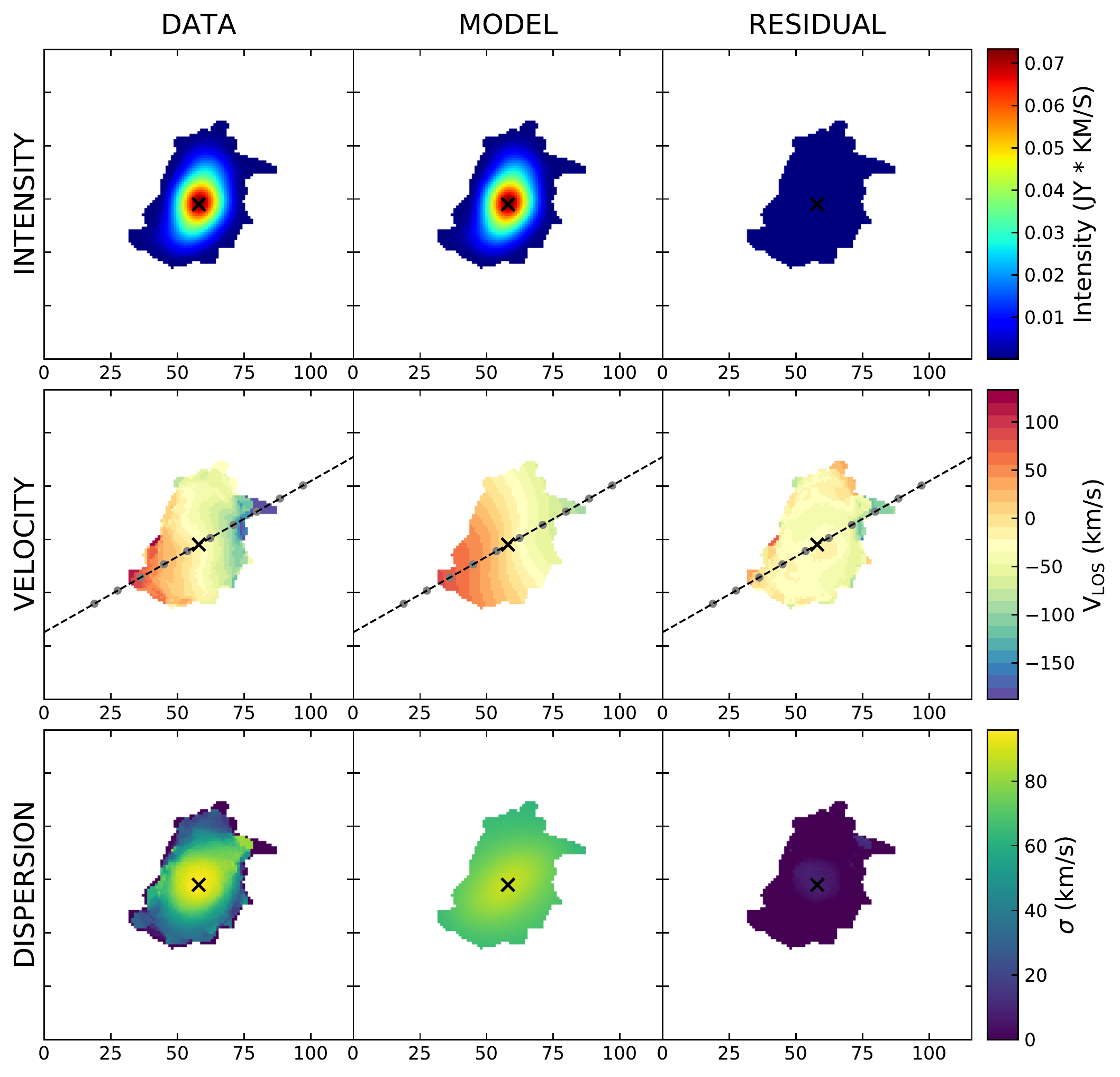}
        \caption{\co~moment maps of the companion galaxy NED02 showing the comparison between the data and the best-fit molecular disk model found by BBarolo.}
        \label{fig:ned02mom}
\end{figure}
\begin{figure}
        \centering
        \includegraphics[clip=true,trim=0.cm 0.cm 0.cm 0.cm, width=.6\columnwidth]{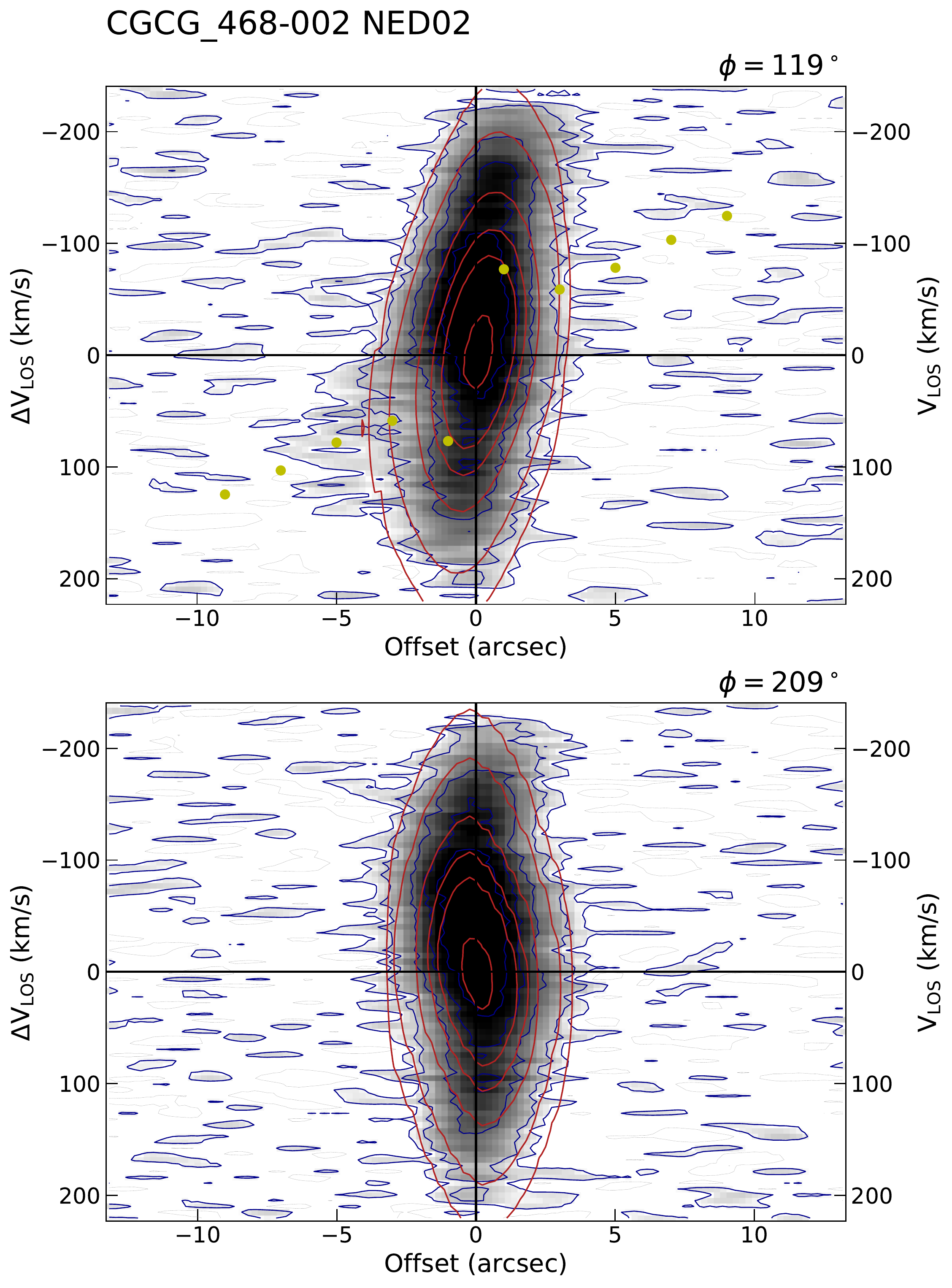}
        \caption{Position velocity diagrams of NED02 showing the comparison between the data (blue contours, plotted at 3, 6, 9, 15, 20, 30, 40, and 50$\sigma$) and the BBarolo best-fit rotating disk model (red contours). The upper panel shows the PV diagram extracted along the major axis, and the lower panel shows the one extracted along the minor axis of rotation.}
        \label{fig:ned02pv}
\end{figure}

\begin{figure*}[tbp]
        \centering 
        \includegraphics[clip=true,trim=0cm 0cm 0cm 0.cm,width=.64\columnwidth]{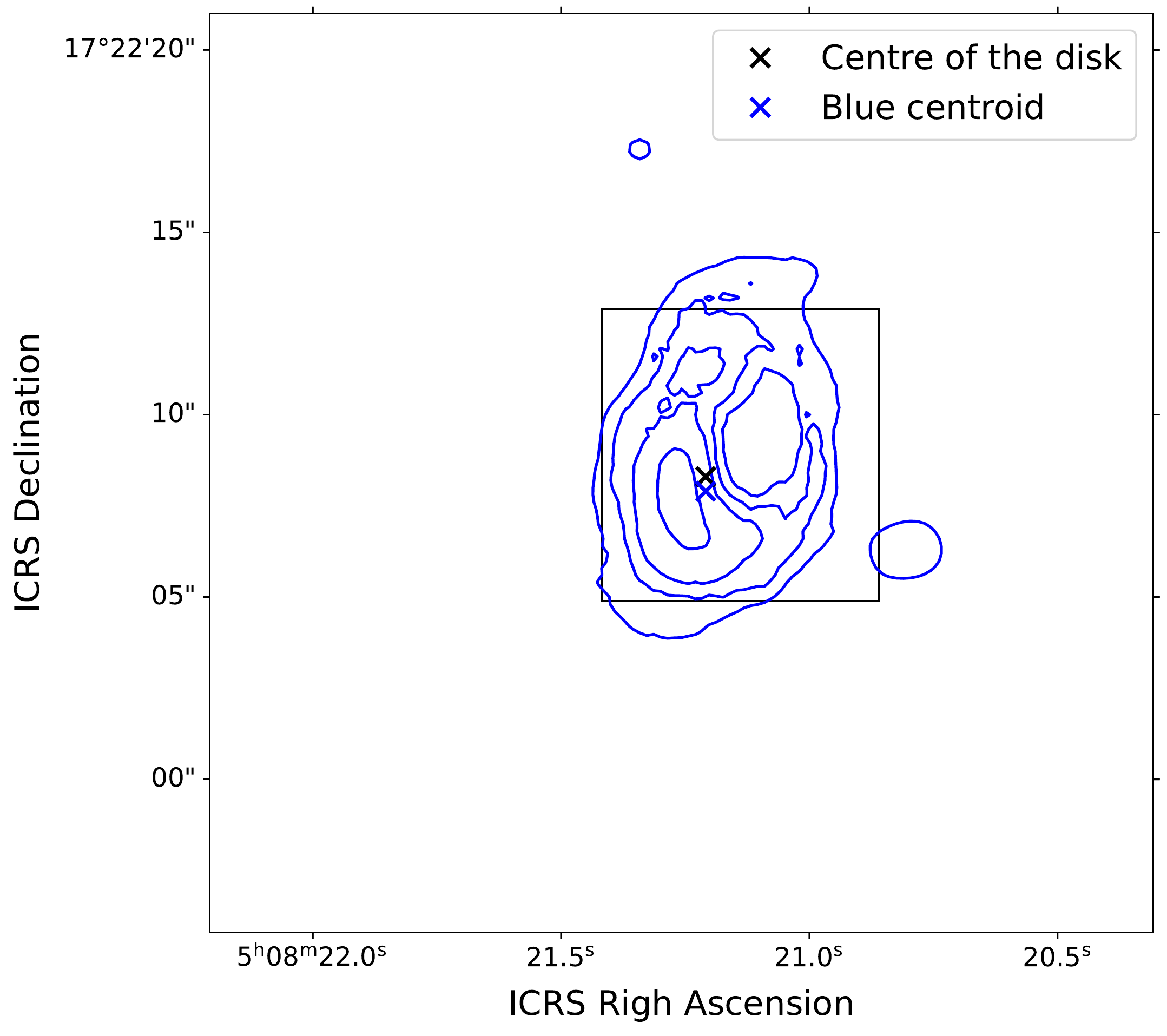}\quad
        \includegraphics[clip=true,trim=0cm 0.2cm 0.5cm 0.cm,width=.68\columnwidth]{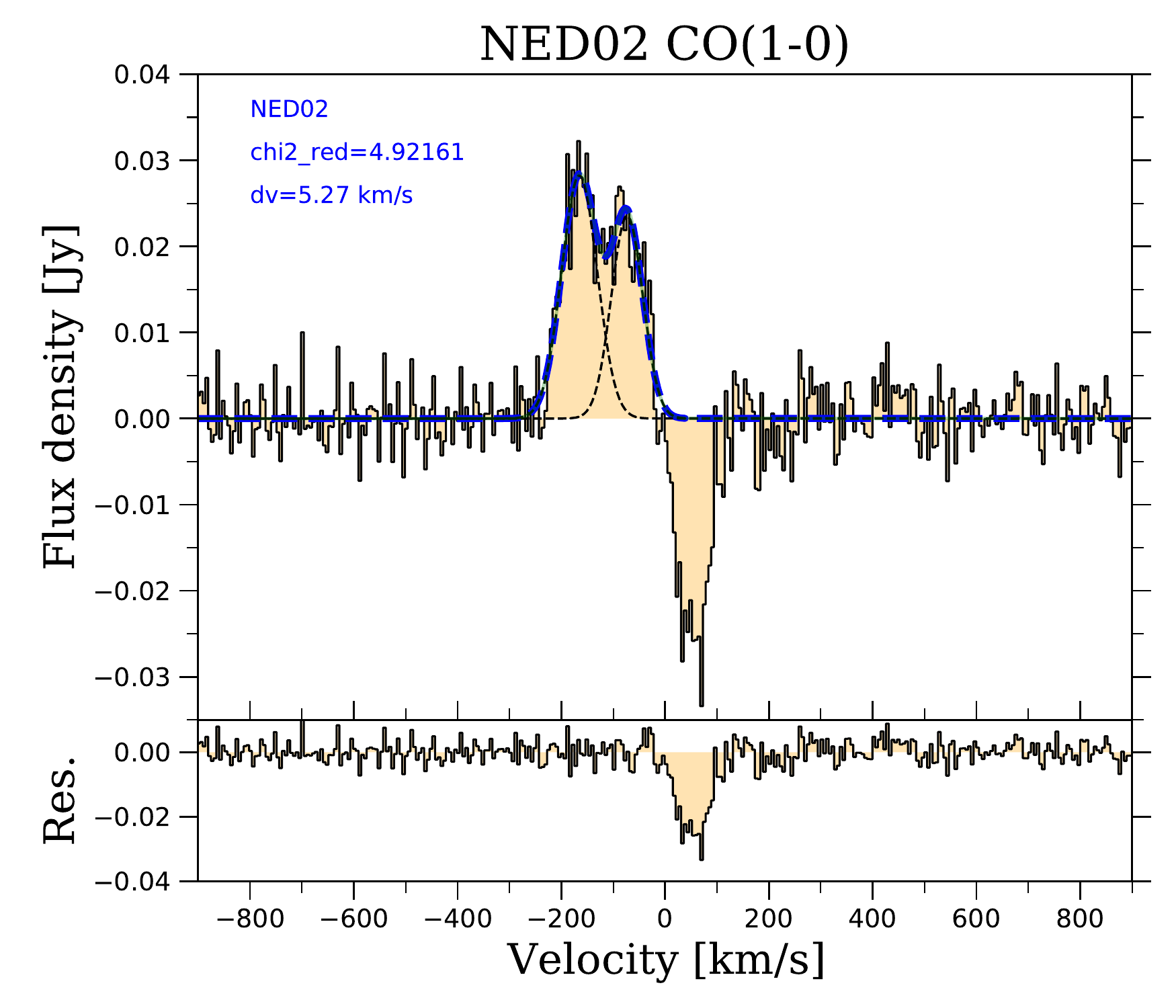}\quad
        \includegraphics[clip=true,trim=0cm 0cm 0cm 0cm,width=.64\columnwidth]{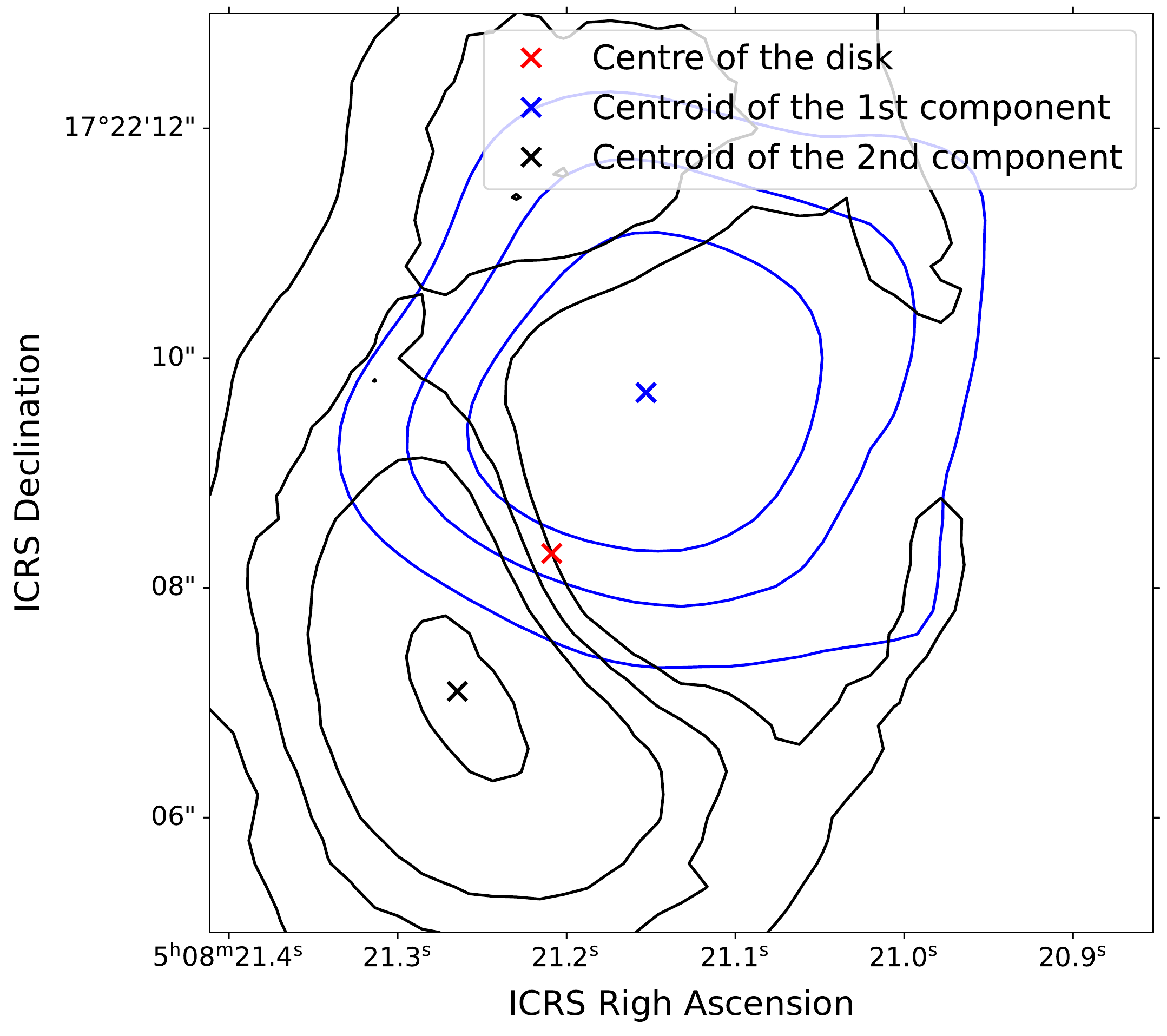}\\
        \caption{Residual \co~emission in NED02 after subtracting the best-fit rotating disk model computed by BBarolo. {\it Left:} Map of the \co~residual emission integrated within  $v\in[-170, 20]$~\kms. Contours are plotted at 3, 6, 9, 15, and 20$\sigma$. The black cross indicates the dynamic centre of the rotating disk and the blue cross shows the centroid of the residual emission. The residual structure can be enclosed within a box of $8''\times8''$ ($\sim3$~kpc, see black box overlayed on the map). {\it Centre:} Spectrum of the \co~residual emission in NED02, extracted from the squared aperture shown in the left panel. The results of the spectral fitting with Gaussian functions are given in Table~\ref{tab:ned02_res}. 
                {\it Right:} Map of the two blueshifted spectral components visible in the residual spectrum shown in the central panel. The first component, integrated within $v\in[-350,-120]$~\kms, is displayed using blue contours. The second component, integrated within  $v\in [-120, 0]$~\kms\ is shown in black. Contours are plotted at 3, 6, 9, 15, and 20$\sigma$.}
        \label{fig:residuals_NED02}
\end{figure*}

Figures~\ref{fig:ned02mom}, ~\ref{fig:ned02pv}, and ~\ref{fig:ned02chann} show the moment maps, PV diagrams, and channel maps of NED02 produced by the fitting code, comparing the data with the best-fit model of a rotating disk. Although BBarolo can reproduce the global rotation pattern, we detect significant residuals, especially at blueshifted velocities, similar to what was found for NED01. The map of such residual \co~emission around NED02, integrated over the spectral range $v\in[-170, 20]$~\kms, is shown in the left panel of Fig.~\ref{fig:residuals_NED02}. The box shown on the map enclosing the 3$\sigma$ level \co~contours has a size of $8''\times8''$ ($\sim3\times3$ kpc). The spectrum of the \co~residual emission extracted from this region (central panel of Fig.~\ref{fig:residuals_NED02}) shows two peaks. The two spectral features peak at different positions (right panel of Fig.~\ref{fig:residuals_NED02}), where the \co~residual emission components integrated between $v\in(-350, -120)$~\kms\ (blue contours) and $v\in(-120, 0)$~\kms\ (black contours) are plotted separately. The two components do not overlap on the map, suggesting they trace physically distinct structures. Table~\ref{tab:ned02_res} lists the best-fit spectral parameters of these two \co~residual features, their corresponding fluxes, luminosities,  molecular gas mass estimates, and their distance from the \co~rotation centre of NED02. 

\begin{table}[tbp]
        \centering
        \caption{Best-fit Gaussian parameters of the \co~residual spectrum of NED02 shown in  Fig.~\ref{fig:residuals_NED02}.}             
        \small 
        \label{tab:ned02_res}      
        \begin{tabular}{lcc}    
                \toprule
                Parameter & Comp. 1 & Comp. 2 \\
                \midrule
                $v_{cen}$ [\kms] & -165 (3) & -73 (3) \\
                $\sigma_v$ [\kms] & 34 (2) & 30 (2) \\
                $S_{\rm CO(1-0)}dv$ [Jy~\kms] & 2.4 (0.2) & 1.8 (0.2) \\
                $L^{\prime}_{\rm CO(1-0)}$ [10$^7$ K km s$^{-1}$ pc$^2$] & 3.7 (0.3) & 2.7 (0.3) \\
                $M_{mol}$$^{\dag}$ [10$^8$~M$_{\odot}$] & 0.8 (0.5) & 0.6 (0.3) \\   
                R [kpc] & 0.8 (0.2) & 1.6 (0.2) \\
                \bottomrule
        \end{tabular}
        
        \begin{flushleft}
                $^{\dag}$ Computed assuming $\alpha_{\rm CO} = 2.1 \pm 1.2$~M$_{\odot}$~(K~\kms~pc$^2$)$^{-1}$; see also Table~\ref{tab:bigres1_fit}.
        \end{flushleft}
\end{table}
\section{X-ray observations}\label{sec:X-ray}
\subsection{Observations and data reduction}

IRAS~05054+1718 was observed three times in the X-ray band. In 2012 it was observed by Nuclear Spectroscopic Telescope Array (\nustar, \citealt{Harrison2013}) for a total of  $\sim 16$ ksec; in  2014  it was the target of a monitoring programme  (totalling $\sim 72$ ksec) performed with the Neil Gehrels Swift Observatory \citep[hereafter \swift,][]{Gehrels2004};  and more recently we obtained a simultaneous deep observation with \xmm\  and   \nustar\ (PI: V. Braito, 2021).  The results of the  \swift\ observations were published by \citet{2015Ballo}, while the 2012 \nustar\ data are reported in \citealt{Ricci2017}.
According to these works, the X-ray spectrum  of  NED01 is that of a moderately absorbed Seyfert 2 galaxy (with photon index $\Gamma\sim 1.7$, column density $\nhsym \sim (1-2) \times 10^{22}$\,\nh, and flux in the 2-10 keV regime $F_{\rm 2-10\, keV}=(0.6-1)\times10^{-11}$ erg cm$^{-2}$ s$^{-1}$) with the  \nustar\  3-10 keV flux in 2012 being a factor 1.6 brighter than the \swift\  value. \citet{2015Ballo} first reported the presence of an   absorption trough  at  $E\sim 7.8$ keV (at $2.1 \sigma$), which was interpreted as the presence of  highly ionised wind,  outflowing with a velocity  of  $v_{\rm out}\sim  0.1\,c$. 

In this work, we re-analyse all X-ray spectra collected for NED01; we re-reduced the \xmm\ and \nustar\ data, while for the \swift\ monitoring we consider the spectrum extracted by \citet{2015Ballo}. In Table~\ref{tab:X-rayobs} we report  the summary of the \xmm\ and \nustar\ observations, while the list of the \swift\ observations can be found in Table 1 of \citet{2015Ballo}.
Although the dominant X-ray source is NED01, the inspection of the soft (0.3-1.5 keV) \XMM\ images revealed a faint X-ray source at the  expected location of the  NED02 as well as possible extended emission. In particular,  for NED02 we detected  $\sim200$ X-ray counts in the 0.3-1.5 keV band with \XMM.  Assuming that they are due to thermal emission ($kT=0.6\pm 0.2$ keV), as generally seen in star-forming galaxies, we derived $F_\mathrm {(0.5-2) keV}\sim 2\times 10^{-14}$ erg cm$^{-2}$ s$^{-1}$ and $L_\mathrm {(0.5-2) keV}\sim 2\times 10^{40}$ erg  s$^{-1}$. 
At higher energies, we cannot assess if there is any X-ray emission from NED02 that could be indicative of a weak or highly obscured AGN   because it falls within the  wings of the PSF of the bright companion. 


\subsection{\xmm}
We processed and cleaned the \xmm\ data     using the Science Analysis Software  (SAS ver. 18.0.0)  and the resulting spectra were analysed  using  standard software packages  (FTOOLS ver. 6.30.1, XSPEC ver. 12.11; \citealt{xspecref}).  The \xmm-EPIC instruments operated in full-frame mode and with the thin filter. 
We first filtered the   EPIC data   for high background, which  only moderately affected  the observations.  
The EPIC-pn,  MOS1, and MOS2  source   spectra  were extracted  using a
circular region  with a   radius of $20''$, while for the   background  we adopted two circular regions  with a radius of $20''$.    
The response matrices and  the ancillary response files at the source position  were generated using the SAS tasks \textit{arfgen} and \textit{rmfgen} and the latest calibration available. After checking for consistency we combined the spectra from each of the individual MOS detectors into a single spectrum. Both the pn and MOS spectra were   then binned to at least  100 counts per bin.

\subsection{\nustar}
\nustar\ observed IRAS~05054+1718 twice;  the second observation was  coordinated with \xmm, starting at the same time but extending for 100 ksec  beyond the \xmm\ observation  (see Table~\ref{tab:X-rayobs}).
We reduced the data following the standard procedure  using the \textsc{heasoft} task \textsc{nupipeline}   (version 0.4.9) of the \nustar\ Data Analysis Software  (\textsc{nustardas}, ver. 1.8.0). We  used the calibration files released with  the CALDB  version 20220426 and applied the standard screening criteria, where we filtered for the passages through the SAS  setting the mode    to  `optimised'  in \textsc{nucalsaa}. For each of the Focal Plane Modules (FPMA and FPMB)  the source spectra were extracted  from a circular region with a radius of $50''$,  while the background spectra were extracted  from two circular regions with a $50''$ radius located on the same detector. Since the second set of observations partially overlap with the \xmm\ exposure,  we also extracted light curves in the 7-10 keV and 10-30 keV band from the same regions  using the \textsc{nuproducts} task to check whether we could use the averaged spectra when performing a joint fit with the \XMM\ spectra. The FPMA and FPMB  background-subtracted  light curves  were then combined into a single curve. The  inspection of  these light curves revealed that in the last 100 ksec, not covered by \xmm, NED01 was slightly brighter; we thus created a 
good time intervals (GTI) file  corresponding to the strictly simultaneous part of the observation. This GTI file was then used to extract  source  and  background    spectra  and  the corresponding  response   files.  

\section{X-ray spectral analysis}\label{sec:X-ray_analysis}

The spectral fits were performed with  XSPEC (ver. 12.11) and in all the models   we included the galactic absorption in the direction of NED01 ($\nhsym=1.93\times 10^{21}$\,\nh, \citealt{nhHI4PICollaboration}), which was modelled with the Tuebingen - Boulder absorption model (tbabs component in XSPEC, \citealt{Wilms2000}). The source spectra were binned to have   at least 20 counts in each energy bin for the \swift\ data, and  100 counts for the EPIC-pn, the EPIC-MOS, and the \nustar\ spectra. For the portion of the \nustar\ spectra that is  strictly simultaneous with {\it XMM} we adopted a lower binning of 50 counts.
We employed $\chi^2$ statistics and errors are quoted at the 90\%\     confidence level for one interesting parameter. All the outflow velocities are relativistically corrected. 

\begin{figure}
        \centering
        \includegraphics[width=0.5\textwidth, angle=0]{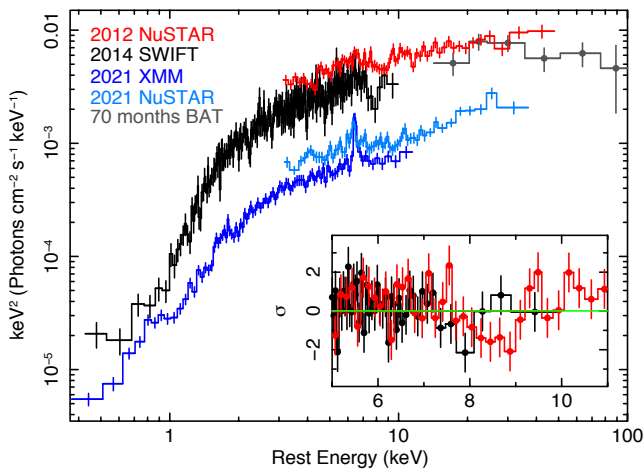}
        \caption{Broad-band  (rest frame 0.3-60 keV) X-ray spectra of all the observations of IRAS~05054+1718. The $E F_{\rm E}$ spectra  are obtained unfolding the data against a simple  power-law model with $\Gamma=2$. The 2012 \nustar\ observation is shown in red, the 2014 \swift\ is in black, while the spectra obtained with the coordinated \xmm\ and \nustar\ observations are shown in dark and light blue, respectively. We also included the averaged \swift-BAT spectrum from the 70-month survey \citep{Baumgartner2013}. The residuals obtained from the joint fit for the \swift\ and \nustar\ spectra using a typical model for a moderately obscured Seyfert 2 as NED01 are shown in the inset. The model, composed of an absorbed  primary power-law component,  a scattered component, with the same photon index ($\Gamma$), and a  reflected component, is able to reproduce the broad-band emission, but it leaves some residuals in absorption (see Sect.~\ref{sec:X-ray_analysis_swiftandnustar}).}
         \label{fig:allxdata}
\end{figure}

In Fig.~\ref{fig:allxdata} we show all the X-ray spectra collected for NED01,  obtained by unfolding the data against a power-law model with photon index $\Gamma=2$. It is clear that  the \swift\ and 
the 2012 \nustar\ observations (black and red data points, respectively) caught  NED01 in a similar bright state, while the most recent \xmm\ and  \nustar\ observations found  the source in a much fainter state (blue and light blue data points, respectively). We therefore  fit separately the 2012-2014 and 2021 epochs. It is also noticeable that the hard ($E>10 $ keV) X-ray emission measured in the first \nustar\ observation is at the same level as the averaged {\it Swift}-BAT spectrum from the 70-month survey\footnote{ \href{http://swift.gsfc.nasa.gov/results/bs70mon/}{http://swift.gsfc.nasa.gov/results/bs70mon/}} suggesting the bright one as the most frequent state (grey data points in Fig.~\ref{fig:allxdata}).

\begin{table*}[tbp]
\centering
\caption{X-ray observations analysed in this work.}\label{tab:X-rayobs} 
\begin{tabular}{lccccc}         
\toprule                  
Observatory & Start Date &  Instrument  &  Elapsed Time &  Exposure$_{\rm(net)}$  & Count rate$_{\rm(net)}$  \\ 
&   [UT time] &       &   [ks]  & [ks]$^{\dag}$  & [count  s$^{-1}$] \\
   \midrule                                
 \nustar\ & 2012-07-23 21:46   & FPMA/B & 26.5& 15.5 & 0.242$\pm$0.003\\     
 {\it XMM } & 2021-09-11 14:38  & MOS1 & 62.2&  58.9 & 0.061$\pm$0.001\\
   {\it XMM } & 2021-09-11 14:38  & MOS2 & 62.1& 59.2 & 0.062$\pm$0.001\\
       \nustar\ & 2021-09-11 14:36   & FPMA/B & 165.6& 81.3 & 0.038$\pm$0.001\\      
    {\it XMM } & 2021-09-11 15:10 & PN & 59.9& 54.2 & 0.174$\pm$0.002\\ 
\bottomrule                               
\end{tabular}

        \begin{flushleft}
        $^{\dag}$ The net exposure times  are  obtained after the screening of the cleaned event files for high background and dead time.
        \end{flushleft}
\end{table*}

\subsection{2012 \nustar\ and 2014 \swift~observations}\label{sec:X-ray_analysis_swiftandnustar}
We performed a joint fit for the \swift\ and \nustar\ spectra and considered the  0.3-10 keV and the 3.5-60 keV  data for the \swift\ and FPM data, respectively. We first tested  the baseline continuum model as reported in \citet{2015Ballo}, which is a typical model for a moderately obscured Seyfert 2 as NED01. 
The model is composed of an absorbed  primary power-law component,  a scattered component, with the same $\Gamma$, and a  reflected component (modelled with the \textsc{xillver} component in XSPEC).  The reflected component  represents the emission  produced by a distant material, such as the putative torus. We tied only the photon index. We allowed the $\nhsym$ of the neutral absorber, and the normalisation of the primary power-law  and of the  reflection components to differ.

This model provides a reasonable fit ($\chi^2=343.9/293$ degrees of freedom) and requires a standard photon index ($\Gamma=1.81\pm 0.06$). The column density of the neutral absorber is   $N_{\mbox{\scriptsize H,  SW}}=(2.4\pm 0.2) \times 10^{22}$\,\nh\  and $N_{\mbox{\scriptsize H,  NU}}=(4.4\pm1.9)\times 10^{22}$\nh\  for the \swift\ and the \nustar\  spectra, respectively. The \nustar\ 3-10 keV flux is a factor of 1.6 higher than the \swift\ value.
Overall, this model is able to reproduce the broad-band emission, but it leaves some residuals in absorption that are shown in the inset of  Fig.~\ref{fig:allxdata}. 
We thus included in the model two  Gaussian absorption lines, and  the fit improved by $\Delta \chi^2/\Delta \nu=19/6$. One feature is detected in the \swift\ spectrum at    $E= 7.8\pm0.3 $ keV ($\Delta \chi^2=6$)    and the second at $E= 8.6\pm 0.2 $ keV ($\Delta \chi^2=13$) in the \nustar\ data. The equivalent widths (EWs) of the absorption troughs are  $220\pm 150$\, eV and $190\pm 100 $ eV in the   \swift\  and \nustar\  data, respectively.

\begin{figure}
        \centering
        \includegraphics[width=0.6\textwidth, angle=0]{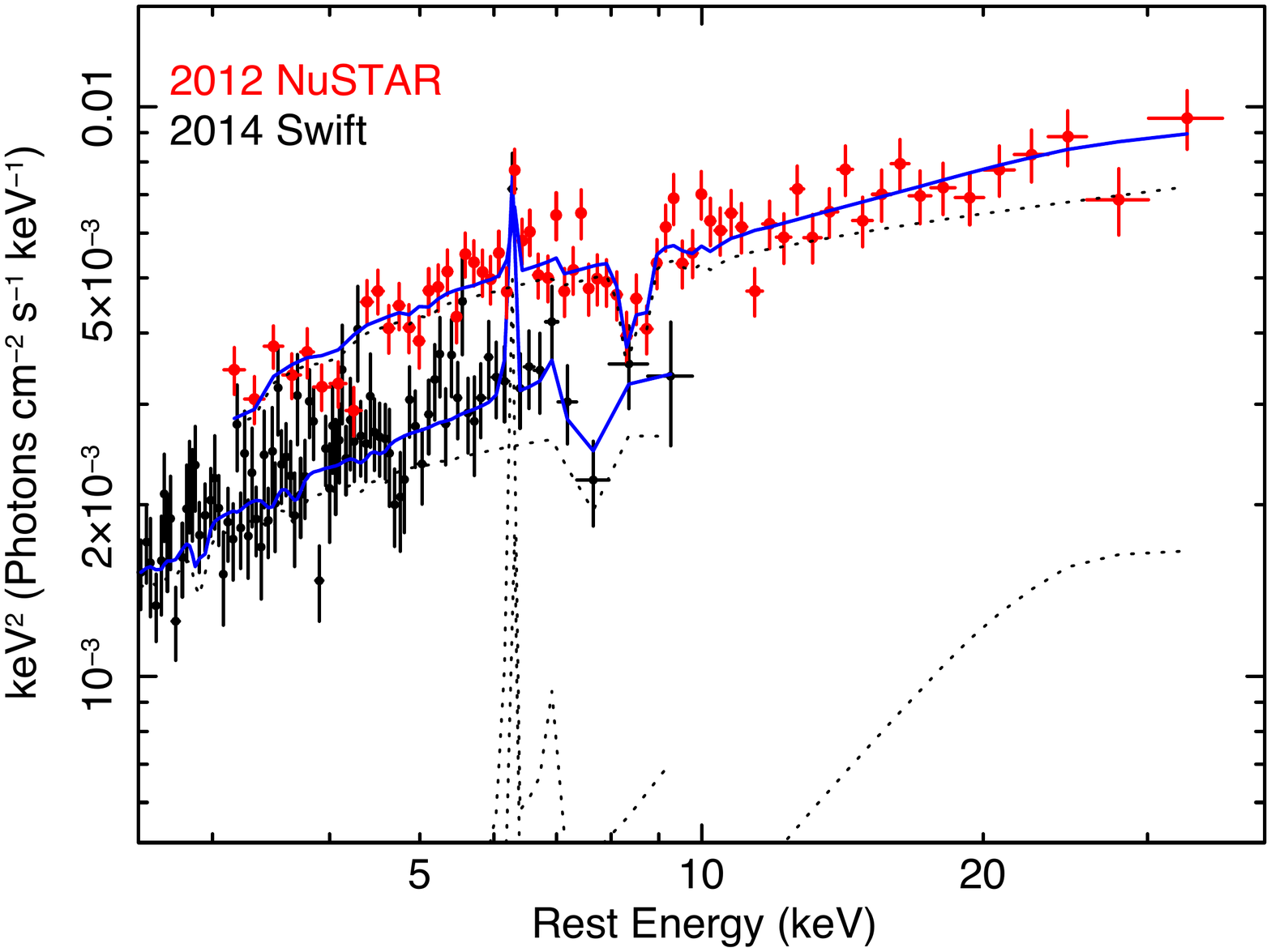}
        \caption{Spectra and best-fit model (blue line) for the   \swift\ (black data points) and 2012 \nustar\ (red) observations. The continuum model   includes  an absorbed power-law component and a reflected component (grey dotted lines). The reflected component (modelled with \textsc{xillver}) also includes the expected   Fe K$\alpha$ emission line at 6.4 keV.  The $\nhsym$ of the  neutral absorber, and the normalisation of the primary power law and of the reflected components are all allowed to vary.  The model also includes an ionised outflowing absorber,   for which     the same ionisation is assumed, but   the column densities  and velocity are allowed to differ in the two spectra. The \swift\  data were rebinned  for plotting purposes.}
         \label{fig:olddata}
\end{figure}

To  self-consistently account for these features, we replaced the two Gaussian absorption lines  with a  multiplicative grid of photoionised absorbers generated with  the \textsc{xstar} photoionisation code (\citealt{xstar}). Since  the lines appear to be broad ($\sigma_{\rm SW} \sim 0.2 $ keV, $\sigma_{\rm NU} =0.3\pm 0.2$ keV), we chose a grid that was generated with a high turbulence velocity  ($v_{\rm turb}=5000$\,km s$^{-1}$) and with  column density  in the range  $\nhsym=10^{21}$  - $3\times10^{24}$\nh\  and  ionisation\footnote{The ionisation parameter is defined as  $\xi=L_{\rm {ion}}/nR^2$,  where $L_{\rm ion}$ is the ionising luminosity in the 1-1000 Rydberg range, $R$ is the distance to the ionising source, and $n$ is the electron density. The units of $\xi$ are \logxi.} log$\,\xi=1-6$.  We assumed   that the ionised absorber  has the same ionisation, but allowed the $\nhsym$ and velocity to be different in the two epochs. 
The inclusion of the fast-outflowing and highly ionised (log$\,\xi=3.4\pm0.5$) absorber improves the fit by $\Delta \chi^2/\nu=20/5$. For the \swift\ observation we found $N_{\rm H, SW}=1.3_{-0.9}^{+1.7}\times10^{23}$\,cm$^{-2}$ and  $v_{\rm{out}}/c=- 0.12_{-0.05}^{+0.02} $, while for the \nustar\ spectrum we found  $N_{\rm H, NU}=1.1\pm 0.6\times10^{23}$\,cm$^{-2}$ and $v_{\rm{out}}/c= -0.23\pm 0.02$. Our independent analysis confirms the results reported by \citet{2015Ballo}, and  also suggests that the wind in IRAS~05054+1718 could be variable or have  multiple velocity  components, as seen for other X-ray disk winds (e.g. PDS~456, \citealt{Matzeu2017,Reeves2018}; IRAS~13224-3809, \citealt{Parker2018}; MCG-03-58-007, \citealt{Braito2022}; HS0810+2554, \citealt{Chartas2016}; and IRAS 17020+4544, \citealt{2015Longinotti}). The best-fit model is given in Table~\ref{tab:bestfits}, while the spectra with their respective best-fit models are shown in Fig.~\ref{fig:olddata}. 
    
\subsection{2021 simultaneous {\it \xmm} and \nustar\ observations}
\begin{figure}
        \centering
        \includegraphics[width=0.55\textwidth, angle=0]{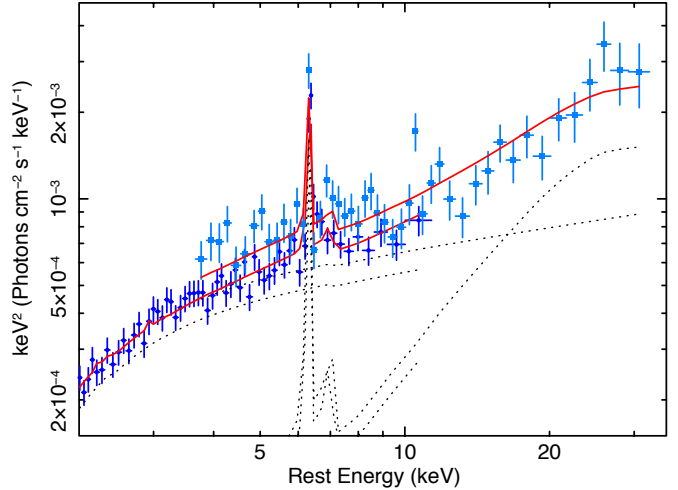}
        \caption{Spectra and best-fit model (red line) for the 2021 \XMM\ (blue data points) and \nustar\  (light blue) observation. The continuum model includes  an absorbed power-law component and a reflected component (grey dotted lines).  }  
         \label{fig:newdata}
 \end{figure}

The 2021 \XMM\ and \nustar\ observation caught NED01  in a faint state (see  Fig.~\ref{fig:allxdata}), where the observed 2-10 keV X-ray flux dropped by a factor of $\sim 4$ with respect to the 2012 and 2014 observations. At this lower flux level the \nustar\ spectrum becomes background-dominated above 40 keV. We thus limited our spectral analysis to the 0.3-10 keV energy band for the \XMM\ data, and the  3.5-40 keV band for the \nustar\ spectrum. 
  Figure~\ref{fig:allxdata} also  reveals that  the average \nustar\ spectrum (light blue) lies well above the EPIC-pn spectrum (blue data points).  If we apply the best-fit continuum model derived in the previous section, we find a cross-normalisation between \XMM\ and \nustar\ of $C=1.36\pm 0.05$, which is well above the current uncertainties in the cross normalisation between the two observatories.  We therefore considered  only the portion of the \nustar\ observation that is strictly simultaneous with \XMM, where we found  $C=1.17\pm 0.06$. We note that the variation is mainly a flux change between the simultaneous spectrum and the last 100 ksec exposure, with no clear spectral variations.
We applied the same baseline continuum model to the new observation, which describes well the overall spectral shape and already provides a reasonable fit ($\chi^2=255.7/205$ d.o.f.) to the spectra with no evident residuals. The broad-band spectra and the best-fit models are presented in Figure~\ref{fig:newdata}. The best-fit parameters are reported in Table~\ref{tab:bestfits}, where we    note that the faint state 
is caused by a drop in the primary emission and not by an increase in absorption, as  is often found for obscured AGN. Here the primary emission drops by a factor of $\sim 4$ with respect to the 2014 \swift\ spectrum and by an order of magnitude with respect to the 2012 \nustar\ observation.

Although there are no residuals that can be associated with the disk wind observed in the past observations, we estimated the upper limits on the column density for a putative ionised absorber outflowing at $v_{\rm{out}} =0.12\;c$ or $ v_{\rm{out}}= 0.23\; c$ for the same ionisation parameter as found previously, of $\log \xi=3.4$. We found that  they are  of the order of $\sim 3\times 10^{22}$\nh\ and $\sim 4\times 10^{22}$\nh~for the slower and the faster phase, respectively. The upper  limits are  only marginally below  the 90\% lower value of  $\nhsym$ estimated in the past observations.

\section{The X-ray disk wind energetics of NED01}\label{sec:X-ray_energetics}

Here we compute a first-order estimate of the energetics of the disk wind detected in NED01. The values were estimated using only the 2012 \nustar\ and the 2014 \swift\ data where the wind was clearly detected.

Following the same arguments presented in many works on  disk winds  (e.g. \citealt{Nardini2015,ReevesBraito2019,Braito2021}), we  estimate the mass outflow rate of the X-ray wind using the equation 
 \begin{equation}\label{eq:xray_ofrate}
 \dot M_{out}^{X-ray}=\Omega~\mu~m_{\mathrm p}~v_{out}~R_{\rm w}~N_{\rm {H}},
\end{equation}
which  assumes a biconical geometry for the flow (\citealt{Krongold2007}). In Eq.~\ref{eq:xray_ofrate}, $\mu$ is  a constant factor  set to $\mu=n_{\rm H}/n_{\rm e}=1.2$ for solar abundances, $\Omega$ is the wind solid angle, $R_{\rm{ w}}$ is the disk wind launching radius,  $\nhsym$  and  $v_{\rm {out}}$ are the column density and the velocity of the disk wind. The main uncertainties in Eq.~\ref{eq:xray_ofrate} are the wind opening angle ($\Omega$) and the launch radius $R_{\rm w}$. We  assume that the wind subtends $\Omega/4\pi = 0.5$.  This is justified by the systematic searches of  ultra-fast disk winds in  bright nearby  AGN (\citealt{Gofford2013}   and \citealt{2010Tombesi}),  which resulted in a detection rate of   about 40\%, thus suggesting that the disk winds have a wide opening angle. A lower limit on the launching radius  can be derived  assuming that  the wind is launched at  its escape radius $R_{\mathrm{min}} = 2\,G\,M_{\rm {BH}}/v^2$.  We note that by adopting this value for $R_{\rm w}$, we obtain the most conservative estimate of the mass outflow rate and energetics (see \citealt{Gofford2015,Tombesi2012}). 
Since the main  uncertainty  in the $R_{\rm min}$ derivation is the black hole mass of NED01, we  decided to normalise the mass outflow rate by the Eddington rate,
\begin{equation}\label{eq:Edd_rate}
 \dot M_{Edd}= \frac{4~\pi~G~m_{\mathrm p}~M_{\rm {BH}}}{\sigma_{\rm {T}}~\eta~c},
\end{equation} 
where $\sigma_{\rm {T}}$ is the Thomson cross section and $\eta=0.1$ is the accretion efficiency. 
Thus, combining Eq.~\ref{eq:xray_ofrate} and Eq.~\ref{eq:Edd_rate}, and substituting for $R_{\rm{min}}$, we obtain
\begin{equation}
\frac{\dot M_{out}^{X-ray}}{\dot M_{Edd}}=2\frac{\Omega}{4\pi}\mu \,N_{\rm {H}}\,\sigma_{\rm {T}}\,\eta\,\left(\frac{v_{\rm{out}}}{c}\right)^{-1}.
\end{equation} 
The X-ray wind kinetic power ($\dot E^{X-ray}_{\rm {kin}}=1/2 \dot M_{out}\, v_{\rm{out}}^2$) normalised by the Eddington luminosity ($L_{\rm{Edd}}=\eta\, \dot M_{\rm {Edd}} \,c^2$) is
\begin{equation}
\frac{\dot E^{X-ray}_{kin}}{L_{Edd}}=\frac{\Omega}{4\pi}\, \mu\, N_{\rm {H}}\,\sigma_{\rm {T}}\frac{v_{\rm{out}}}{c}.
\end{equation} 
For the nuclear wind  detected by \swift, we measure $\nhsym=1.3\errUD{1.7}{0.9}\times 10^{23}$\,\nh\ and $v_{\rm {out}}/c = -0.12\errUD{0.02}{0.05}$. The ({\it Swift}) mass outflow rate  is  thus $\dot M_{ out}^{X-ray}/\dot{M}_{\rm Edd}\sim 0.09$ and the wind kinetic power is then $\dot E^{X-ray}_{out} \sim 0.6 $\% of Eddington. Conversely, for the 2012 \nustar\ spectrum  we derive  $\dot M^{X-ray}_{out}/\dot{M}_{\rm Edd}\sim0.04$ and  $\dot E^{X-ray}_{out}\sim 1 $\% of Eddington. Considering a BH mass of $\sim 10^8 M_\odot$,  as estimated by \citet[][]{Alonso13} in NED01, 
these values correspond  to $\dot E_{out}^{\rm Swift}\sim 8\times 10^{43}$ erg s$^{-1}$ and  $\dot E_{out}^{\rm Nustar}\sim 1.3\times 10^{44}$ erg s$^{-1}$, where the higher value measured in the 2012 observation is driven by the higher velocity of the wind. We note that these estimates are clearly affected by large errors;   only considering the 90\% errors on the column density of the wind, we derive that the kinetic power of the X-ray wind detected in the \swift\ observations could range between $2.5\times 10^{43}$ erg s$^{-1}$ and $1.7\times 10^{44}$  erg s$^{-1}$, while for the  \nustar\ wind the range is $0.6-2\times 10^{44}$ erg s$^{-1}$. The momentum rate  of the X-ray wind  is $\dot p_{out}^{X-ray}\sim (4\pm2)\times 10^{34}$~g~cm~s$^{-2}$  from \nustar, which is consistent with the value obtained from the \swift~data, although the latter has a larger uncertainty.
  
\begin{table*}[tbp]
\centering
  \caption{Summary of the best-fit spectral models for all the X-ray observations.}
   \begin{tabular}{llccc}
\toprule
 Model Component  &  Parameter  &  \swift\ & 2012 \nustar\  & 2021\XMM\ \& \nustar\ \\ 
 \midrule
Primary Power law &$\Gamma$ & $1.80\pm 0.06$ & 1.80$^t$ & $1.75\pm0.06$ \\
& Norm.$^{\dag}$ & $1.9\pm0.2$&  $3.7\pm0.5$ & $0.33\pm0.03$  \\
\midrule
Neutral absorber &$N_\mathrm{H}(\times 10^{22}$ \nh)& $2.3\pm0.2$ & $4.8\pm 1.9$  & $2.7\pm0.2$ \\
\midrule
Disk wind & $N_\mathrm{H} $($\times 10^{23}$ \nh)&$1.3\errUD{1.7}{0.9}$ &  $1.1\pm0.6$&-  \\
              &$log \xi_1$&  $3.4\pm 0.5$ &  $3.4^{t}$ & -\\
               &$v_\mathrm{out}/c$&$-0.12\errUD{0.02}{0.05}$ & $-0.23\pm0.02$  &- \\
&  $  \chi^2/\nu$    &\multicolumn{2}{c}{$323.9/288$} & $257.5/205$   \\  
 \midrule
  &$F_{(2-10)\,\mathrm {keV}}\times 10^{-12}$ (erg~cm$^{-2}$~s$^{-1}$) &   6.4 &  $9.9$ & 1.5 \\
 &$F_{(10-30)\,\mathrm {keV}}\times 10^{-11}$ (erg~cm$^{-2}$~s$^{-1}$) &   -&  $1.3$ & 0.3\\
  &$L_{(2-10)\,\mathrm {keV}}\times 10^{42}$ (erg~s$^{-1}$) &   5.1 &  $9.0$ & 1.2 \\  
   \bottomrule
    \label{tab:bestfits}
\end{tabular}

\begin{flushleft}
    {\it Notes:} $^{\dag}$: The normalisation units are $10^{-3}$ ph keV$^{-1}$\,cm$^{-2}$.
 $^{t}$: Denotes that the  parameter was tied.
\end{flushleft}
\end{table*}

\section{Discussion}\label{sec:discussion}

Since the large amount of energy released by AGN and the associated feedback processes affect the ISM on different physical scales, studying outflows in different ISM phases using multi-wavelength data is the only way to probe the AGN--ISM interplay. A multi-phase analysis is the only method that leads to a complete description of the outflows,    allowing us to determine their physical parameters (e.g. their full extent and mass) and to obtain  an overview of their driving mechanisms to test their theoretical models.
The goal of our ALMA \co~observations was to investigate the presence of a high-velocity molecular outflow in NED01, which is expected to be experiencing strong AGN feedback since it hosts a powerful X-ray wind. At first look, the ALMA \co~data did not show any evidence for a massive powerful molecular outflow in this source, and instead presented clear evidence for \co~rotating disks in both NED01 and its companion galaxy NED02. Our BBarolo modelling confirmed that the bulk of the \co~emissions from the two galaxies can be fitted using a rotating disk model. We then focused on studying the presence of any \co~residual emission, not accounted for by the disk modelling, following a methodology commonly adopted in the literature (e.g. \citealt{2019Sirressi, 2021BewketuBelete, 2022RamosAlmeida}).

By subtracting the best-fit BBarolo disk model from the ALMA datacube, we detected \co~residual emission at the $\sim10$\% level around NED01, for all three models constructed by applying BBarolo on regions of different sizes (see also Appendix~\ref{sec:appendix}). The fit that leaves the least residuals in NED01, with flux corresponding to $\sim8$~\% of its total \co~emission, is the one described in Sect.~\ref{sec:NED01_BBarolo} and  performed on the {\it Big region}. The subtraction of this fit leaves mainly blueshifted residual emission within $v\in (-400, -100)$~\kms, whose contours are enclosed in a squared box of 3~kpc in size, which is rather compact compared to the main rotating disk that has a diameter of $\sim10$~kpc. Quite strikingly, blueshifted residual \co~emission was detected consistently in the data after subtracting all three BBarolo best-fit disk models, with similar morphology and spectral properties, and so we believe it traces structures that may be truly disconnected from the main disk and whose origin is worth exploring further.
On the other hand, redshifted residual emission was not detected in the fit performed on the {\it Big region} around NED01, although it was detected in the other two fits performed on smaller regions, which did not properly account for the extended redshifted \co~components (labelled B and C in the central panel of Fig~\ref{fig:NED01_mom_maps}). 

For NED02, after subtracting the best-fit BBarolo disk model from the ALMA data, we also detected residual \co~emission, corresponding to $\sim7$~\% of the total line flux from this galaxy. Similar to NED01, the residuals are blueshifted with respect to the \co~systemic velocity and have a compact morphology, being enclosed within the central 3~kpc of the source.
In the following, we discuss possible interpretations for the blueshifted \co~residual emission detected in the IRAS~05054+1718 galaxy pair.

\subsubsection*{Hypothesis I: Molecular outflows}

The first hypothesis we consider is that the \co~residuals trace a molecular outflow, which was the scenario we aimed to test with the acquisition of these new ALMA data. 

Under the hypothesis that the \co~residual emission in NED01 traces a molecular outflow, we used the best-fit Gaussian parameters reported in Table~\ref{tab:bigres1_fit} to estimate its energetics, by summing the contribution from the two blueshifted \co~spectral components. For each spectral feature, we set $v_{out}$ equal to the Gaussian central velocity, the outflow radius $R$ as the projected distance from the disk centre (see Table~\ref{tab:bigres1_fit} and \ref{tab:ned02_res} for NED01 and NED02, respectively), and compute the dynamical timescale of the outflow as $\tau_{dyn}^{mol}=R/v_{out}$. Using the molecular gas mass inferred from the \co~luminosity reported in Table~\ref{tab:bigres1_fit}, we computed the mass-loss rate ($\dot{M}^{mol}_{out}=M^{mol}_{out}/\tau_{dyn}$), momentum rate ($v\dot{M}^{mol}_{out}$) and kinetic power ($0.5\dot{M}^{mol}_{out}v^2$) for each of the two CO residual spectral features, and report these numbers in Table~\ref{tab:ned01_ned02_energetics}.
A similar exercise was performed for NED02, with results listed in Table~\ref{tab:ned01_ned02_energetics}.

\begin{table*}[tbp]
        \centering
        \caption{Energetics parameters for the putative molecular outflows in IRAS~05054+1718 NED01 and NED02. } 
        \label{tab:ned01_ned02_energetics}      
        \begin{tabular}{lccc}    
                \toprule
                \multicolumn{4}{c}{IRAS~05054+1718 NED01} \\
                \midrule
                Parameter & Comp. 1 & Comp. 2 & Total \\
                \midrule
                $\tau^{mol}_{dyn}$ [Myr] & 3.0 (0.8) & 10.0 (1.4) & \\
                $\dot{M}^{mol}_{out}$$^{\dag}$ [$M_{\odot}~yr^{-1}$] & 16 (12) & 3 (2) &  19 (14) \\
                $\dot{p}^{mol}_{out}$ [$10^{34}g~cm~s^{-2}$] & 2.4 (2.2) & 0.3 (0.2) & 2.7 (2.4) \\
                $\dot{E}^{mol}_{out}$ [$10^{41}~erg~s^{-1}$] & 3 (2) & 0.22 (0.15) & 3.2 (2.2) \\
                \midrule
                \multicolumn{4}{c}{IRAS~05054+1718 NED02} \\
                \midrule 
                Parameter & Comp. 1 & Comp. 2 & Total \\
                \midrule
                $\tau^{mol}_{dyn}$ [Myr] & 4.7 (1.3) & 21 (4) &\\
                $\dot{M}^{mol}_{out}$$^{\dag}$ [$M_{\odot}~yr^{-1}$] & 17 (15) & 3 (2) &  20 (16) \\
            $\dot{p}^{mol}_{out}$ [$10^{34}~g~cm~s^{-2}$] & 1.8 (1.6) & 0.13 (0.09) & 2.0 (1.7) \\
                $\dot{E}^{mol}_{out}$ [$10^{41}~erg~s^{-1}$] & 1.5 (1.3) & 0.05 (0.04) & 1.6 (1.3) \\
            \bottomrule
        \end{tabular}

        \begin{flushleft}
                {\it Notes:} $^{\dag}$ Computed assuming $\alpha_{\rm CO} = 2.1 \pm 1.2$~M$_{\odot}$~(K~\kms~pc$^2$)$^{-1}$ \citep{2018Cicone2}.
        \end{flushleft}
\end{table*}

The total properties of the putative outflow in NED01 were computed by summing the contribution of all the residual components obtaining a total mass-loss rate of $\dot M^{mol}_{out}=\sum_{i} M^{mol}_{out,i}=19\pm14$~M$_{\odot}$~yr$^{-1}$ (including a systematic uncertainty on the $\alpha_{\rm CO}$ factor), a total momentum rate  of
$\dot p^{mol}_{out}=\sum_{i} v_i\dot{M}^{mol}_{out,i}=(2.7\pm2.4)\times10^{34}$~g~cm~s$^{-1}$, and a total kinetic power of $\dot E^{mol}_{out}=\sum_{i} 0.5\dot{M}^{mol}_{out,i}v_i^2=(3.2\pm2.2)\times10^{41}$~erg~s$^{-1}$.
For NED02, we get a putative outflow mass-loss rate of $20\pm16$~M$_{\odot}$~yr$^{-1}$. However, a molecular outflow, if present, cannot be brighter than the residuals, implying all computed values as upper limits on the outflow energetics parameters.

In a blast-wave AGN feedback scenario \citep[see models by][]{2012Faucher-Giguere,1999Fabian}, the energetics of the molecular outflow detected on galactic scales should be compared with that of the X-ray disk wind, since the former is theoretically expected to be triggered by the latter. According 
to \citet{2015Ballo}, and to the new analysis based on the 2012 \nustar\ data (see Sect. 6),
the X-ray wind in NED01 has a velocity of $v\sim0.2c$ 
with corresponding X-ray wind outflow and momentum rates values, although highly uncertain,\footnote{The errors on the amount of the absorption are of the order of $\sim70$\%} of $\sim0.13$~M$_{\odot}~yr^{-1}$ and $\sim 4 \times 10^{34}$~g~cm~s$^{-1}$, respectively. The kinetic power is estimated to range between 2.5$\times 10^{43}$~erg~s$^{-1}$ and 1.7$\times 10^{44}$~erg~s$^{-1}$ and $0.6-2\times 10^{43}$~erg~s$^{-1}$ for the \swift \, and the \nustar \, winds, respectively. This is two orders of magnitude higher than that inferred for the putative molecular outflow in this source. 
The corresponding ratio of the momentum rates of the molecular and X-ray winds is 
$\dot{p}^{mol}_{out}/\dot{p}^{X-ray}_{out}\sim0.67$, consistent with unity within the large uncertainties. Even in the (unlikely) scenario that all of the \co~residual emission detected in NED01 can be ascribed to a molecular outflow, there is no evidence for a boost in the momentum rate of the molecular outflow with respect to the X-ray wind, which would be an expectation of a `classic' energy-driven outflow model \citep{2012Faucher-Giguere}. 
We can robustly conclude that the current data do not require any momentum rate boost for the large-scale outflow. 


\begin{figure*}
        \centering
        \includegraphics[width=0.8\textwidth]{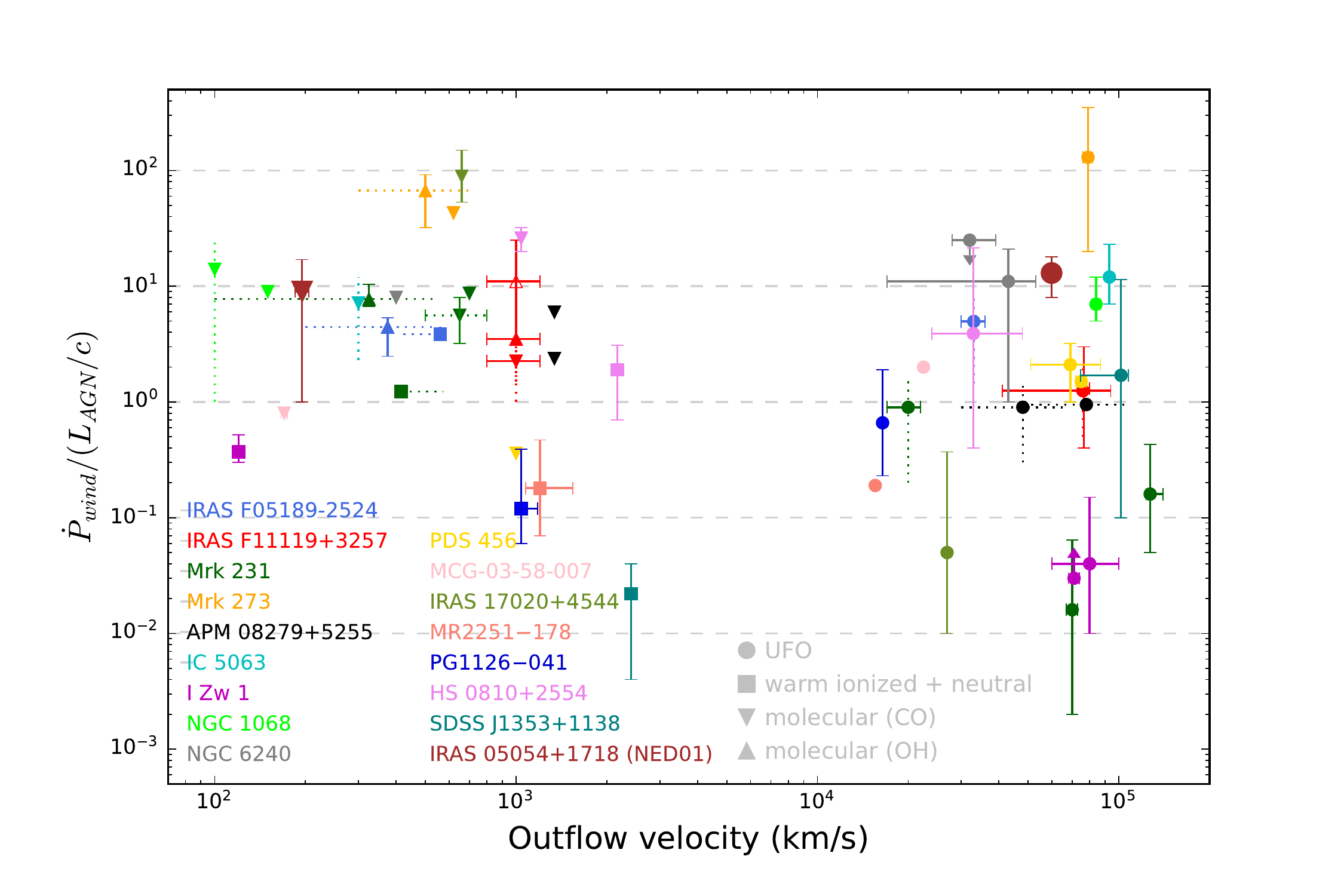}
        \caption{Outflow momentum rate plotted against its velocity for twelve objects with both ultra-fast winds and large-scale galactic outflows, adapted from \citet{2019Smith}. Different colors and shapes identify different objects and outflows. Solid error bars are used for values with upper and lower errors calculated, dotted bars if only a range of value was available, and arrows for limits. Values reported in this plot taken from this work, \citet{2019Bischetti}, \citet{2020Chartas}, \citet{2014Cicone}, \citet{2015Feruglio}, \citet{2017Feruglio}, \citet{2019Fleutsch}, \citet{2014Garcia-Burillo}, \citet{2017Gonzalez-Alfonso},\citet{2018Luminari}, \citet{2018Longinotti}, \citet{2020Marasco}, \citet{2019Mizumoto}, \citet{2015Longinotti}, \citet{2017Rupke}, \citet{2019Sirressi}, \citet{2019Smith}, \citet{2015Tombesi}, \cite{2017Tombesi}, \citet{2021Tozzi}, and \citet{2017Veilleux}.}
        \label{fig:smithplot}
\end{figure*}

The position of NED01 in the outflow momentum rate ratio versus velocity diagram proposed by \cite{2015Feruglio,2015Tombesi} is shown in brown  in Fig.~\ref{fig:smithplot}, compared to other galaxies studied in the literature. One of these sources is the Seyfert 2 galaxy MCG-03-58-007, whose ALMA \co~observations were analysed by \cite{2019Sirressi}. With a similar methodology to this work, based on the subtraction from ALMA \co~data of a rotating disk model obtained with the BBarolo code, \cite{2019Sirressi} detected compact \co~residual emission within $\sim2$ kpc of the AGN, presenting two symmetric components at blueshifted velocities  and one at  redshifted velocities ($v= \pm170$~\kms). These authors cautiously interpreted such \co~residuals as possible evidence for either a compact molecular outflow or a kiloparsec-scale rotating structure disconnected from the main disk. In the outflow scenario, the momentum rate of the putative molecular outflow in MCG-03-58-007 would be $\dot{p}^{mol}_{out}\sim6\times10^{34}$~g~cm~s$^{-2}$, corresponding to 40\% of the momentum rate of the X-ray UFO, consistent with our new findings on IRAS~05054+1718 NED01.
Overall, Fig.~\ref{fig:smithplot} shows that the interplay between nuclear disk winds and large-scale molecular outflows is more complex than   predicted by theoretical blast-wave feedback models, and the data suggest a range of values for the outflow momentum rates on small and large scales.

\begin{figure*}
        \centering
        \includegraphics[width=1\textwidth]{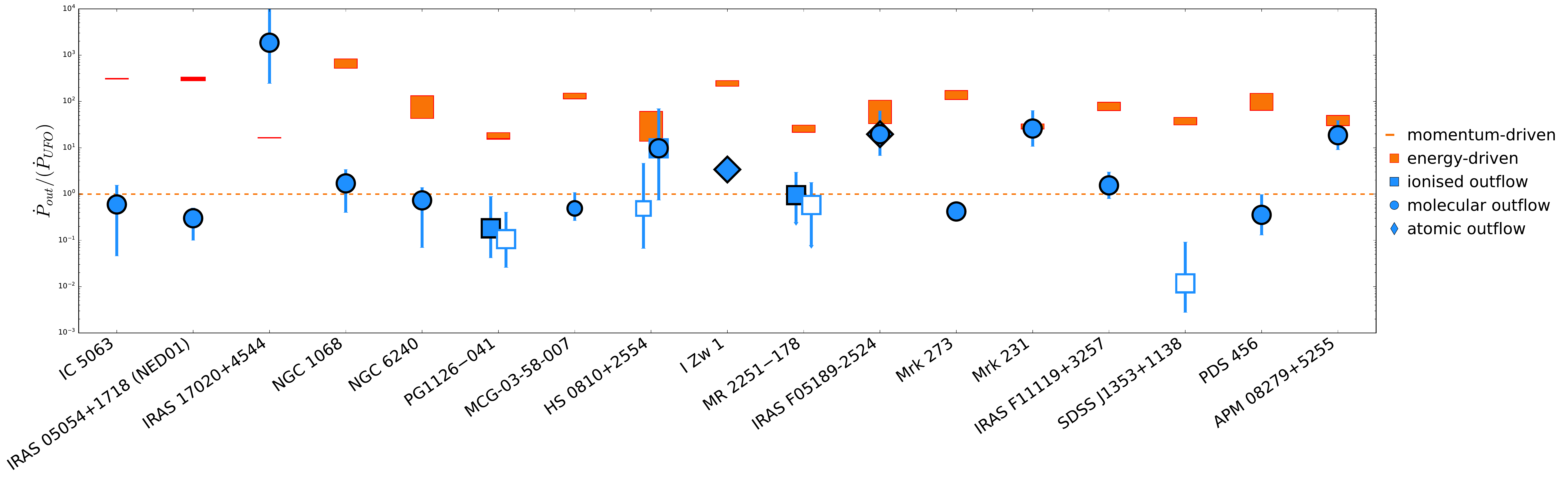}
        \caption{Ratio of the large-scale outflow rate to the UFO momentum rate (adapted from \citealt{2020Marasco,2021Tozzi}). Different shapes identify different gas phases of the winds. The theoretical prediction for momentum-driven ($\dot{P}_{out}/\dot{P}_{UFO}=1$) and energy-driven winds are shown by a dashed line and orange rectangles, respectively. Empty and filled squares indicate H$\alpha$ and [OIII]-based measurements. Values reported in this plot are taken from this work, \citet{2019Bischetti}, \citet{2018Braito}, \citet{2009Chartas}, \citet{2014Cicone}, \citet{2015Feruglio}, \citet{2017Feruglio}, \citet{2014Garcia-Burillo}, \citet{2017Gonzalez-Alfonso}, \citet{2015Longinotti}, \citet{2018Longinotti}, \citet{2020Lutz}, \citet{2020Marasco}, \citet{2019Mizumoto}, \citet{ReevesBraito2019}, \citet{2017Rupke}, \citet{2019Sirressi}, \citet{2019Smith}, \citet{2015Tombesi}, \citet{2021Tozzi}, and \citet{2017Veilleux}.}
        \label{fig:marascoplot}
\end{figure*}

In order to directly compare observational values with predictions of the blast-wave model, Fig.~\ref{fig:marascoplot} shows the ratio of the outflow rate to the X-ray UFO momentum rate, and the ratio of the estimated values of the related energy- and momentum-driven regimes. Following \citet{2020Marasco,2021Tozzi}, we extended this analysis including our targets and some sources from \citet{2019Mizumoto}. Since \citet{2020Marasco,2021Tozzi} adopted $\alpha_{\rm CO} = 0.8 \pm 1.2$~M$_{\odot}$~(K~\kms~pc$^2$)$^{-1}$ \citep{1998Downes,2013Carilli,2013Bolatto} in their analysis, we re-derived the energetics applying this conversion factor, obtaining $\dot{p}^{mol}_{out}/\dot{p}^{X-ray}_{out}\sim0.3$. Overall, the energetics of the outflows appears to be consistent in most of the targets (15 out of 17 objects) with either a momentum-driven (12) or an energy-driven (3) scenario. There are only two cases where the outflow energetics seems completely unrelated to the nuclear source. The momentum ratio of the quasar SDSS~J1353+1138 is $\sim$100 times smaller than the momentum-driven prediction. According to \citet{2021Tozzi}, this low value may be associated with a massive molecular outflow not considered in that work or to a high variability among the quasar activity. The other exception is the Seyfert 1 galaxy IRAS~17020+4544 studied by \citet{2015Longinotti,2018Longinotti}, with its extremely high molecular outflow momentum rate, possibly related to an uncertain estimate of $\dot p^{mol}_{out}$ or again to the high AGN variability.
Overall, according to \citet{2020Marasco}, Fig.~\ref{fig:marascoplot} shows that the blast-wave scenario either in a momentum- or an energy-driven regime appears to describe the interplay between nuclear winds and large-scale outflows, although the majority of measurements suggests the absence of a significant momentum boost. 

Many of the local galaxies with a molecular outflow detection shown in Fig.~\ref{fig:smithplot} (but also see the review by \citealt{2020Veilleux}) are intensely star forming, and so their large-scale outflows may be driven by star formation. In comparison to the extreme ULIRGs that have been the focus of previous investigations, the SFRs of both members of the IRAS~05054+1718 galaxy pair are rather modest. From its total \co~line luminosity (see Table~\ref{tab:mol_masses}), by assuming an $L^{\prime}_{CO(1-0)}$--SFR relation typical of local star-forming galaxies \cite[see e.g.][]{2017Cicone}, we estimate a SFR$_{NED01}$($L^{\prime}_{CO}$)=$2.5\pm0.3$~M$_{\odot}$~yr$^{-1}$ for NED01, which is lower than the range reported in the literature obtained from the IR luminosity and SED modelling (SFR=5-10~M$_{\odot}$~yr$^{-1}$; see Sect.~\ref{sec:targets}), possibly because of AGN contamination affecting photometric measurements of this source. 
For NED02, using the same procedure, we estimate from the total \co~luminosity SFR$_{NED02}$($L^{\prime}_{CO}$) =  $3.6\pm0.5$~M$_{\odot}$~yr$^{-1}$, which is also lower than the values reported in the literature based on IR luminosity. Since the presence of an AGN has not been confirmed in this source, the discrepancy could be  due to a particularly high star formation efficiency in this source (SFE=SFR/M$_{mol}$) or to the presence of additional molecular gas undetected by our ALMA \co~line observations. As shown in Fig.~\ref{fig:source}, the \co~emission from NED02 is rather compact compared to its optical extent.
According to \cite{2005Veilleux}, assuming a maximum coupling efficiency of $\sim10\%$, star formation feedback in galaxies with the SFRs inferred for NED01 and NED02 would drive weak outflows, with mass loss rates of $\dot{M}_{out}<1$~M$_{\odot}$~yr$^{-1}$, which would be undetectable in our data and as not consistent with the blueshifted CO residuals investigated here. 

\subsubsection*{Hypothesis II: Compact rotating structures}

An alternative hypothesis that can account for the residual emission is the presence of a compact structure, similar to the circumnuclear disks (CNDs) detected in some local AGNs \cite[see e.g.][]{2014Garcia-Burillo, 2019Combes}, whose rotation departs from the rotating molecular disk modelled by BBarolo. Although this was a viable hypothesis for the \co~residual components detected in MCG-03-58-007 by \cite{2019Sirressi}, it seems an unlikely scenario in the case of the galaxy pair IRAS~05054+1718 studied in this work. The \co~residuals in both NED01 and NED02 have a complex, asymmetrical, spectral profile, with multiple peaks at blueshifted velocities, and lack a reliable redshifted counterpart. 

\subsubsection*{Hypothesis III: ISM disk asymmetries}

Finally, the blueshifted CO residuals detected in both NED01 and NED02 could be due to other types of asymmetries of the molecular disks that are not properly captured by the BBarolo model. The negative residuals seen around the systemic velocity in both sources would also be consistent with this hypothesis. As discussed by \cite{BAROLO}, although BBarolo manages quite well minor asymmetries such as the disk warp in NGC~5055, the code always favours the model that generates the lowest residuals, which necessarily results in an average rotation curve that may follow one side of the disk more closely than the other. In the case of our target, the ongoing gravitational interaction between the two members of the pair could have already slightly perturbed their ISM, hence producing perturbations of the molecular disk that are not properly captured by BBarolo, causing the residual emission. We note that in this case, the residuals obtained after subtracting the best-fit BBarolo model may not correspond directly to the perturbed components of the ISM, as it could be that the model is forcing the fit to adapt better to the perturbations than to the rest of the (unperturbed) disk, hence producing artificial residuals with spectral and spatial characteristics that are unrelated to the underlying physical process. Therefore, this third scenario can only be tested using higher resolution data.

\section{Summary and conclusions}\label{sec:summary}

In this study we presented new high-sensitivity ALMA \co~observations of the interacting galaxy pair IRAS 05054+1718 (projected distance between the components $\sim29.6''\sim11$ kpc).

We investigated the presence of a galactic-scale molecular outflow in the main target NED01 ($z=0.0178\pm0.0004$, this work) and its possible relation with the powerful X-ray wind detected in its nucleus \citep[see Sect.~\ref{sec:X-ray_analysis} and][]{2015Ballo}. 
Given its moderate SFR \citep[$\sim5 M_\odot /$yr,][]{2014Delooze,2015Pereira}, NED01 is a suitable object to study AGN feedback mechanisms and the blast-wave scenario, usually tested in environments that are extremely bright in the IR and with high SFRs where the AGN contribution is hard to isolate from the stellar contribution.
The companion NED02 ($z=0.016812\pm 0.000003$, this work)  was included in the field of view of our compact-configuration ALMA data. This source is  a  LIRG, classified as composite by \citet{2015Pereira}, but there is  no clear evidence of the presence of an AGN \citep{2012Alonso-Herrero}.

We investigated the distribution and kinematics of the \co~emission at a spatial resolution of $\sim1$ kpc, which appears dominated by the rotating molecular gas disk for both targets. We modelled the disk using 3D-BAROLO and studied the  residual emission to identify structures deviating from the rotation.
We detected residual emission (S/N $\ge15$) at $v  \in  (-400,-100)$ \kms~corresponding to the 8\% of its total \co~emission in the main target NED01, characterised by a compact morphology and extending within 3 kpc of the galaxy centre. We also detected   a similar blueshifted residual \co~emission ($\sim7$ \% of the total \co~flux from this galaxy) in the companion NED02. A molecular outflow in NED01, if present, cannot be brighter than this residual, implying an upper limit on its mass-loss rate of $19\pm14$~M$_{\odot}$~yr$^{-1}$, on its momentum rate of $(2.7\pm2.4)\times10^{34}$~g~cm~s$^{-1}$, and on its kinetic power of $(3.2\pm2.2)\times10^{41}$~erg~s$^{-1}$. For NED02, we obtain a putative outflow mass-loss rate of $20\pm16$~M$_{\odot}$~yr$^{-1}$.

Regarding the X-ray UFO, the analysis performed on the 2012 \nustar\  data confirmed the results already found by \cite{2015Ballo} regarding the presence of a highly ionised  and high-velocity outflowing disk wind in the NED01 nucleus. Although no disk wind features have been detected in the more recent  \xmm\ and \nustar\ data, the upper limits estimated for the  column density of a putative ionised absorber outflowing are only marginally below the 90\% lower value of $N_{\rm H}$  estimated in the past observations. This suggests that a small change in the opacity of the wind can explain a non-detection of the wind at the low-flux state during 2021. This, combined with the factor of 4 lower intrinsic flux of the 2021 X-ray observations, can account for the non-detection in the most recent data.

As for the companion NED02, we derive a faint soft (0.5-2 keV) flux of the  order of $\sim$2$\times
10^{-14}$ erg cm$^{-2}$ s$^{-1}$, corresponding to $L_\mathrm {(0.5-2) keV}\sim 2\times 10^{40}$ erg  s$^{-1}$. 
According to  the $\rm L_X-SFR$ relations derived from several surveys of LIRGs and ULIRGs (see e.g. \citealt{2017Cicone}) this would correspond to a $\rm SFR \sim 7\,M_\odot/$~yr. 
At higher energies, only  X-ray observations with the high spatial resolution, such as that offered by the {\it Chandra} observatory, will allow us to properly separate the X-ray emission of NED02 from the brighter companion NED01 and reach a definitive answer on the possible presence of a weak and/or  obscured AGN in NED02.

We tested the AGN feedback blast-wave scenario in NED01 by comparing the energetics of the UFO and the putative molecular outflow. The results we obtained are not consistent with an energy-driven scenario since there is no evidence for a boost in the momentum rate of the molecular outflow with respect to the X-ray wind. 
We   also considered the hypothesis of an outflow triggered by star formation activity. The SFR of NED01, equal to $\rm SFR_{NED01}(L'_{CO})=2.5\pm$0.3 M$_\odot$ s$^{-1}$, refined using its \co~luminosity and a local tight $\rm L_{CO}-SFR$ relation, is too low to account for a putative star formation-driven molecular outflow.

Since the presence of an AGN has not been confirmed in the companion NED02, the hypothesis of AGN feedback cannot apply to this case. Following the hypothesis of a star formation-driven outflow, we estimated \co~luminosity in NED02 as $\rm SFR_{NED01}(L'_{CO})=3.6\pm$0.5 M$_\odot$ s$^{-1}$, lower than the values reported in the literature based on IR luminosity. As for the main target, the estimated SFR is too low to account for the observed residual emission. 

Because of the ambiguous interpretation of the CO residual emission obtained after subtracting the best-fit disks modelled by BBarolo, the residuals observed in NED01 and NED02 may have a different explanation from that of a molecular outflow. We considered the alternative hypothesis that the detected residual CO emission is due to an additional compact structure similar to a circumnuclear disk whose rotation cannot be associated with the central molecular disk. This hypothesis seems unlikely since the residuals detected in NED01 and NED02 show a complex asymmetrical profile without a redshifted counterpart. 
We   also considered a third hypothesis, according to which the residuals may be due to disk asymmetries that are not properly captured by the BBarolo model. In this case the position and spectral properties of such residuals may not be highly informative of the underlying physical process responsible for the disk perturbations, and only higher resolution data could reveal their origin.

Our study, together with other recent works in the literature (e.g. \citealt{2019Smith}, \citealt{2019Sirressi}, \citealt{2020Veilleux}), has increased the sample of  AGN hosting an ultra-fast X-ray wind with observational constraints on the presence and properties of a molecular outflow, on which it is possible to test the predictions of the AGN-driven feedback scenario. 
The increased statistics of observations appears  to complicate the scenario of AGN-driven feedback mechanisms more and more, drawing a much more complex picture than expected. Furthermore, the high sensitivity to extended structures enabled by the use of a compact array configuration has revealed important information on the close environment of the target galaxy. These data have indeed allowed us to study NED01 simultaneously with its nearby companion NED02, and to investigate the low surface brightness components of their respective molecular disks, hence to compare their molecular ISM morphologies. Our understanding of diffuse low surface brightness and extended molecular gas components due to feedback processes and gravitational interactions will be transformed by a facility such as the Atacama Large Aperture Submillimetre Telescope (AtLAST)\footnote{\href{https://www.atlast.uio.no}{https://www.atlast.uio.no}}.

\begin{acknowledgements}
        This paper makes use of the following ALMA data: ADS/JAO.ALMA$\#$2016.1.00694.S. ALMA is a partnership of ESO (representing its member states), NSF (USA) and NINS (Japan), together with NRC (Canada), MOST and ASIAA (Taiwan), and KASI (Republic of Korea), in cooperation with the Republic of Chile. The Joint ALMA Observatory is operated by ESO, AUI/NRAO and NAOJ. This project has received funding from the European Union’s Horizon 2020 research and innovation programme under grant agreement No 951815 (AtLAST).
    FB acknowledges funding from the European Research Council (ERC) under the European Union’s Horizon 2020 research and innovation programme (Grant agreement No. 851435).
        CV acknowledges funding from PRIN MIUR 2017PH3WAT (`Black hole winds and the baryon life cycle of galaxies'). 
    CC, PS, VB, CV, RDC, LB and MD acknowledge financial contributions from Bando Ricerca Fondamentale INAF 2022 Large Grant “Dual and binary supermassive black holes in the multi-messenger era: from galaxy mergers to gravitational waves” and from the agreement ASI-INAF n.2017-14-H.O.
    VB acknowledges financial support through the NASA grant 80NSSC22K0220.
    VB and JR acknowledge support from NASA grant 80NSSC22K0474.
\end{acknowledgements}

\bibliographystyle{aa} 
\bibliography{Mybibliography} 

\begin{thebibliography}{90}
\expandafter\ifx\csname natexlab\endcsname\relax\def\natexlab#1{#1}\fi

\bibitem[{{Aalto} {et~al.}(2012){Aalto}, {Garcia-Burillo}, {Muller}, {Winters},
  {van der Werf}, {Henkel}, {Costagliola}, \& {Neri}}]{2012Aalto}
{Aalto}, S., {Garcia-Burillo}, S., {Muller}, S., {et~al.} 2012, \aap, 537, A44

\bibitem[{{Alonso-Herrero} {et~al.}(2013{\natexlab{a}}){Alonso-Herrero},
  {Pereira-Santaella}, {Rieke}, {Diamond-Stanic}, {Wang},
  {Hern{\'a}n-Caballero}, \& {Rigopoulou}}]{2013Alonso-Herrero}
{Alonso-Herrero}, A., {Pereira-Santaella}, M., {Rieke}, G.~H., {et~al.}
  2013{\natexlab{a}}, \apj, 765, 78

\bibitem[{{Alonso-Herrero} {et~al.}(2013{\natexlab{b}}){Alonso-Herrero},
  {Pereira-Santaella}, {Rieke}, {Diamond-Stanic}, {Wang},
  {Hern{\'a}n-Caballero}, \& {Rigopoulou}}]{Alonso13}
{Alonso-Herrero}, A., {Pereira-Santaella}, M., {Rieke}, G.~H., {et~al.}
  2013{\natexlab{b}}, \apj, 765, 78

\bibitem[{{Alonso-Herrero} {et~al.}(2012){Alonso-Herrero}, {Pereira-Santaella},
  {Rieke}, \& {Rigopoulou}}]{2012Alonso-Herrero}
{Alonso-Herrero}, A., {Pereira-Santaella}, M., {Rieke}, G.~H., \& {Rigopoulou},
  D. 2012, \apj, 744, 2

\bibitem[{{Andr{\'e}} {et~al.}(2014){Andr{\'e}}, {Di Francesco},
  {Ward-Thompson}, {Inutsuka}, {Pudritz}, \& {Pineda}}]{2014Andre}
{Andr{\'e}}, P., {Di Francesco}, J., {Ward-Thompson}, D., {et~al.} 2014, in
  Protostars and Planets VI, ed. H.~{Beuther}, R.~S. {Klessen}, C.~P.
  {Dullemond}, \& T.~{Henning}, 27

\bibitem[{{Arnaud}(1996)}]{xspecref}
{Arnaud}, K.~A. 1996, in Astronomical Society of the Pacific Conference Series,
  Vol. 101, Astronomical Data Analysis Software and Systems V, ed. G.~H.
  {Jacoby} \& J.~{Barnes}, 17

\bibitem[{{Ballo} {et~al.}(2015){Ballo}, {Severgnini}, {Braito}, {Campana},
  {Della Ceca}, {Moretti}, \& {Vignali}}]{2015Ballo}
{Ballo}, L., {Severgnini}, P., {Braito}, V., {et~al.} 2015, \aap, 581, A87

\bibitem[{{Baumgartner} {et~al.}(2013){Baumgartner}, {Tueller}, {Markwardt},
  {Skinner}, {Barthelmy}, {Mushotzky}, {Evans}, \& {Gehrels}}]{Baumgartner2013}
{Baumgartner}, W.~H., {Tueller}, J., {Markwardt}, C.~B., {et~al.} 2013, \apjs,
  207, 19

\bibitem[{{Bewketu Belete} {et~al.}(2021){Bewketu Belete}, {Andreani},
  {Fern{\'a}ndez-Ontiveros}, {Hatziminaoglou}, {Combes}, {Sirressi}, {Slater},
  {Ricci}, {Dasyra}, {Cicone}, {Aalto}, {Spinoglio}, {Imanishi}, \& {De
  Medeiros}}]{2021BewketuBelete}
{Bewketu Belete}, A., {Andreani}, P., {Fern{\'a}ndez-Ontiveros}, J.~A.,
  {et~al.} 2021, \aap, 654, A24

\bibitem[{{Bigiel} {et~al.}(2008){Bigiel}, {Leroy}, {Walter}, {Brinks}, {de
  Blok}, {Madore}, \& {Thornley}}]{2008Bigiel}
{Bigiel}, F., {Leroy}, A., {Walter}, F., {et~al.} 2008, \aj, 136, 2846

\bibitem[{{Bischetti} {et~al.}(2019){Bischetti}, {Piconcelli}, {Feruglio},
  {Fiore}, {Carniani}, {Brusa}, {Cicone}, {Vignali}, {Bongiorno}, {Cresci},
  {Mainieri}, {Maiolino}, {Marconi}, {Nardini}, \&
  {Zappacosta}}]{2019Bischetti}
{Bischetti}, M., {Piconcelli}, E., {Feruglio}, C., {et~al.} 2019, \aap, 628,
  A118

\bibitem[{{Bolatto} {et~al.}(2013){Bolatto}, {Wolfire}, \&
  {Leroy}}]{2013Bolatto}
{Bolatto}, A.~D., {Wolfire}, M., \& {Leroy}, A.~K. 2013, \araa, 51, 207

\bibitem[{{Bower} {et~al.}(2012){Bower}, {Benson}, \& {Crain}}]{2012Bower}
{Bower}, R.~G., {Benson}, A.~J., \& {Crain}, R.~A. 2012, \mnras, 422, 2816

\bibitem[{{Braito} {et~al.}(2022){Braito}, {Reeves}, {Matzeu}, {Severgnini},
  {Ballo}, {Cicone}, {Ceca}, {Giustini}, \& {Sirressi}}]{Braito2022}
{Braito}, V., {Reeves}, J.~N., {Matzeu}, G., {et~al.} 2022, \apj, 926, 219

\bibitem[{{Braito} {et~al.}(2018){Braito}, {Reeves}, {Matzeu}, {Severgnini},
  {Ballo}, {Caccianiga}, {Campana}, {Cicone}, {Della Ceca}, \&
  {Turner}}]{2018Braito}
{Braito}, V., {Reeves}, J.~N., {Matzeu}, G.~A., {et~al.} 2018, \mnras, 479,
  3592

\bibitem[{{Braito} {et~al.}(2021){Braito}, {Reeves}, {Severgnini}, {Della
  Ceca}, {Ballo}, {Cicone}, {Matzeu}, {Serafinelli}, \&
  {Sirressi}}]{Braito2021}
{Braito}, V., {Reeves}, J.~N., {Severgnini}, P., {et~al.} 2021, \mnras, 500,
  291

\bibitem[{{Carilli} \& {Walter}(2013)}]{2013Carilli}
{Carilli}, C.~L. \& {Walter}, F. 2013, \araa, 51, 105

\bibitem[{{Chambers} {et~al.}(2016){Chambers}, {Magnier}, {Metcalfe},
  {Flewelling}, {Huber}, {Waters}, {Denneau}, {Draper}, {Farrow}, {Finkbeiner},
  {Holmberg}, {Koppenhoefer}, {Price}, {Rest}, {Saglia}, {Schlafly}, {Smartt},
  {Sweeney}, {Wainscoat}, {Burgett}, {Chastel}, {Grav}, {Heasley}, {Hodapp},
  {Jedicke}, {Kaiser}, {Kudritzki}, {Luppino}, {Lupton}, {Monet}, {Morgan},
  {Onaka}, {Shiao}, {Stubbs}, {Tonry}, {White}, {Ba{\~n}ados}, {Bell},
  {Bender}, {Bernard}, {Boegner}, {Boffi}, {Botticella}, {Calamida},
  {Casertano}, {Chen}, {Chen}, {Cole}, {Deacon}, {Frenk}, {Fitzsimmons},
  {Gezari}, {Gibbs}, {Goessl}, {Goggia}, {Gourgue}, {Goldman}, {Grant},
  {Grebel}, {Hambly}, {Hasinger}, {Heavens}, {Heckman}, {Henderson}, {Henning},
  {Holman}, {Hopp}, {Ip}, {Isani}, {Jackson}, {Keyes}, {Koekemoer}, {Kotak},
  {Le}, {Liska}, {Long}, {Lucey}, {Liu}, {Martin}, {Masci}, {McLean}, {Mindel},
  {Misra}, {Morganson}, {Murphy}, {Obaika}, {Narayan}, {Nieto-Santisteban},
  {Norberg}, {Peacock}, {Pier}, {Postman}, {Primak}, {Rae}, {Rai}, {Riess},
  {Riffeser}, {Rix}, {R{\"o}ser}, {Russel}, {Rutz}, {Schilbach}, {Schultz},
  {Scolnic}, {Strolger}, {Szalay}, {Seitz}, {Small}, {Smith}, {Soderblom},
  {Taylor}, {Thomson}, {Taylor}, {Thakar}, {Thiel}, {Thilker}, {Unger},
  {Urata}, {Valenti}, {Wagner}, {Walder}, {Walter}, {Watters}, {Werner},
  {Wood-Vasey}, \& {Wyse}}]{Panstar}
{Chambers}, K.~C., {Magnier}, E.~A., {Metcalfe}, N., {et~al.} 2016, arXiv
  e-prints, arXiv:1612.05560

\bibitem[{{Chartas} {et~al.}(2016){Chartas}, {Cappi}, {Hamann}, {Eracleous},
  {Strickland}, {Giustini}, \& {Misawa}}]{Chartas2016}
{Chartas}, G., {Cappi}, M., {Hamann}, F., {et~al.} 2016, \apj, 824, 53

\bibitem[{{Chartas} {et~al.}(2020){Chartas}, {Davidson}, {Brusa}, {Vignali},
  {Cappi}, {Dadina}, {Cresci}, {Paladino}, {Lanzuisi}, \&
  {Comastri}}]{2020Chartas}
{Chartas}, G., {Davidson}, E., {Brusa}, M., {et~al.} 2020, \mnras, 496, 598

\bibitem[{{Chartas} {et~al.}(2009){Chartas}, {Saez}, {Brandt}, {Giustini}, \&
  {Garmire}}]{2009Chartas}
{Chartas}, G., {Saez}, C., {Brandt}, W.~N., {Giustini}, M., \& {Garmire}, G.~P.
  2009, \apj, 706, 644

\bibitem[{{Cicone} {et~al.}(2017){Cicone}, {Bothwell}, {Wagg}, {M{\o}ller}, {De
  Breuck}, {Zhang}, {Mart{\'\i}n}, {Maiolino}, {Severgnini}, {Aravena},
  {Belfiore}, {Espada}, {Fl{\"u}tsch}, {Impellizzeri}, {Peng}, {Raj},
  {Ram{\'\i}rez-Olivencia}, {Riechers}, \& {Schawinski}}]{2017Cicone}
{Cicone}, C., {Bothwell}, M., {Wagg}, J., {et~al.} 2017, \aap, 604, A53

\bibitem[{{Cicone} {et~al.}(2012){Cicone}, {Feruglio}, {Maiolino}, {Fiore},
  {Piconcelli}, {Menci}, {Aussel}, \& {Sturm}}]{2012Cicone}
{Cicone}, C., {Feruglio}, C., {Maiolino}, R., {et~al.} 2012, \aap, 543, A99

\bibitem[{{Cicone} {et~al.}(2020){Cicone}, {Maiolino}, {Aalto}, {Muller}, \&
  {Feruglio}}]{2020Cicone}
{Cicone}, C., {Maiolino}, R., {Aalto}, S., {Muller}, S., \& {Feruglio}, C.
  2020, \aap, 633, A163

\bibitem[{{Cicone} {et~al.}(2014){Cicone}, {Maiolino}, {Sturm},
  {Graci{\'a}-Carpio}, {Feruglio}, {Neri}, {Aalto}, {Davies}, {Fiore},
  {Fischer}, {Garc{\'\i}a-Burillo}, {Gonz{\'a}lez-Alfonso}, {Hailey-Dunsheath},
  {Piconcelli}, \& {Veilleux}}]{2014Cicone}
{Cicone}, C., {Maiolino}, R., {Sturm}, E., {et~al.} 2014, \aap, 562, A21

\bibitem[{{Cicone} {et~al.}(2018){Cicone}, {Severgnini}, {Papadopoulos},
  {Maiolino}, {Feruglio}, {Treister}, {Privon}, {Zhang}, {Della Ceca}, {Fiore},
  {Schawinski}, \& {Wagg}}]{2018Cicone2}
{Cicone}, C., {Severgnini}, P., {Papadopoulos}, P.~P., {et~al.} 2018, \apj,
  863, 143

\bibitem[{{Combes} {et~al.}(2019){Combes}, {Garc{\'\i}a-Burillo}, {Audibert},
  {Hunt}, {Eckart}, {Aalto}, {Casasola}, {Boone}, {Krips}, {Viti}, {Sakamoto},
  {Muller}, {Dasyra}, {van der Werf}, \& {Martin}}]{2019Combes}
{Combes}, F., {Garc{\'\i}a-Burillo}, S., {Audibert}, A., {et~al.} 2019, \aap,
  623, A79

\bibitem[{{De Looze} {et~al.}(2014){De Looze}, {Cormier}, {Lebouteiller},
  {Madden}, {Baes}, {Bendo}, {Boquien}, {Boselli}, {Clements}, {Cortese},
  {Cooray}, {Galametz}, {Galliano}, {Graci{\'a}-Carpio}, {Isaak}, {Karczewski},
  {Parkin}, {Pellegrini}, {R{\'e}my-Ruyer}, {Spinoglio}, {Smith}, \&
  {Sturm}}]{2014Delooze}
{De Looze}, I., {Cormier}, D., {Lebouteiller}, V., {et~al.} 2014, \aap, 568,
  A62

\bibitem[{{Di Teodoro} \& {Fraternali}(2015)}]{BAROLO}
{Di Teodoro}, E.~M. \& {Fraternali}, F. 2015, \mnras, 451, 3021

\bibitem[{{Downes} \& {Solomon}(1998)}]{1998Downes}
{Downes}, D. \& {Solomon}, P.~M. 1998, \apj, 507, 615

\bibitem[{{Fabian}(1999)}]{1999Fabian}
{Fabian}, A.~C. 1999, \mnras, 308, L39

\bibitem[{{Faucher-Gigu{\`e}re} \& {Quataert}(2012)}]{2012Faucher-Giguere}
{Faucher-Gigu{\`e}re}, C.-A. \& {Quataert}, E. 2012, \mnras, 425, 605

\bibitem[{{Feruglio} {et~al.}(2017){Feruglio}, {Ferrara}, {Bischetti},
  {Downes}, {Neri}, {Ceccarelli}, {Cicone}, {Fiore}, {Gallerani}, {Maiolino},
  {Menci}, {Piconcelli}, {Vietri}, {Vignali}, \& {Zappacosta}}]{2017Feruglio}
{Feruglio}, C., {Ferrara}, A., {Bischetti}, M., {et~al.} 2017, \aap, 608, A30

\bibitem[{{Feruglio} {et~al.}(2015){Feruglio}, {Fiore}, {Carniani},
  {Piconcelli}, {Zappacosta}, {Bongiorno}, {Cicone}, {Maiolino}, {Marconi},
  {Menci}, {Puccetti}, \& {Veilleux}}]{2015Feruglio}
{Feruglio}, C., {Fiore}, F., {Carniani}, S., {et~al.} 2015, \aap, 583, A99

\bibitem[{{Feruglio} {et~al.}(2010){Feruglio}, {Maiolino}, {Piconcelli},
  {Menci}, {Aussel}, {Lamastra}, \& {Fiore}}]{2010Feruglio}
{Feruglio}, C., {Maiolino}, R., {Piconcelli}, E., {et~al.} 2010, \aap, 518,
  L155

\bibitem[{{Fischer} {et~al.}(2010){Fischer}, {Sturm}, {Gonz{\'a}lez-Alfonso},
  {Graci{\'a}-Carpio}, {Hailey-Dunsheath}, {Poglitsch}, {Contursi}, {Lutz},
  {Genzel}, {Sternberg}, {Verma}, \& {Tacconi}}]{2010Fisher}
{Fischer}, J., {Sturm}, E., {Gonz{\'a}lez-Alfonso}, E., {et~al.} 2010, \aap,
  518, L41

\bibitem[{{Fluetsch} {et~al.}(2019){Fluetsch}, {Maiolino}, {Carniani},
  {Marconi}, {Cicone}, {Bourne}, {Costa}, {Fabian}, {Ishibashi}, \&
  {Venturi}}]{2019Fleutsch}
{Fluetsch}, A., {Maiolino}, R., {Carniani}, S., {et~al.} 2019, \mnras, 483,
  4586

\bibitem[{{Garc{\'\i}a-Burillo} {et~al.}(2014){Garc{\'\i}a-Burillo}, {Combes},
  {Usero}, {Aalto}, {Krips}, {Viti}, {Alonso-Herrero}, {Hunt}, {Schinnerer},
  {Baker}, {Boone}, {Casasola}, {Colina}, {Costagliola}, {Eckart}, {Fuente},
  {Henkel}, {Labiano}, {Mart{\'\i}n}, {M{\'a}rquez}, {Muller}, {Planesas},
  {Ramos Almeida}, {Spaans}, {Tacconi}, \& {van der Werf}}]{2014Garcia-Burillo}
{Garc{\'\i}a-Burillo}, S., {Combes}, F., {Usero}, A., {et~al.} 2014, \aap, 567,
  A125

\bibitem[{{Gehrels} {et~al.}(2004){Gehrels}, {Chincarini}, {Giommi}, {Mason},
  {Nousek}, {Wells}, {White}, {Barthelmy}, {Burrows}, {Cominsky}, {Hurley},
  {Marshall}, {M{\'e}sz{\'a}ros}, {Roming}, {Angelini}, {Barbier}, {Belloni},
  {Campana}, {Caraveo}, {Chester}, {Citterio}, {Cline}, {Cropper}, {Cummings},
  {Dean}, {Feigelson}, {Fenimore}, {Frail}, {Fruchter}, {Garmire}, {Gendreau},
  {Ghisellini}, {Greiner}, {Hill}, {Hunsberger}, {Krimm}, {Kulkarni}, {Kumar},
  {Lebrun}, {Lloyd-Ronning}, {Markwardt}, {Mattson}, {Mushotzky}, {Norris},
  {Osborne}, {Paczynski}, {Palmer}, {Park}, {Parsons}, {Paul}, {Rees},
  {Reynolds}, {Rhoads}, {Sasseen}, {Schaefer}, {Short}, {Smale}, {Smith},
  {Stella}, {Tagliaferri}, {Takahashi}, {Tashiro}, {Townsley}, {Tueller},
  {Turner}, {Vietri}, {Voges}, {Ward}, {Willingale}, {Zerbi}, \&
  {Zhang}}]{Gehrels2004}
{Gehrels}, N., {Chincarini}, G., {Giommi}, P., {et~al.} 2004, \apj, 611, 1005

\bibitem[{{Gofford} {et~al.}(2015){Gofford}, {Reeves}, {McLaughlin}, {Braito},
  {Turner}, {Tombesi}, \& {Cappi}}]{Gofford2015}
{Gofford}, J., {Reeves}, J.~N., {McLaughlin}, D.~E., {et~al.} 2015, \mnras,
  451, 4169

\bibitem[{{Gofford} {et~al.}(2013){Gofford}, {Reeves}, {Tombesi}, {Braito},
  {Turner}, {Miller}, \& {Cappi}}]{Gofford2013}
{Gofford}, J., {Reeves}, J.~N., {Tombesi}, F., {et~al.} 2013, \mnras, 430, 60

\bibitem[{{Gonz{\'a}lez-Alfonso} {et~al.}(2017){Gonz{\'a}lez-Alfonso},
  {Fischer}, {Spoon}, {Stewart}, {Ashby}, {Veilleux}, {Smith}, {Sturm},
  {Farrah}, {Falstad}, {Mel{\'e}ndez}, {Graci{\'a}-Carpio}, {Janssen}, \&
  {Lebouteiller}}]{2017Gonzalez-Alfonso}
{Gonz{\'a}lez-Alfonso}, E., {Fischer}, J., {Spoon}, H.~W.~W., {et~al.} 2017,
  \apj, 836, 11

\bibitem[{{Harrison} {et~al.}(2013){Harrison}, {Craig}, {Christensen},
  {Hailey}, {Zhang}, {Boggs}, {Stern}, {Cook}, {Forster}, {Giommi},
  {Grefenstette}, {Kim}, {Kitaguchi}, {Koglin}, {Madsen}, {Mao}, {Miyasaka},
  {Mori}, {Perri}, {Pivovaroff}, {Puccetti}, {Rana}, {Westergaard}, {Willis},
  {Zoglauer}, {An}, {Bachetti}, {Barri{\`e}re}, {Bellm}, {Bhalerao},
  {Brejnholt}, {Fuerst}, {Liebe}, {Markwardt}, {Nynka}, {Vogel}, {Walton},
  {Wik}, {Alexander}, {Cominsky}, {Hornschemeier}, {Hornstrup}, {Kaspi},
  {Madejski}, {Matt}, {Molendi}, {Smith}, {Tomsick}, {Ajello}, {Ballantyne},
  {Balokovi{\'c}}, {Barret}, {Bauer}, {Blandford}, {Brandt}, {Brenneman},
  {Chiang}, {Chakrabarty}, {Chenevez}, {Comastri}, {Dufour}, {Elvis}, {Fabian},
  {Farrah}, {Fryer}, {Gotthelf}, {Grindlay}, {Helfand}, {Krivonos}, {Meier},
  {Miller}, {Natalucci}, {Ogle}, {Ofek}, {Ptak}, {Reynolds}, {Rigby},
  {Tagliaferri}, {Thorsett}, {Treister}, \& {Urry}}]{Harrison2013}
{Harrison}, F.~A., {Craig}, W.~W., {Christensen}, F.~E., {et~al.} 2013, \apj,
  770, 103

\bibitem[{{HI4PI Collaboration} {et~al.}(2016){HI4PI Collaboration}, {Ben
  Bekhti}, {Fl{\"o}er}, {Keller}, {Kerp}, {Lenz}, {Winkel}, {Bailin},
  {Calabretta}, {Dedes}, {Ford}, {Gibson}, {Haud}, {Janowiecki}, {Kalberla},
  {Lockman}, {McClure-Griffiths}, {Murphy}, {Nakanishi}, {Pisano}, \&
  {Staveley-Smith}}]{nhHI4PICollaboration}
{HI4PI Collaboration}, {Ben Bekhti}, N., {Fl{\"o}er}, L., {et~al.} 2016, \aap,
  594, A116

\bibitem[{{Hopkins} {et~al.}(2014){Hopkins}, {Kere{\v{s}}}, {O{\~n}orbe},
  {Faucher-Gigu{\`e}re}, {Quataert}, {Murray}, \& {Bullock}}]{2014Hopkins}
{Hopkins}, P.~F., {Kere{\v{s}}}, D., {O{\~n}orbe}, J., {et~al.} 2014, \mnras,
  445, 581

\bibitem[{{Howell} {et~al.}(2010){Howell}, {Armus}, {Mazzarella}, {Evans},
  {Surace}, {Sanders}, {Petric}, {Appleton}, {Bothun}, {Bridge}, {Chan},
  {Charmandaris}, {Frayer}, {Haan}, {Inami}, {Kim}, {Lord}, {Madore},
  {Melbourne}, {Schulz}, {U}, {Vavilkin}, {Veilleux}, \& {Xu}}]{2010Howell}
{Howell}, J.~H., {Armus}, L., {Mazzarella}, J.~M., {et~al.} 2010, \apj, 715,
  572

\bibitem[{{Huchra} {et~al.}(2012){Huchra}, {Macri}, {Masters}, {Jarrett},
  {Berlind}, {Calkins}, {Crook}, {Cutri}, {Erdo{\v{g}}du}, {Falco}, {George},
  {Hutcheson}, {Lahav}, {Mader}, {Mink}, {Martimbeau}, {Schneider},
  {Skrutskie}, {Tokarz}, \& {Westover}}]{2012Huchra}
{Huchra}, J.~P., {Macri}, L.~M., {Masters}, K.~L., {et~al.} 2012, \apjs, 199,
  26

\bibitem[{{Kallman} {et~al.}(2004){Kallman}, {Palmeri}, {Bautista}, {Mendoza},
  \& {Krolik}}]{xstar}
{Kallman}, T.~R., {Palmeri}, P., {Bautista}, M.~A., {Mendoza}, C., \& {Krolik},
  J.~H. 2004, \apjs, 155, 675

\bibitem[{{King}(2010)}]{2010King}
{King}, A.~R. 2010, \mnras, 402, 1516

\bibitem[{{Krongold} {et~al.}(2007){Krongold}, {Nicastro}, {Elvis},
  {Brickhouse}, {Binette}, {Mathur}, \&
  {Jim{\'e}nez-Bail{\'o}n}}]{Krongold2007}
{Krongold}, Y., {Nicastro}, F., {Elvis}, M., {et~al.} 2007, \apj, 659, 1022

\bibitem[{{Lada} {et~al.}(2010){Lada}, {Lombardi}, \& {Alves}}]{2010Lada}
{Lada}, C.~J., {Lombardi}, M., \& {Alves}, J.~F. 2010, \apj, 724, 687

\bibitem[{{Leroy} {et~al.}(2008){Leroy}, {Walter}, {Brinks}, {Bigiel}, {de
  Blok}, {Madore}, \& {Thornley}}]{2008Leroy}
{Leroy}, A.~K., {Walter}, F., {Brinks}, E., {et~al.} 2008, \aj, 136, 2782

\bibitem[{{Longinotti} {et~al.}(2015){Longinotti}, {Krongold}, {Guainazzi},
  {Giroletti}, {Panessa}, {Costantini}, {Santos-Lleo}, \&
  {Rodriguez-Pascual}}]{2015Longinotti}
{Longinotti}, A.~L., {Krongold}, Y., {Guainazzi}, M., {et~al.} 2015, \apjl,
  813, L39

\bibitem[{{Longinotti} {et~al.}(2018){Longinotti}, {Vega}, {Krongold},
  {Aretxaga}, {Yun}, {Chavushyan}, {Feruglio}, {G{\'o}mez-Ruiz}, {Monta{\~n}a},
  {Le{\'o}n-Tavares}, {Olgu{\'\i}n-Iglesias}, {Giroletti}, {Guainazzi},
  {Kotilainen}, {Panessa}, {Zapata}, {Cruz-Gonzalez}, {Pati{\~n}o-{\'A}lvarez},
  {Rosa-Gonzalez}, {Carrami{\~n}ana}, {Carrasco}, {Costantini}, {Dultzin},
  {Guichard}, {Puerari}, \& {Santos-Lleo}}]{2018Longinotti}
{Longinotti}, A.~L., {Vega}, O., {Krongold}, Y., {et~al.} 2018, \apjl, 867, L11

\bibitem[{{Luminari} {et~al.}(2018){Luminari}, {Piconcelli}, {Tombesi},
  {Zappacosta}, {Fiore}, {Piro}, \& {Vagnetti}}]{2018Luminari}
{Luminari}, A., {Piconcelli}, E., {Tombesi}, F., {et~al.} 2018, \aap, 619, A149

\bibitem[{{Lutz} {et~al.}(2020){Lutz}, {Sturm}, {Janssen}, {Veilleux}, {Aalto},
  {Cicone}, {Contursi}, {Davies}, {Feruglio}, {Fischer}, {Fluetsch},
  {Garcia-Burillo}, {Genzel}, {Gonz{\'a}lez-Alfonso}, {Graci{\'a}-Carpio},
  {Herrera-Camus}, {Maiolino}, {Schruba}, {Shimizu}, {Sternberg}, {Tacconi}, \&
  {Wei{\ss}}}]{2020Lutz}
{Lutz}, D., {Sturm}, E., {Janssen}, A., {et~al.} 2020, \aap, 633, A134

\bibitem[{{Makarov} {et~al.}(2014){Makarov}, {Prugniel}, {Terekhova},
  {Courtois}, \& {Vauglin}}]{2014Makarov}
{Makarov}, D., {Prugniel}, P., {Terekhova}, N., {Courtois}, H., \& {Vauglin},
  I. 2014, \aap, 570, A13

\bibitem[{{Marasco} {et~al.}(2020){Marasco}, {Cresci}, {Nardini}, {Mannucci},
  {Marconi}, {Tozzi}, {Tozzi}, {Amiri}, {Venturi}, {Piconcelli}, {Lanzuisi},
  {Tombesi}, {Mingozzi}, {Perna}, {Carniani}, {Brusa}, \& {di Serego
  Alighieri}}]{2020Marasco}
{Marasco}, A., {Cresci}, G., {Nardini}, E., {et~al.} 2020, \aap, 644, A15

\bibitem[{{Matzeu} {et~al.}(2017){Matzeu}, {Reeves}, {Braito}, {Nardini},
  {McLaughlin}, {Lobban}, {Tombesi}, \& {Costa}}]{Matzeu2017}
{Matzeu}, G.~A., {Reeves}, J.~N., {Braito}, V., {et~al.} 2017, \mnras, 472, L15

\bibitem[{{McMullin} {et~al.}(2007){McMullin}, {Waters}, {Schiebel}, {Young},
  \& {Golap}}]{CASA}
{McMullin}, J.~P., {Waters}, B., {Schiebel}, D., {Young}, W., \& {Golap}, K.
  2007, in Astronomical Society of the Pacific Conference Series, Vol. 376,
  Astronomical Data Analysis Software and Systems XVI, ed. R.~A. {Shaw},
  F.~{Hill}, \& D.~J. {Bell}, 127

\bibitem[{{Mizumoto} {et~al.}(2019){Mizumoto}, {Izumi}, \&
  {Kohno}}]{2019Mizumoto}
{Mizumoto}, M., {Izumi}, T., \& {Kohno}, K. 2019, \apj, 871, 156

\bibitem[{{Nardini} {et~al.}(2015){Nardini}, {Reeves}, {Gofford}, {Harrison},
  {Risaliti}, {Braito}, {Costa}, {Matzeu}, {Walton}, {Behar}, {Boggs},
  {Christensen}, {Craig}, {Hailey}, {Matt}, {Miller}, {O'Brien}, {Stern},
  {Turner}, \& {Ward}}]{Nardini2015}
{Nardini}, E., {Reeves}, J.~N., {Gofford}, J., {et~al.} 2015, Science, 347, 860

\bibitem[{{Omont}(2007)}]{2007Omont}
{Omont}, A. 2007, Reports on Progress in Physics, 70, 1099

\bibitem[{{Parker} {et~al.}(2018){Parker}, {Matzeu}, {Guainazzi},
  {Kalfountzou}, {Miniutti}, {Santos-Lle{\'o}}, \& {Schartel}}]{Parker2018}
{Parker}, M.~L., {Matzeu}, G.~A., {Guainazzi}, M., {et~al.} 2018, \mnras, 480,
  2365

\bibitem[{{Pereira-Santaella} {et~al.}(2015){Pereira-Santaella},
  {Alonso-Herrero}, {Colina}, {Miralles-Caballero}, {P{\'e}rez-Gonz{\'a}lez},
  {Arribas}, {Bellocchi}, {Cazzoli}, {D{\'\i}az-Santos}, \& {Piqueras
  L{\'o}pez}}]{2015Pereira}
{Pereira-Santaella}, M., {Alonso-Herrero}, A., {Colina}, L., {et~al.} 2015,
  \aap, 577, A78

\bibitem[{{Planck Collaboration} {et~al.}(2016){Planck Collaboration}, {Ade},
  {Aghanim}, {Arnaud}, {Ashdown}, {Aumont}, {Baccigalupi}, {Banday},
  {Barreiro}, {Bartlett}, {Bartolo}, {Battaner}, {Battye}, {Benabed},
  {Beno{\^\i}t}, {Benoit-L{\'e}vy}, {Bernard}, {Bersanelli}, {Bielewicz},
  {Bock}, {Bonaldi}, {Bonavera}, {Bond}, {Borrill}, {Bouchet}, {Boulanger},
  {Bucher}, {Burigana}, {Butler}, {Calabrese}, {Cardoso}, {Catalano},
  {Challinor}, {Chamballu}, {Chary}, {Chiang}, {Chluba}, {Christensen},
  {Church}, {Clements}, {Colombi}, {Colombo}, {Combet}, {Coulais}, {Crill},
  {Curto}, {Cuttaia}, {Danese}, {Davies}, {Davis}, {de Bernardis}, {de Rosa},
  {de Zotti}, {Delabrouille}, {D{\'e}sert}, {Di Valentino}, {Dickinson},
  {Diego}, {Dolag}, {Dole}, {Donzelli}, {Dor{\'e}}, {Douspis}, {Ducout},
  {Dunkley}, {Dupac}, {Efstathiou}, {Elsner}, {En{\ss}lin}, {Eriksen},
  {Farhang}, {Fergusson}, {Finelli}, {Forni}, {Frailis}, {Fraisse},
  {Franceschi}, {Frejsel}, {Galeotta}, {Galli}, {Ganga}, {Gauthier}, {Gerbino},
  {Ghosh}, {Giard}, {Giraud-H{\'e}raud}, {Giusarma}, {Gjerl{\o}w},
  {Gonz{\'a}lez-Nuevo}, {G{\'o}rski}, {Gratton}, {Gregorio}, {Gruppuso},
  {Gudmundsson}, {Hamann}, {Hansen}, {Hanson}, {Harrison}, {Helou},
  {Henrot-Versill{\'e}}, {Hern{\'a}ndez-Monteagudo}, {Herranz}, {Hildebrandt},
  {Hivon}, {Hobson}, {Holmes}, {Hornstrup}, {Hovest}, {Huang}, {Huffenberger},
  {Hurier}, {Jaffe}, {Jaffe}, {Jones}, {Juvela}, {Keih{\"a}nen}, {Keskitalo},
  {Kisner}, {Kneissl}, {Knoche}, {Knox}, {Kunz}, {Kurki-Suonio}, {Lagache},
  {L{\"a}hteenm{\"a}ki}, {Lamarre}, {Lasenby}, {Lattanzi}, {Lawrence}, {Leahy},
  {Leonardi}, {Lesgourgues}, {Levrier}, {Lewis}, {Liguori}, {Lilje},
  {Linden-V{\o}rnle}, {L{\'o}pez-Caniego}, {Lubin}, {Mac{\'\i}as-P{\'e}rez},
  {Maggio}, {Maino}, {Mandolesi}, {Mangilli}, {Marchini}, {Maris}, {Martin},
  {Martinelli}, {Mart{\'\i}nez-Gonz{\'a}lez}, {Masi}, {Matarrese}, {McGehee},
  {Meinhold}, {Melchiorri}, {Melin}, {Mendes}, {Mennella}, {Migliaccio},
  {Millea}, {Mitra}, {Miville-Desch{\^e}nes}, {Moneti}, {Montier}, {Morgante},
  {Mortlock}, {Moss}, {Munshi}, {Murphy}, {Naselsky}, {Nati}, {Natoli},
  {Netterfield}, {N{\o}rgaard-Nielsen}, {Noviello}, {Novikov}, {Novikov},
  {Oxborrow}, {Paci}, {Pagano}, {Pajot}, {Paladini}, {Paoletti}, {Partridge},
  {Pasian}, {Patanchon}, {Pearson}, {Perdereau}, {Perotto}, {Perrotta},
  {Pettorino}, {Piacentini}, {Piat}, {Pierpaoli}, {Pietrobon}, {Plaszczynski},
  {Pointecouteau}, {Polenta}, {Popa}, {Pratt}, {Pr{\'e}zeau}, {Prunet},
  {Puget}, {Rachen}, {Reach}, {Rebolo}, {Reinecke}, {Remazeilles}, {Renault},
  {Renzi}, {Ristorcelli}, {Rocha}, {Rosset}, {Rossetti}, {Roudier},
  {Rouill{\'e} d'Orfeuil}, {Rowan-Robinson}, {Rubi{\~n}o-Mart{\'\i}n},
  {Rusholme}, {Said}, {Salvatelli}, {Salvati}, {Sandri}, {Santos},
  {Savelainen}, {Savini}, {Scott}, {Seiffert}, {Serra}, {Shellard}, {Spencer},
  {Spinelli}, {Stolyarov}, {Stompor}, {Sudiwala}, {Sunyaev}, {Sutton},
  {Suur-Uski}, {Sygnet}, {Tauber}, {Terenzi}, {Toffolatti}, {Tomasi},
  {Tristram}, {Trombetti}, {Tucci}, {Tuovinen}, {T{\"u}rler}, {Umana},
  {Valenziano}, {Valiviita}, {Van Tent}, {Vielva}, {Villa}, {Wade}, {Wandelt},
  {Wehus}, {White}, {White}, {Wilkinson}, {Yvon}, {Zacchei}, \&
  {Zonca}}]{2016PlanckCollaboration}
{Planck Collaboration}, {Ade}, P.~A.~R., {Aghanim}, N., {et~al.} 2016, \aap,
  594, A13

\bibitem[{{Ramos Almeida} {et~al.}(2022){Ramos Almeida}, {Bischetti},
  {Garc{\'\i}a-Burillo}, {Alonso-Herrero}, {Audibert}, {Cicone}, {Feruglio},
  {Tadhunter}, {Pierce}, {Pereira-Santaella}, \& {Bessiere}}]{2022RamosAlmeida}
{Ramos Almeida}, C., {Bischetti}, M., {Garc{\'\i}a-Burillo}, S., {et~al.} 2022,
  \aap, 658, A155

\bibitem[{{Reeves} \& {Braito}(2019)}]{ReevesBraito2019}
{Reeves}, J.~N. \& {Braito}, V. 2019, \apj, 884, 80

\bibitem[{{Reeves} {et~al.}(2018){Reeves}, {Braito}, {Nardini}, {Hamann},
  {Chartas}, {Lobban}, {O'Brien}, \& {Turner}}]{Reeves2018}
{Reeves}, J.~N., {Braito}, V., {Nardini}, E., {et~al.} 2018, \apj, 867, 38

\bibitem[{{Reeves} {et~al.}(2003){Reeves}, {O'Brien}, \& {Ward}}]{2003Reeves}
{Reeves}, J.~N., {O'Brien}, P.~T., \& {Ward}, M.~J. 2003, \apjl, 593, L65

\bibitem[{{Ricci} {et~al.}(2017){Ricci}, {Bauer}, {Treister}, {Schawinski},
  {Privon}, {Blecha}, {Arevalo}, {Armus}, {Harrison}, {Ho}, {Iwasawa},
  {Sanders}, \& {Stern}}]{Ricci2017}
{Ricci}, C., {Bauer}, F.~E., {Treister}, E., {et~al.} 2017, \mnras, 468, 1273

\bibitem[{{Rupke} {et~al.}(2017){Rupke}, {G{\"u}ltekin}, \&
  {Veilleux}}]{2017Rupke}
{Rupke}, D. S.~N., {G{\"u}ltekin}, K., \& {Veilleux}, S. 2017, \apj, 850, 40

\bibitem[{{Sanders} {et~al.}(2003){Sanders}, {Mazzarella}, {Kim}, {Surace}, \&
  {Soifer}}]{2003Sanders}
{Sanders}, D.~B., {Mazzarella}, J.~M., {Kim}, D.~C., {Surace}, J.~A., \&
  {Soifer}, B.~T. 2003, \aj, 126, 1607

\bibitem[{{Silk} \& {Rees}(1998)}]{1998Silk-Rees}
{Silk}, J. \& {Rees}, M.~J. 1998, \aap, 331, L1

\bibitem[{{Sirressi} {et~al.}(2019){Sirressi}, {Cicone}, {Severgnini},
  {Braito}, {Dotti}, {Della Ceca}, {Reeves}, {Matzeu}, {Vignali}, \&
  {Ballo}}]{2019Sirressi}
{Sirressi}, M., {Cicone}, C., {Severgnini}, P., {et~al.} 2019, \mnras, 489,
  1927

\bibitem[{{Smith} {et~al.}(2019){Smith}, {Tombesi}, {Veilleux}, {Lohfink}, \&
  {Luminari}}]{2019Smith}
{Smith}, R.~N., {Tombesi}, F., {Veilleux}, S., {Lohfink}, A.~M., \& {Luminari},
  A. 2019, \apj, 887, 69

\bibitem[{{Solomon}(1997)}]{1997Solomon}
{Solomon}, P.~M. 1997, IAU Symposium, 170, 289

\bibitem[{{Stierwalt} {et~al.}(2013){Stierwalt}, {Armus}, {Surace}, {Inami},
  {Petric}, {Diaz-Santos}, {Haan}, {Charmandaris}, {Howell}, {Kim}, {Marshall},
  {Mazzarella}, {Spoon}, {Veilleux}, {Evans}, {Sanders}, {Appleton}, {Bothun},
  {Bridge}, {Chan}, {Frayer}, {Iwasawa}, {Kewley}, {Lord}, {Madore},
  {Melbourne}, {Murphy}, {Rich}, {Schulz}, {Sturm}, {Vavilkin}, \&
  {Xu}}]{2013Stierwalt}
{Stierwalt}, S., {Armus}, L., {Surace}, J.~A., {et~al.} 2013, \apjs, 206, 1

\bibitem[{{Sturm} {et~al.}(2011){Sturm}, {Gonz{\'a}lez-Alfonso}, {Veilleux},
  {Fischer}, {Graci{\'a}-Carpio}, {Hailey-Dunsheath}, {Contursi}, {Poglitsch},
  {Sternberg}, {Davies}, {Genzel}, {Lutz}, {Tacconi}, {Verma}, {Maiolino}, \&
  {de Jong}}]{2011Sturm}
{Sturm}, E., {Gonz{\'a}lez-Alfonso}, E., {Veilleux}, S., {et~al.} 2011, \apjl,
  733, L16

\bibitem[{{Tombesi} {et~al.}(2012){Tombesi}, {Cappi}, {Reeves}, \&
  {Braito}}]{Tombesi2012}
{Tombesi}, F., {Cappi}, M., {Reeves}, J.~N., \& {Braito}, V. 2012, \mnras, 422,
  L1

\bibitem[{{Tombesi} {et~al.}(2010){Tombesi}, {Cappi}, {Reeves}, {Palumbo},
  {Yaqoob}, {Braito}, \& {Dadina}}]{2010Tombesi}
{Tombesi}, F., {Cappi}, M., {Reeves}, J.~N., {et~al.} 2010, \aap, 521, A57

\bibitem[{{Tombesi} {et~al.}(2015){Tombesi}, {Mel{\'e}ndez}, {Veilleux},
  {Reeves}, {Gonz{\'a}lez-Alfonso}, \& {Reynolds}}]{2015Tombesi}
{Tombesi}, F., {Mel{\'e}ndez}, M., {Veilleux}, S., {et~al.} 2015, \nat, 519,
  436

\bibitem[{{Tombesi} {et~al.}(2017){Tombesi}, {Veilleux}, {Mel{\'e}ndez},
  {Lohfink}, {Reeves}, {Piconcelli}, {Fiore}, \& {Feruglio}}]{2017Tombesi}
{Tombesi}, F., {Veilleux}, S., {Mel{\'e}ndez}, M., {et~al.} 2017, \apj, 850,
  151

\bibitem[{{Tozzi} {et~al.}(2021){Tozzi}, {Cresci}, {Marasco}, {Nardini},
  {Marconi}, {Mannucci}, {Chartas}, {Rizzo}, {Amiri}, {Brusa}, {Comastri},
  {Dadina}, {Lanzuisi}, {Mainieri}, {Mingozzi}, {Perna}, {Venturi}, \&
  {Vignali}}]{2021Tozzi}
{Tozzi}, G., {Cresci}, G., {Marasco}, A., {et~al.} 2021, \aap, 648, A99

\bibitem[{{Veilleux} {et~al.}(2017){Veilleux}, {Bolatto}, {Tombesi},
  {Mel{\'e}ndez}, {Sturm}, {Gonz{\'a}lez-Alfonso}, {Fischer}, \&
  {Rupke}}]{2017Veilleux}
{Veilleux}, S., {Bolatto}, A., {Tombesi}, F., {et~al.} 2017, \apj, 843, 18

\bibitem[{{Veilleux} {et~al.}(2005){Veilleux}, {Cecil}, \&
  {Bland-Hawthorn}}]{2005Veilleux}
{Veilleux}, S., {Cecil}, G., \& {Bland-Hawthorn}, J. 2005, \araa, 43, 769

\bibitem[{{Veilleux} {et~al.}(2020){Veilleux}, {Maiolino}, {Bolatto}, \&
  {Aalto}}]{2020Veilleux}
{Veilleux}, S., {Maiolino}, R., {Bolatto}, A.~D., \& {Aalto}, S. 2020, \aapr,
  28, 2

\bibitem[{{Veilleux} {et~al.}(2013){Veilleux}, {Mel{\'e}ndez}, {Sturm},
  {Gracia-Carpio}, {Fischer}, {Gonz{\'a}lez-Alfonso}, {Contursi}, {Lutz},
  {Poglitsch}, {Davies}, {Genzel}, {Tacconi}, {de Jong}, {Sternberg}, {Netzer},
  {Hailey-Dunsheath}, {Verma}, {Rupke}, {Maiolino}, {Teng}, \&
  {Polisensky}}]{2013Veilleux}
{Veilleux}, S., {Mel{\'e}ndez}, M., {Sturm}, E., {et~al.} 2013, \apj, 776, 27

\bibitem[{{Wilms} {et~al.}(2000){Wilms}, {Allen}, \& {McCray}}]{Wilms2000}
{Wilms}, J., {Allen}, A., \& {McCray}, R. 2000, \apj, 542, 914

\bibitem[{{Wong} \& {Blitz}(2002)}]{2002Wong}
{Wong}, T. \& {Blitz}, L. 2002, \apj, 569, 157

\end{thebibliography}

\begin{appendix}

\section{ALMA map of the 3mm continuum}\label{sec:appendix_continuum}

\begin{figure}[tbp]
        \centering
        \includegraphics[clip=true,trim=1cm 10cm 3cm 5cm, width=.95\columnwidth]{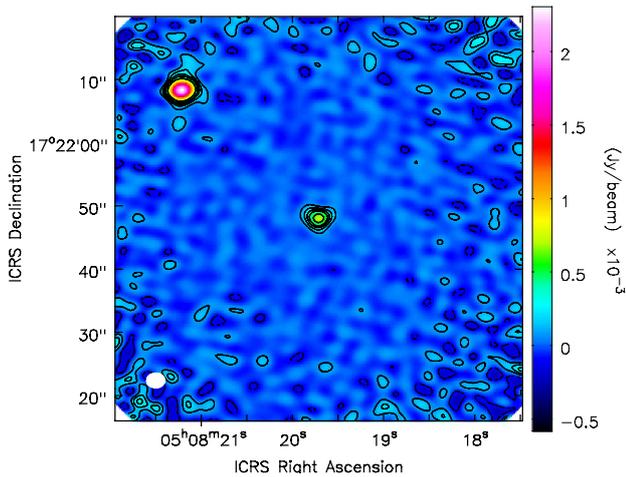}\\
        \caption{ALMA       3~mm continuum map computed at an average frequency of $100~$GHz. The map was obtained by averaging the data over a bandwidth of 4~GHz. Contours are plotted at $(-3, 3, 6, 9, 12, 18)\times \sigma_{cont}$, with $\sigma_{cont}=0.03$~mJy/beam being the average 1$\sigma$ rms noise level within the central 10$\arcsec$ portion of the FoV.}
        \label{fig:cont_pbcor}
\end{figure}

\begin{figure}[tbp]
        \centering
        \includegraphics[clip=true,trim=0.5cm 9cm 3cm 6cm, width=.95\columnwidth]{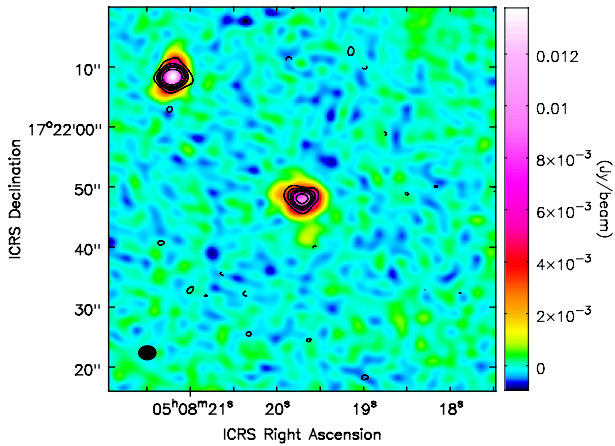}\\
        \caption{Overlay of ALMA 3~mm continuum contours (in black) on the CO(1-0) line map. This highlights the compactness of the continuum emission compared to the line emission in both members of the IRAS~05054+1718 galaxy pair. Neither the continuum contours nor the CO(1-0) line map is corrected for the PB, to ease the visualisation of the companion galaxy NED02. }
        \label{fig:cont_overCO}
\end{figure}

The ALMA interferometric map of the 3~mm continuum emission from the target is shown in Figure~\ref{fig:cont_pbcor}. The map was obtained by combining the data (total bandwidth $\Delta\nu=4$~GHz) from the two low-resolution spectral windows centred at 101.190~GHz and 99.387~GHz, which are not contaminated by line emission. A 2D Gaussian fit performed on the PB-corrected map shown in Fig.~\ref{fig:cont_pbcor} delivers best-fit peak flux densities of $S_{cont}^{NED01}=0.71\pm0.03$~mJy~beam$^{-1}$ and $S_{cont}^{NED02}=2.26\pm0.05$~mJy~beam$^{-1}$, indicating that the 3~mm continuum of NED02 is approximately three times brighter than NED01. The fit also provides   information about the size and position of the two sources. For NED01, we get a peak centre at RA(ICRS) =05:08:19.713 and Dec(ICRS)=17.21.48.11, with a deconvolved size of $3.0\arcsec\times2.4\arcsec$ (PA=84.5~deg). For NED02, the ALMA continuum peaks at 
RA(ICRS)=05:08:21.211 and Dec(ICRS)=17.22.08.40 (PA=98.4~deg), with a deconvolved size of $3.0\arcsec\times2.5\arcsec$. The source-integrated continuum flux obtained through the fit is $S_{3mm}=0.76$~mJy for NED01 and $S_{3mm}=2.4$~mJy for NED02. These results indicate that the 3~mm continuum emission from the targets is only marginally resolved at the resolution of our ALMA data, and much more compact than the CO(1-0) line emission. This is further shown by Figure~\ref{fig:cont_overCO}, which displays the overlay of the ALMA 3~mm continuum contours on the CO(1-0) line emission map.

\section{Alternative BBarolo CO(1-0) disk models for NED01}\label{sec:appendix}

\begin{figure}[tbp]
        \centering
        \includegraphics[clip=true,trim=0cm 0cm 0cm 0.cm,width=\columnwidth]{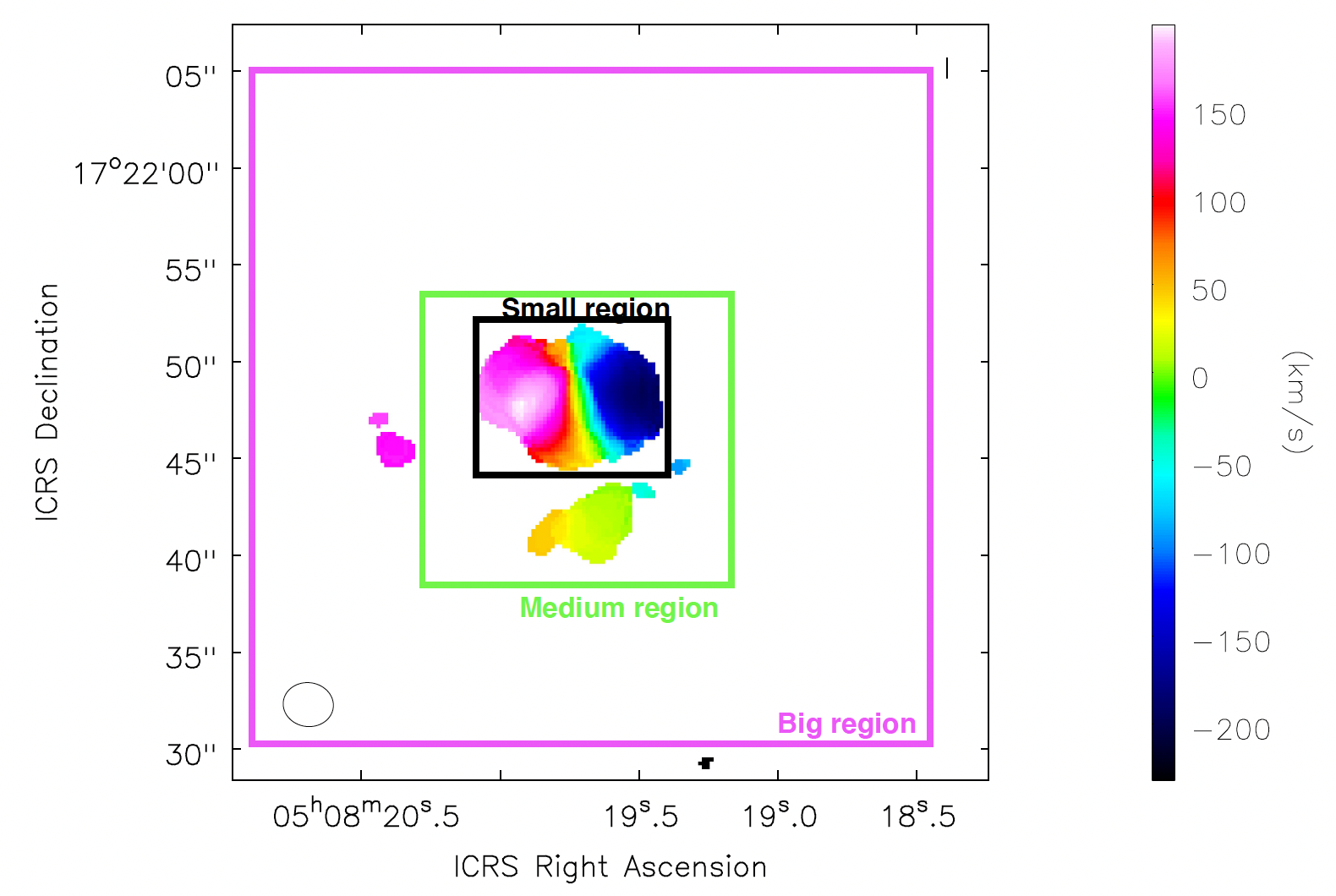}
        \caption{Velocity map of NED01 (see also Fig.~\ref{fig:NED01_mom_maps}), with the three regions used for the BBarolo analysis overplotted. }
        \label{fig:mommap1}
\end{figure}

\begin{figure*}
        \centering
        \includegraphics[clip=true,trim=0.5cm 0.5cm 0.5cm 0.cm,width=\textwidth]{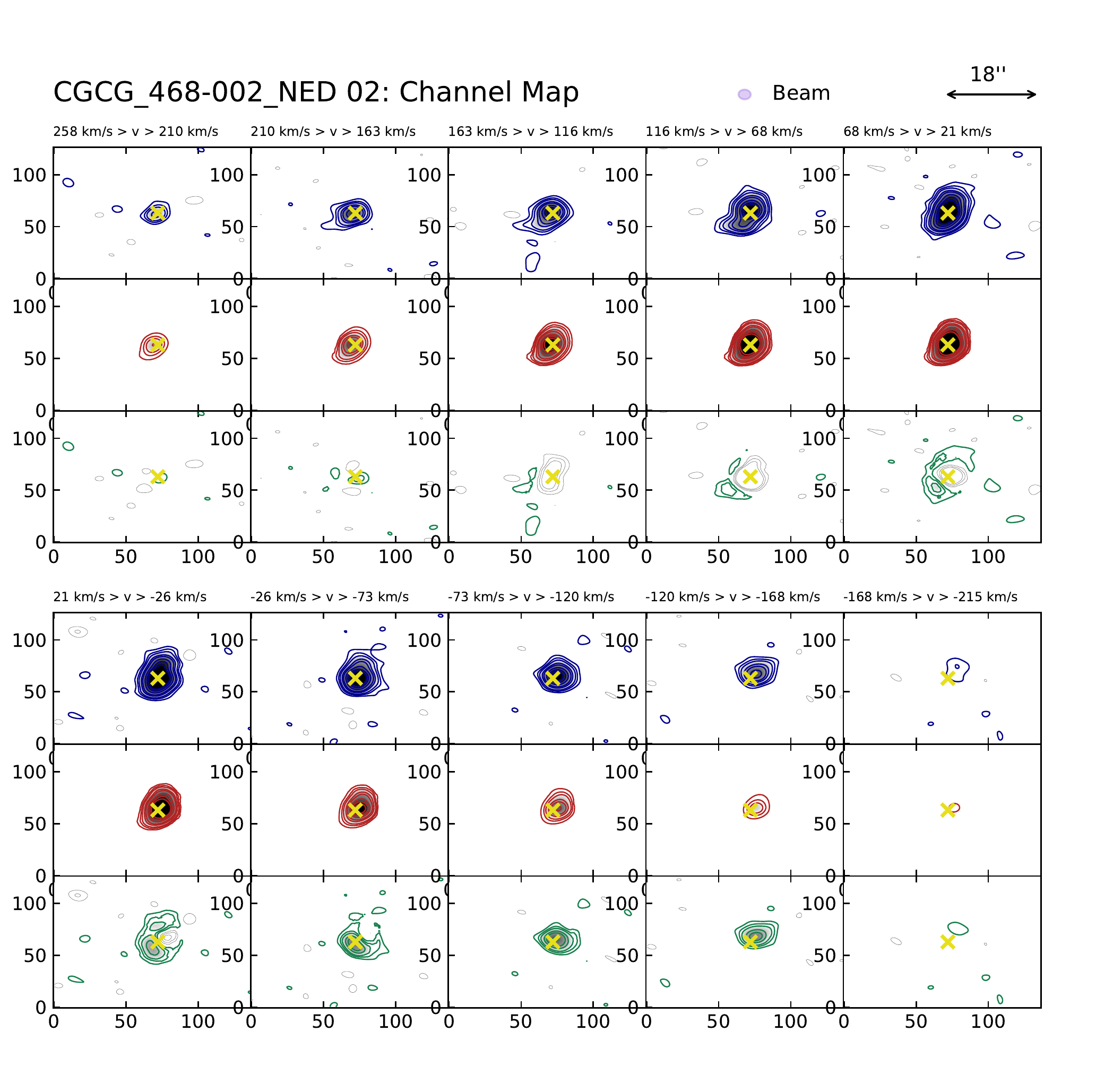}
        \caption{CO channel maps of NED02, computed by integrating   \co~emission in channels of $dv = 47$ km/s. This  shows the comparison between the ALMA data and the BBarolo disk model. The contours are plotted at 3, 6, 9, 15, 20, 30, 40, 50$\sigma$.}
        \label{fig:ned02chann}
\end{figure*}

\begin{figure}[tbp]
        \centering
        \includegraphics[clip=true,trim=0.cm 0.cm 0.cm 0.cm, width=.95\columnwidth]{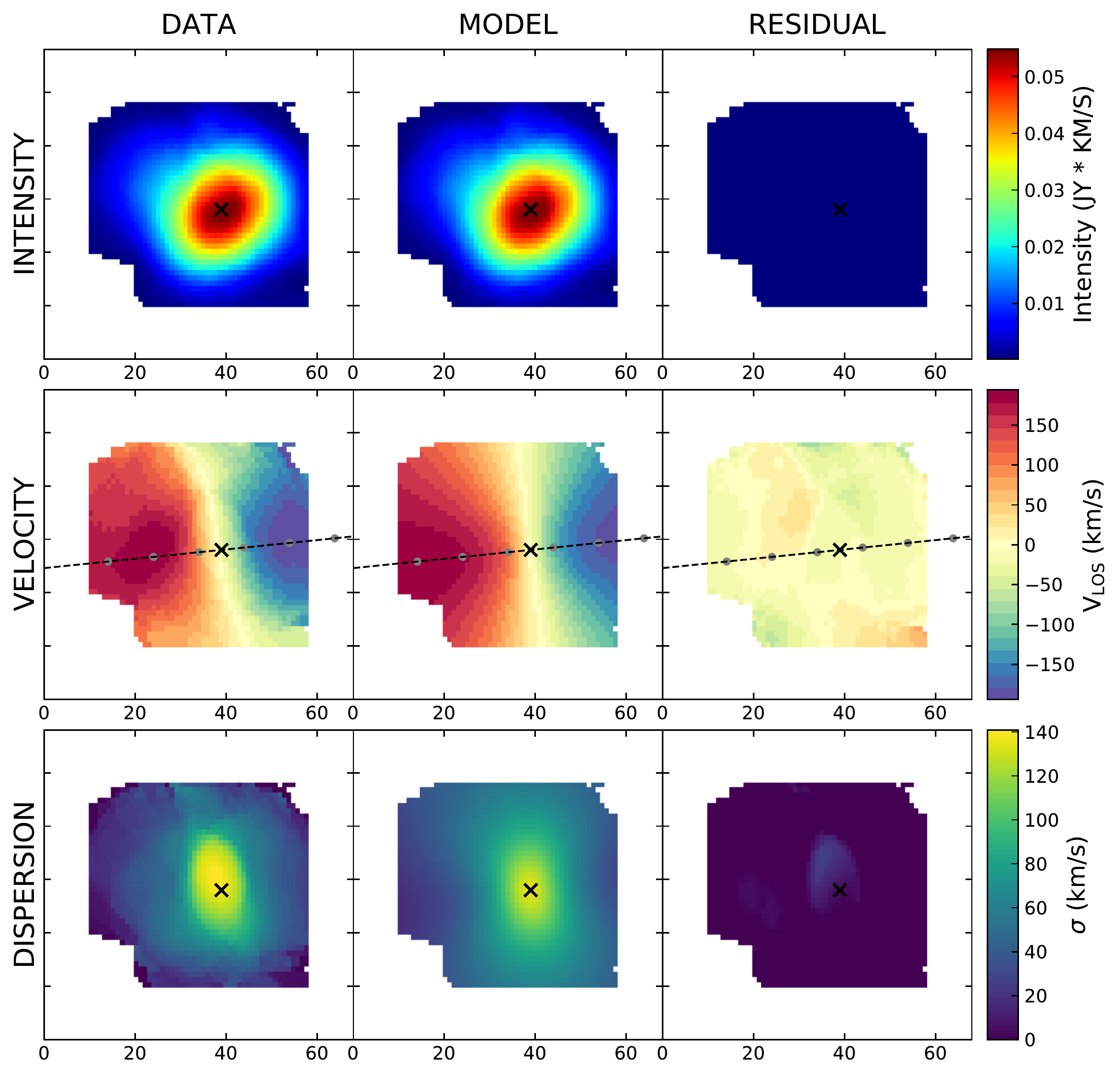}\quad
        \includegraphics[clip=true,trim=0.cm 0.cm 0.cm 0.cm, width=.95\columnwidth]{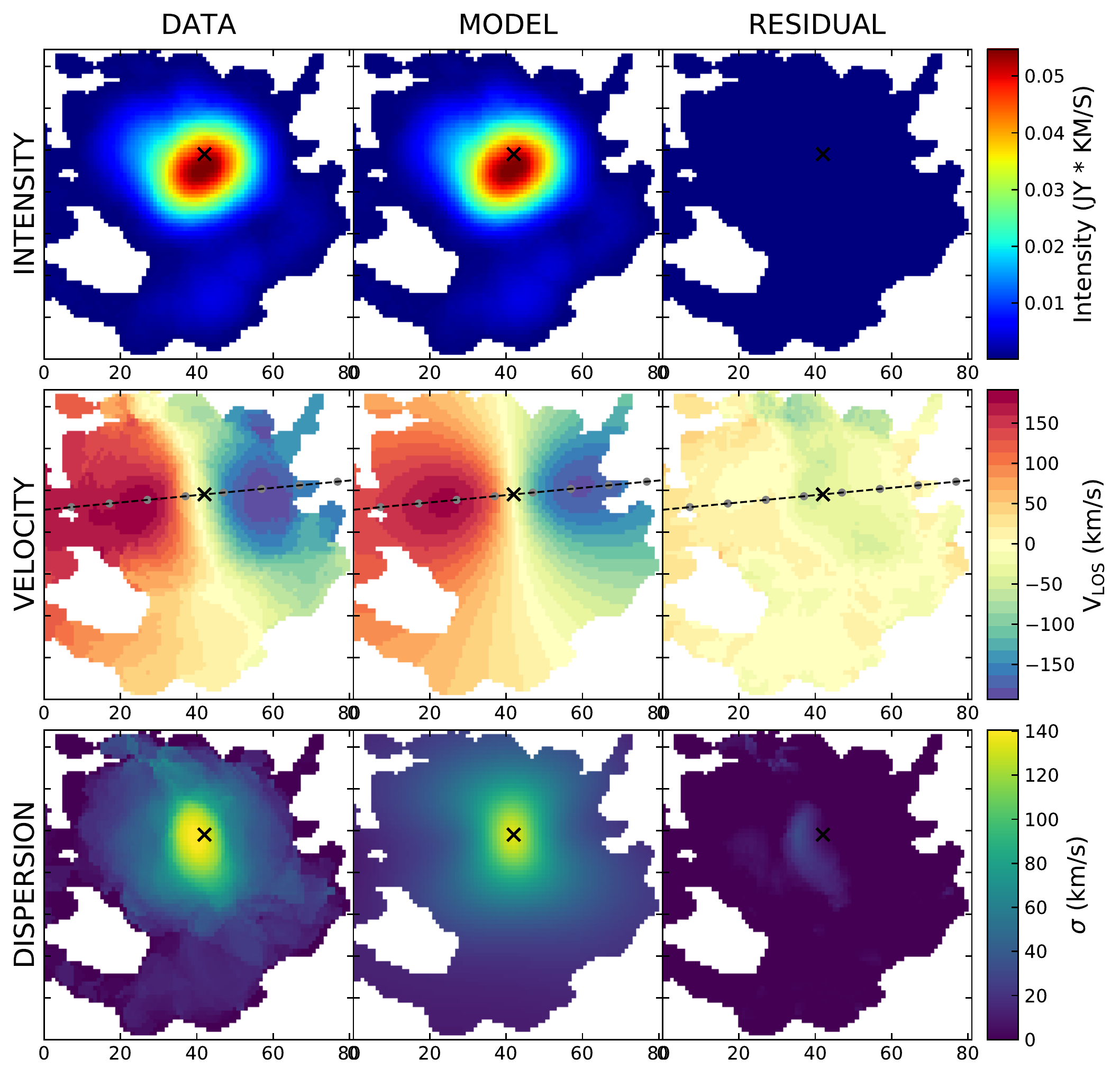}\\
        \caption{\co~moment maps showing   comparison between data and BBarolo best-fit molecular disk model performed on the {\it Small} (upper panel) and {\it Medium} (lower panel) regions around NED01.}
        \label{fig:small_medium_mom}
\end{figure}

\begin{figure*}[tbp]
        \centering
        \includegraphics[clip=true,trim=0.cm 0cm 0cm 0cm,width=.56\columnwidth]{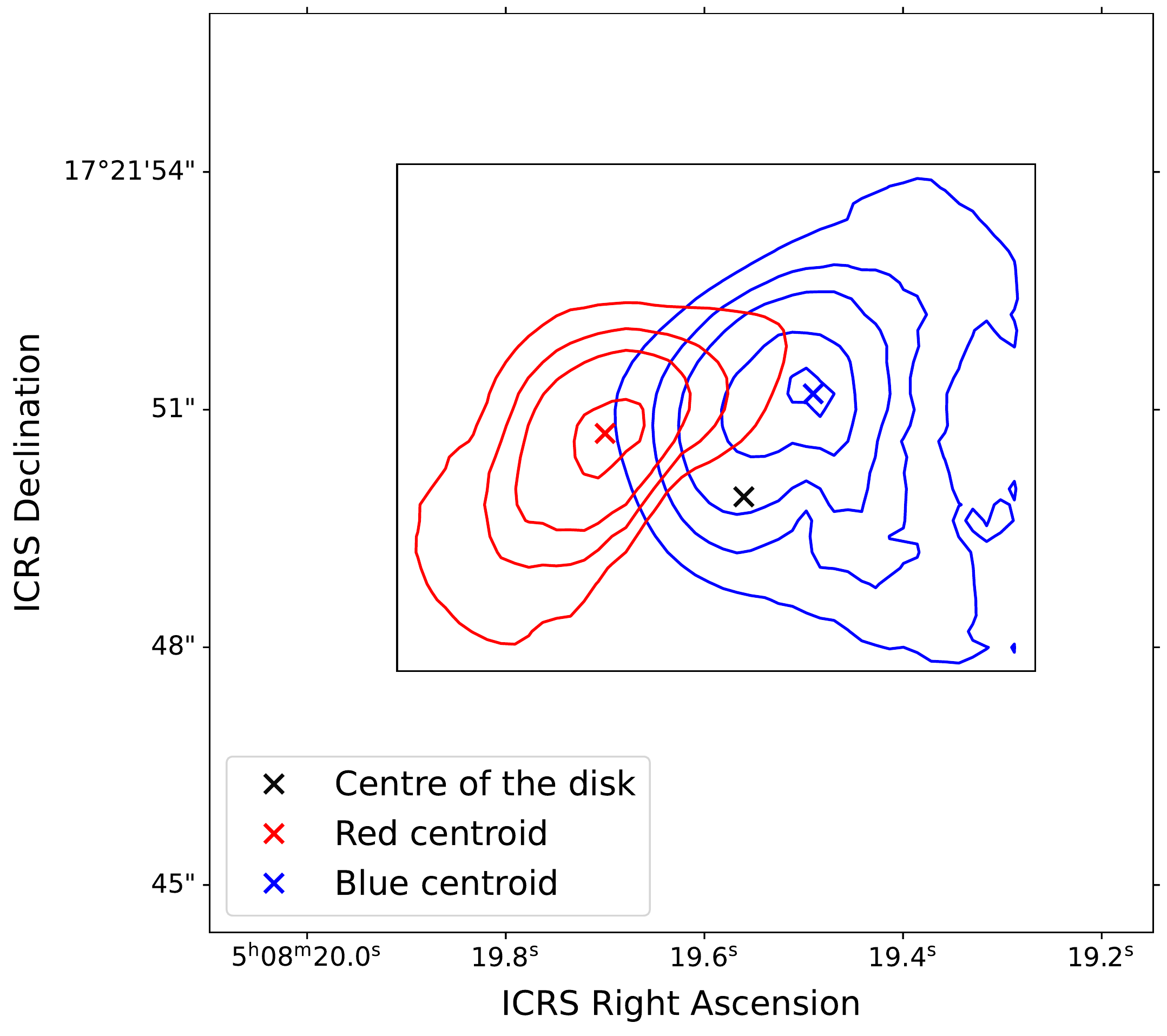}\quad
        \includegraphics[clip=true,trim=-1cm 0.cm 0cm 0.cm,width=.65\columnwidth]{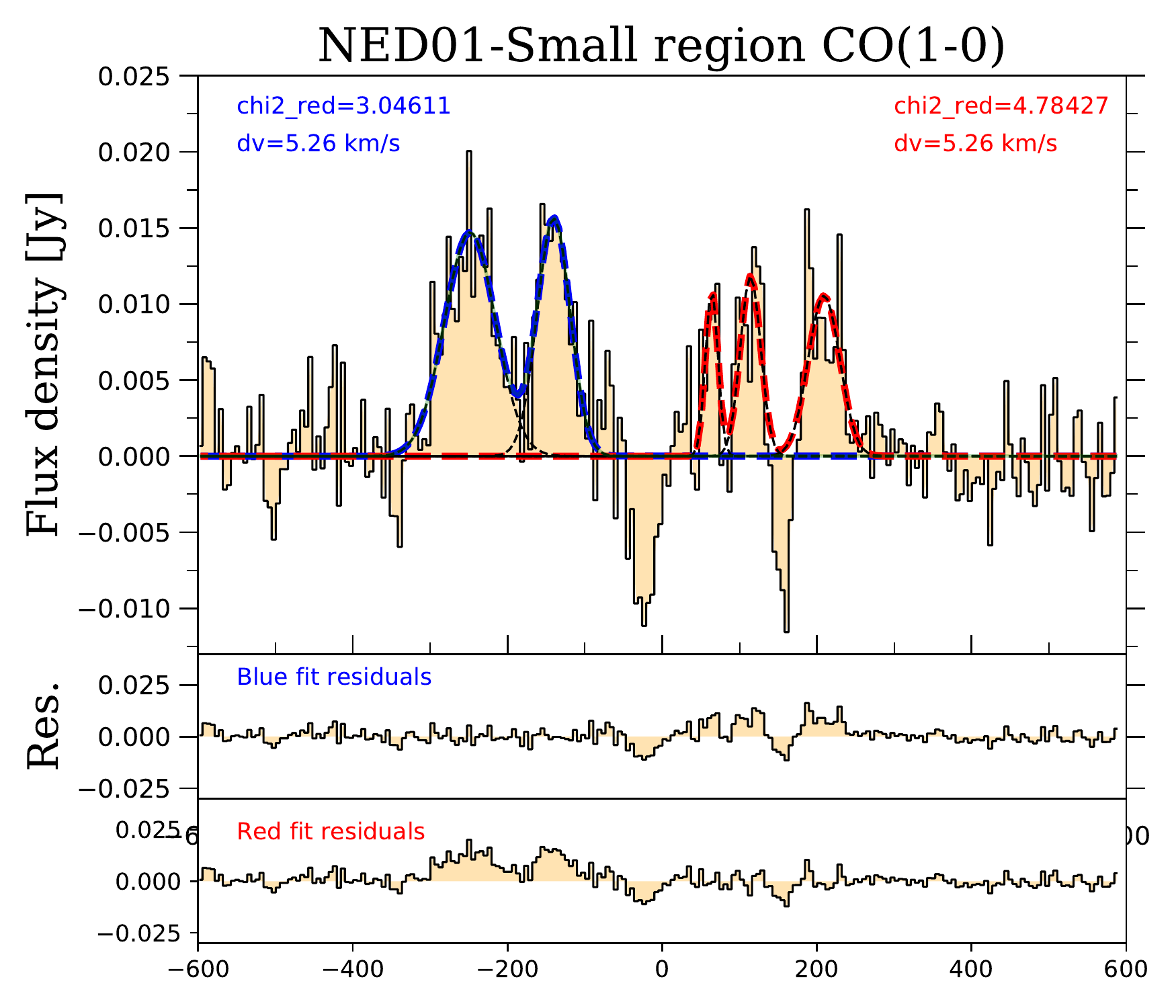}\quad
        \includegraphics[clip=true,trim=0cm 0cm 0cm 0cm,width=.56\columnwidth]{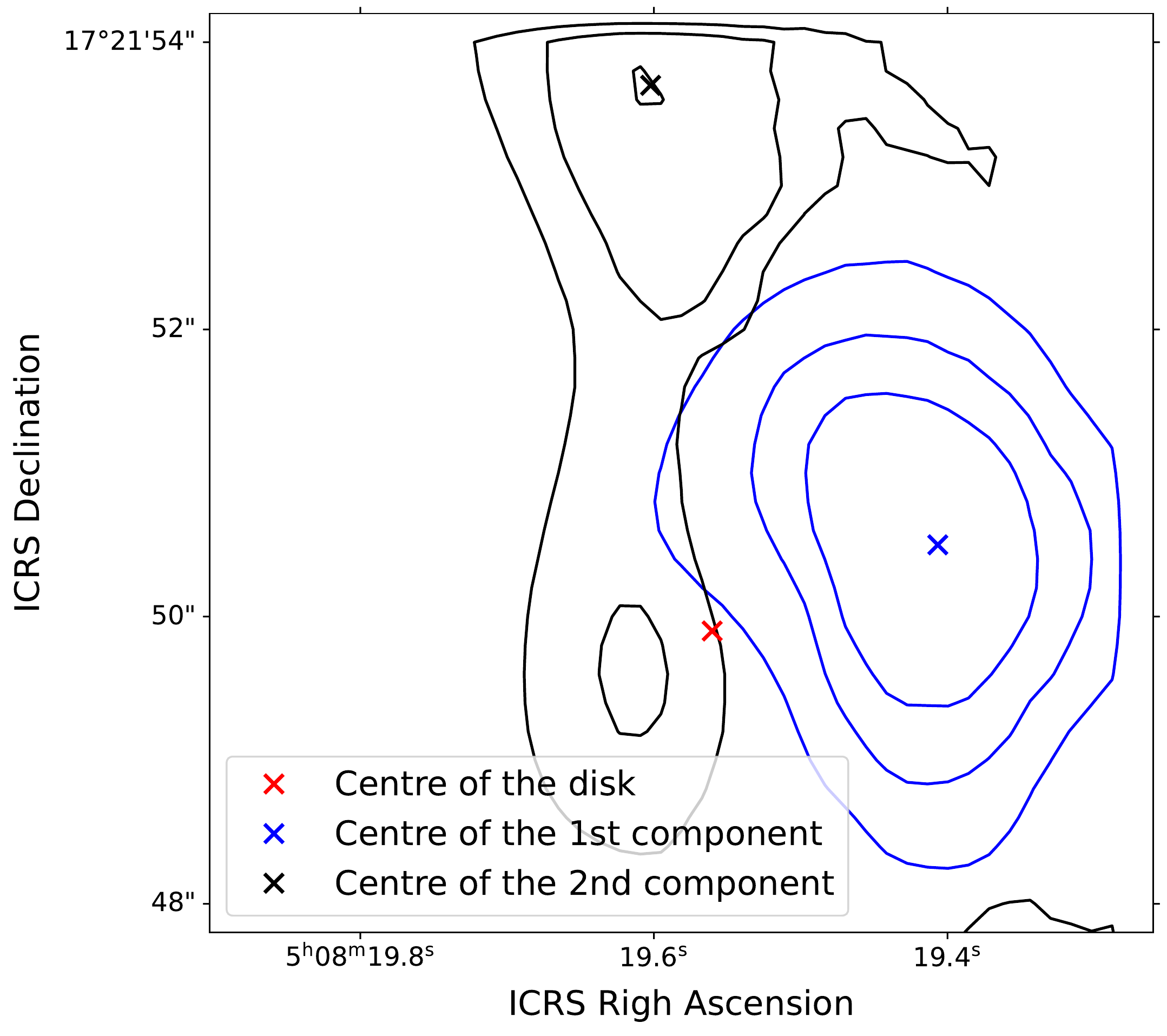}\\
        \includegraphics[clip=true,trim=0cm 0cm 0cm 0cm,width=.56\columnwidth]{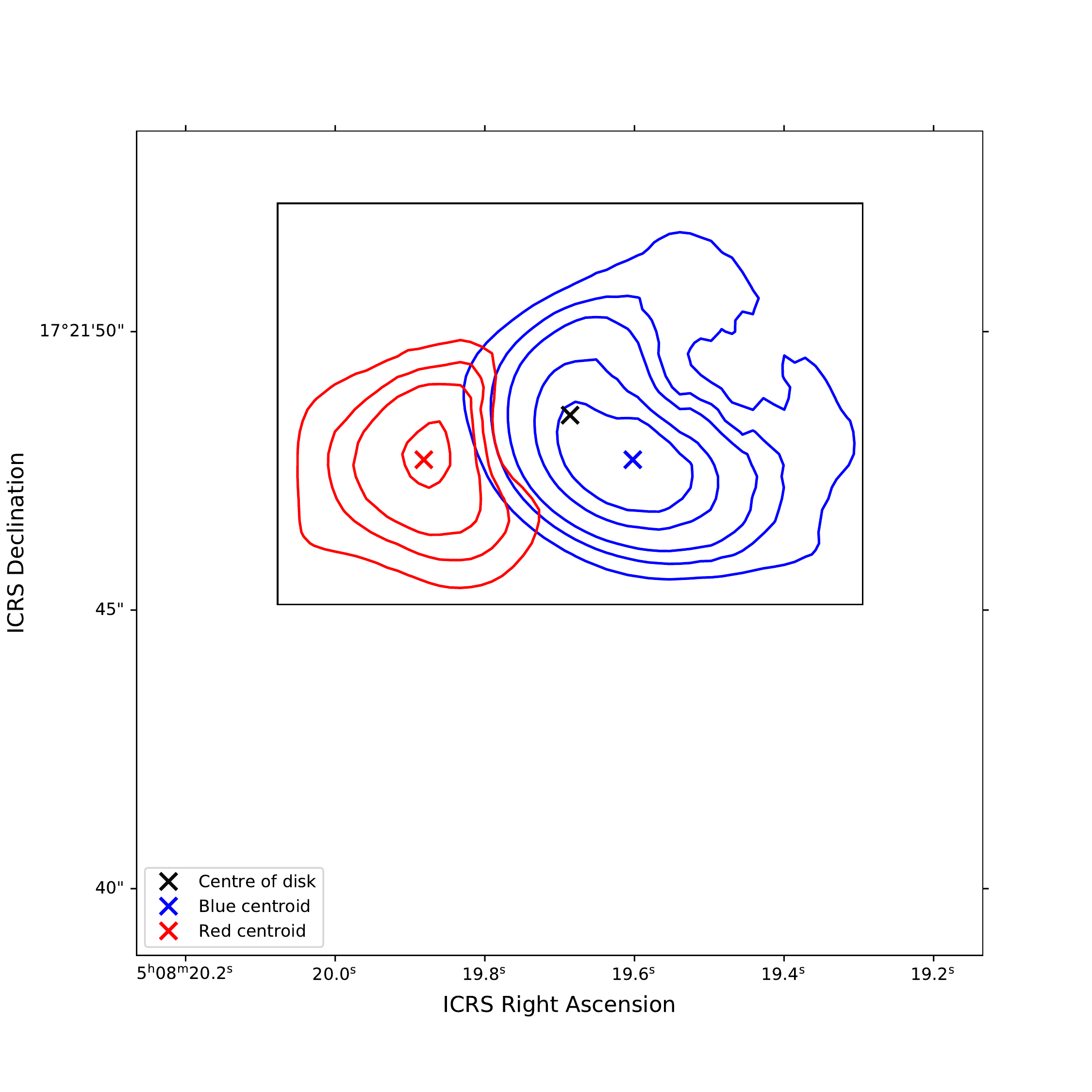}\quad
        \includegraphics[clip=true,trim=-1cm 0.cm 0cm 0.cm,width=.65\columnwidth]{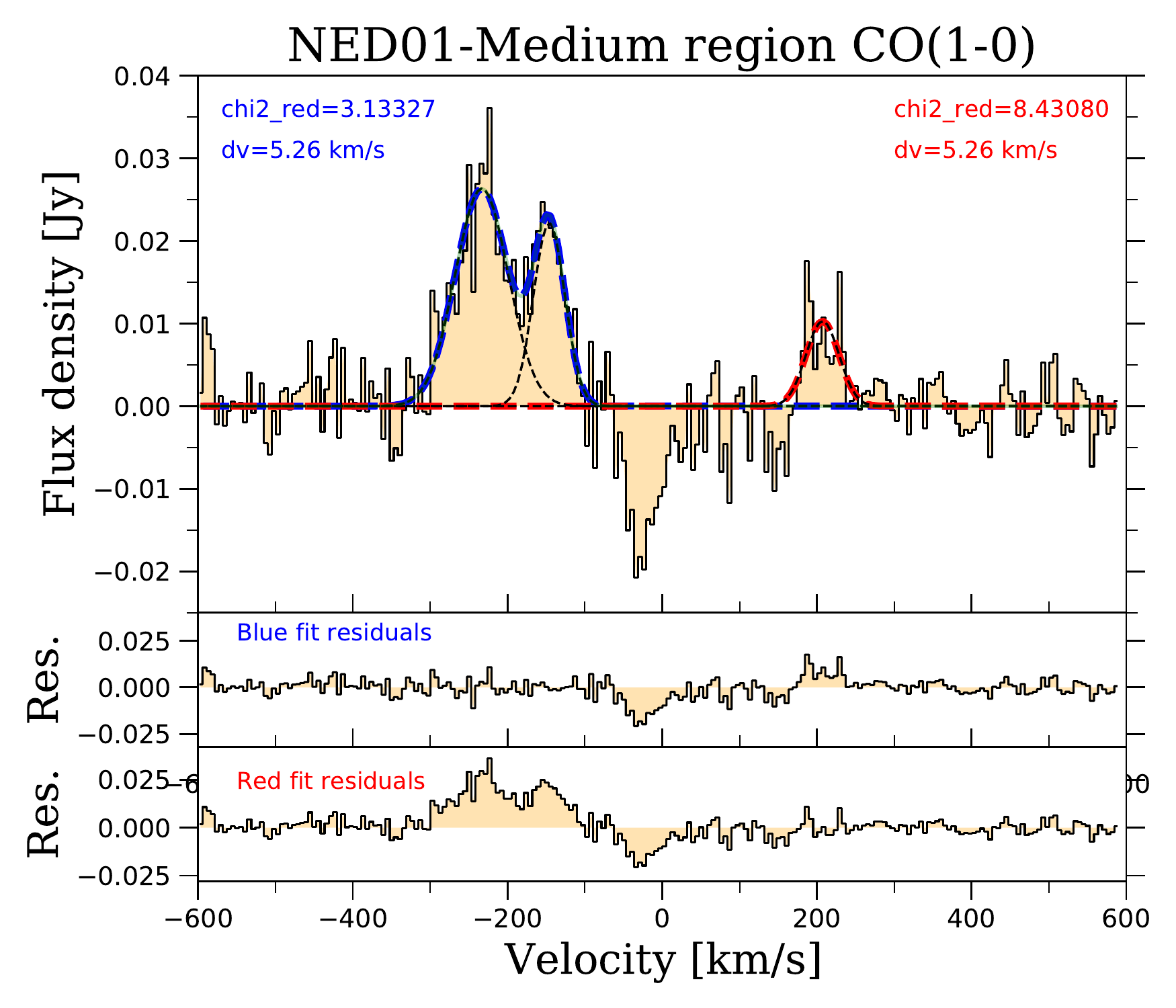}\quad
        \includegraphics[clip=true,trim=0cm 0cm 0cm 0.cm,width=.56\columnwidth]{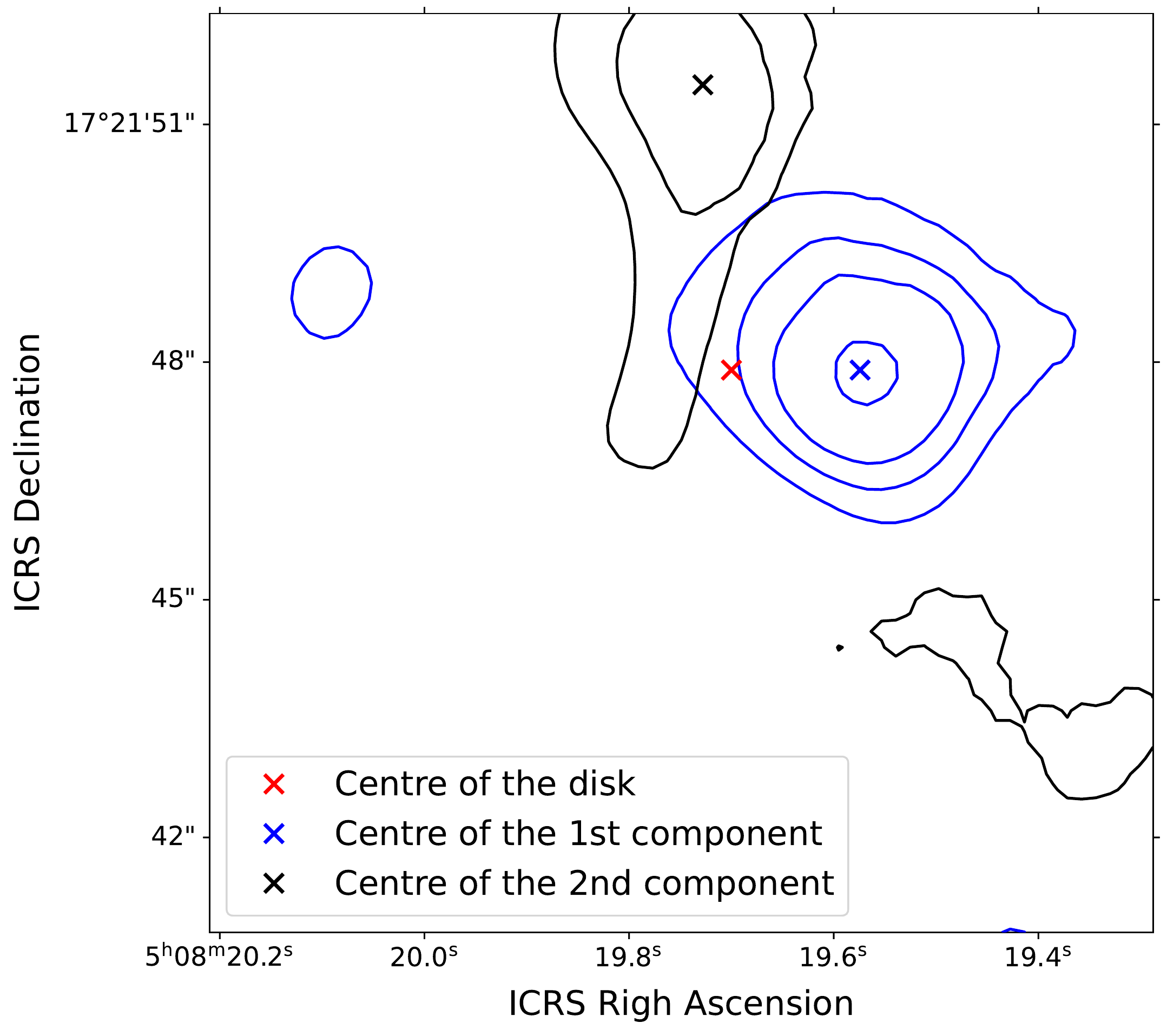}\\
        \caption{\co~residual emission around NED01 obtained after subtracting the best-fit disk model computed by BBarolo on the {\it Small region} (top panels) and on the {\it Medium} region (bottom panels). 
        {\it Top left}: Map of the \co~residual emission integrated within $v\in[170, 220]$\kms\ (red contours) and $v\in[-290,-200]$~\kms\ (blue contours). Contours are plotted at 3, 6, 9, 15, 20$\sigma$. The black cross indicates the centre of the best-fit disk model, while the red and blue crosses indicate the centroids of the red- and blueshifted residuals, respectively. 
        {\it Top centre:} Spectrum of the \co~residual emission extracted from the black box (size $9.2''\times6.4''$) in the left panel. 
        {\it Top right:} Imaging of the blueshifted \co~residual components identified in the spectrum.
        {\it Bottom left}: Map of the \co~residual emission integrated within $v\in[170, 220]$\kms\ (red contours) and $v\in[-290, -120]$~\kms\ (blue contours). 
        {\it Bottom centre:} Spectrum of the \co~residual emission extracted from the black box (size of $11.2''\times7.2''$) in the left panel. 
        {\it Bottom right:} Imaging of the blueshifted \co~spectral components identified in the central panel.}
        \label{fig:residuals_NED01_small_medium}
\end{figure*}

The \textit{Small} region, shown in Fig.~\ref{fig:mommap1}, includes only the central molecular disk of NED01 (i.e. component A in Fig.~\ref{fig:NED01_mom_maps}), whereas the {\it Medium} region also embeds   the southern extension (component C) of the \co~emission from NED01.
We ran the BBarolo disk modelling for NED01  on these two additional regions, and the results are reported in Fig.~\ref{fig:small_medium_mom}. For the {\it Small} region we used 
three rings, each $2\arcsec$ in width, while for the {\it Medium} region we increased the number of rings to seven.
The rotation patterns resulting from these two fits does not differ significantly from the fit shown in the main text, which was performed by selecting the {\it Big region} around NED01. 
However, these two fits, performed on smaller areas, produce more significant residuals, at both blue- and redshifted velocities, as shown in Fig.~\ref{fig:residuals_NED01_small_medium}. 
The best-fit parameters of the \co~residuals obtained in these two fits performed on the {\it Small} and {\it Medium} regions are listed in Table~\ref{tab:ned01_small_medium_res}. 

\begin{figure*}[tbp]
        \centering
        \includegraphics[clip=true,trim=0cm 4.2cm 0.2cm 4.8cm,width=.65\textwidth]{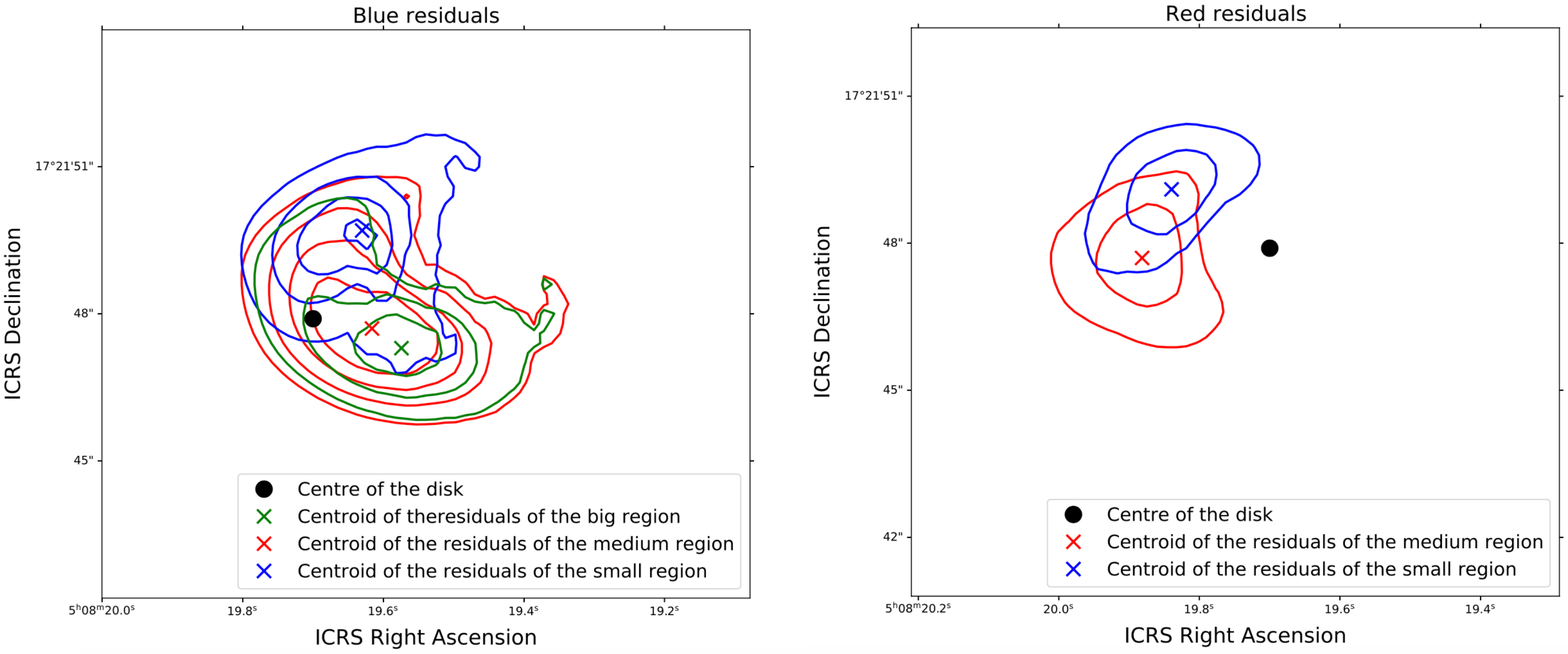}
        \caption{Comparison between   \co~residual emission revealed in NED01 after subtracting the different BBarolo disk models. The green, red, and blue contours show respectively the \co~residuals obtained in the {\it Big}, {\it Medium}, and {\it Small} regions fits, and they are plotted at 5, 10, 15, 20$\sigma$. {\it Left panel}: Blueshifted residuals. {\it Right panel}: Redshifted residuals.}
        \label{fig:comparison_residuals_NED01}
 \end{figure*}

Quite strikingly, the spectral profile and morphology of the blueshifted residuals are consistent with those produced by the fit performed on the {\it Big region}, shown in the main body of the paper shown in Sect.~\ref{sec:NED01_BBarolo}. 
This is best visualised in the left panel of  Fig.~\ref{fig:comparison_residuals_NED01}, which compares on the same map the blueshifted \co~residual emission obtained from all three fits performed on NED01. 
Our analysis focuses on these blueshifted residuals, which are persistent in all three fits. 
In addition, the two fits reported here produce residuals at receding velocities, east of the NED01 galaxy centre, which were not detected in the {\it Big region} fit. The right panel of Fig.~\ref{fig:comparison_residuals_NED01} compares the morphology of this redshifted \co~residual emission obtained in the {\it Small} and {\it Medium} region fits. The contours from the two fits overlap only partially. These redshifted residuals must be due to the additional extended \co~features that are not captured in these smaller regions, but were instead well modelled by the fit performed on the {\it Big region}, where  they displayed redshifted velocities on the eastern side of the galaxy consistent with components that follow the same pattern as the main disk rotation.

\begin{table*}[tbp]
        \centering
        \caption{Best-fit Gaussian parameters of the \co~residual spectrum of NED01 obtained after subtracting the best-fit BBarolo disk models computed on the {\it Small} and {\it Medium} regions.}             
        \small 
        \label{tab:ned01_small_medium_res}      
        \begin{tabular}{lccccc}    
                \toprule
                \multicolumn{6}{c}{\it Small Region} \\
                \midrule
                Parameter & Comp. 1 & Comp. 2 & Comp. 3 & Comp. 4 & Comp. 5 \\
                \midrule
                $v_{cen}$ [\kms] & -248 (3) & -140 (2) &  64 (2) & 114 (2) & 209 (2) \\
                $\sigma_v$ [\kms] & 33 (3) & 21 (2) &  8.0 (1.7) & 13 (2) & 20 (3) \\
                $S_{\rm CO(1-0)}dv$ [Jy~\kms] & 1.21 (0.13) & 0.84 (0.10) & 0.22 (0.06) & 0.39 (0.08) & 0.54 (0.09) \\
                $L^{\prime}_{\rm CO(1-0)}$ [10$^7$ K km s$^{-1}$ pc$^2$] & 1.9 (0.2) & 1.3 (0.2) & 0.34 (0.09) & 0.60 (0.12) & 0.83 (0.15) \\
                $M_{mol}$$^{\dag}$ [10$^7$~M$_{\odot}$] & 4 (2) & 3 (2) & 0.7 (0.4) &  1.3 (0.8) & 1.7 (1.0) \\   
                R [kpc] & 0.9 (0.2) & 1.2 (0.2) & 1.8 (0.2) & 1.8 (0.2) & 1.0 (0.2) \\
                \midrule
                \multicolumn{6}{c}{\it Medium Region} \\
                \midrule 
                Parameter & Comp. 1 & Comp. 2 & Comp. 3 &  &  \\
                \midrule
                                $v_{cen}$ [\kms] & -233 (2)     & -146.1 (1.9) & 207 (3) & & \\
                                $\sigma_v$ [\kms] & 34 (2) & 20 (2) &  21 (3) & & \\
                                $S_{\rm CO(1-0)}dv$ [Jy~\kms] & 2.26 (0.18) & 1.13 (0.13) &  0.55 (0.12) & & \\
                                $L^{\prime}_{\rm CO(1-0)}$ [10$^7$ K km s$^{-1}$ pc$^2$]  & 3.5 (0.3)  & 1.7 (0.2) & 0.8 (0.2) & & \\
                                $M_{mol}$$^{\dag}$ [10$^7$~M$_{\odot}$] &7 (4)   & 4 (2)  & 1.8 (1.1) & & \\
                                R [kpc] & 0.7 (0.2) & 1.4 (0.2) &  1.5 (0.2) & & \\
                \bottomrule
        \end{tabular}

        \begin{flushleft}
                $^{\dag}$ Computed assuming $\alpha_{\rm CO} = 2.1 \pm 1.2$~M$_{\odot}$~(K~\kms~pc$^2$)$^{-1}$ \citep{2018Cicone2}.
        \end{flushleft}
\end{table*}
 
 \end{appendix}

\end{document}